\documentclass[longauth]{aa}

\usepackage[colorlinks=true, linkcolor=blue, citecolor=blue, filecolor=blue, urlcolor=blue]{hyperref}

\usepackage{txfonts}
\usepackage[T1]{fontenc}

\usepackage{caption}
\usepackage{subcaption}

\usepackage{graphicx}
\usepackage{natbib}
\bibpunct{(}{)}{;}{a}{}{,} 

\usepackage{amsmath,color,xspace,bigints,comment}
\usepackage{pdflscape}

\newcommand\textlcsc[1]{\textsc{\MakeLowercase{#1}}}

\newcommand{\angstrom}{\textup{\AA}\xspace}

\newcommand{\hg}{H$\gamma$\xspace}
\newcommand{\hb}{H$\beta$\xspace}
\newcommand{\oiiiA}{[OIII]$\lambda4959$\xspace}
\newcommand{\oiiiB}{[OIII]$\lambda5007$\xspace}
\newcommand{\niiA}{[NII]$\lambda6548$\xspace}
\newcommand{\ha}{H$\alpha$\xspace}
\newcommand{\niiB}{[NII]$\lambda6584$\xspace}
\newcommand{\siiA}{[SII]$\lambda6716$\xspace}
\newcommand{\siiB}{[SII]$\lambda6731$\xspace}
\newcommand{\siii}{[SIII]$\lambda9532$\xspace}
\newcommand{\hei}{HeI$\lambda10829$\xspace}
\newcommand{\siiAB}{[SII]$\lambda\lambda6716,6731$\xspace}
\newcommand{\oiiiAB}{[OIII]$\lambda\lambda4959,5007$\xspace}
\newcommand{\niiAB}{[NII]$\lambda\lambda6548,6584$\xspace}

\newcommand{\niiMAYBE}{[NII]$\lambda5755$\xspace}

\newcommand{\lmfit}{\textlcsc{lmfit}\xspace}
\newcommand{\PAL}{\textlcsc{PyAutoLens}\xspace}

\begin{document}

\title{GA-NIFS: JWST/NIRSpec IFU observations of HFLS3 reveal a dense galaxy group at $z\sim6.3$}
\author{
Gareth C. Jones\inst{1}\thanks{E-mail: gareth.jones@physics.ox.ac.uk}
\and
Hannah \"{U}bler\inst{2,3}\and
Michele Perna\inst{4}\and
Santiago Arribas\inst{4}\and
Andrew J. Bunker\inst{1}\and
Stefano Carniani\inst{5}\and
Stephane Charlot\inst{6}\and
Roberto Maiolino\inst{2,3,7}\and
Bruno Rodríguez Del Pino\inst{4}\and
Chris Willott\inst{8}\and
Rebecca A. A. Bowler\inst{9}\and
Torsten B\"{o}ker\inst{10}\and
Alex J. Cameron\inst{1}\and
Jacopo Chevallard\inst{1}\and
Giovanni Cresci\inst{11}\and
Mirko Curti\inst{12}\and
Francesco D'Eugenio\inst{2,3}\and
Nimisha Kumari\inst{13}\and
Aayush Saxena\inst{1,7}\and
Jan Scholtz\inst{2,3}\and
Giacomo Venturi\inst{5}\and
Joris Witstok\inst{2,3}
}

\institute{
Department of Physics, University of Oxford, Denys Wilkinson Building, Keble Road, Oxford OX1 3RH, UK\label{1}
\and
Kavli Institute for Cosmology, University of Cambridge, Madingley Road, Cambridge CB3 0HA, UK\label{2}
\and
Cavendish Laboratory, University of Cambridge, 19 JJ Thomson Avenue, Cambridge CB3 0HE, UK\label{3}
\and
Centro de Astrobiolog\'{i}a (CAB), CSIC–INTA, Cra. de Ajalvir Km.~4, 28850- Torrej\'{o}n de Ardoz, Madrid, Spain\label{4}
\and
Scuola Normale Superiore, Piazza dei Cavalieri 7, I-56126 Pisa, Italy\label{5}
\and
Sorbonne Universit\'{e}, CNRS, UMR 7095, Institut d'Astrophysique de Paris, 98 bis bd Arago, 75014 Paris, France\label{6}
\and
Department of Physics and Astronomy, University College London, Gower Street, London WC1E 6BT, UK\label{7}
\and
NRC Herzberg, 5071 West Saanich Rd, Victoria, BC V9E 2E7, Canada\label{8}
\and
Jodrell Bank Centre for Astrophysics, Department of Physics and Astronomy, School of Natural Sciences, The University of Manchester, Manchester, M13 9PL, UK\label{9}
\and
European Space Agency, c/o STScI, 3700 San Martin Drive, Baltimore, MD 21218, USA\label{10}
\and
INAF - Osservatorio Astrofisco di Arcetri, largo E. Fermi 5, 50127 Firenze, Italy\label{11}
\and
European Southern Observatory, Karl-Schwarzschild-Strasse 2, 85748 Garching, Germany\label{12}
\and
AURA for European Space Agency, Space Telescope Science Institute, 3700 San Martin Drive. Baltimore, MD, 21210, USA\label{13}
}

\date{Received X / Accepted Y}

\abstract
{Massive, starbursting galaxies in the early Universe represent some of the most extreme objects in the study of galaxy evolution. One such source is HFLS3 ($z\sim6.34$), which was originally identified as an extreme starburst galaxy with mild gravitational magnification ($\mu\sim2.2$). Here, we present new observations of HFLS3 with the JWST/NIRSpec IFU in both low (PRISM/CLEAR; $R\sim100$) and high spectral resolution (G395H/290LP; $R\sim2700$), with high spatial resolution ($\sim0.1''$) and sensitivity. Thanks to the combination of the NIRSpec data and a new lensing model with accurate spectroscopic redshifts, we find that the $3''\times3''$ field is crowded, with a lensed arc (C, $z=6.3425\pm0.0002$), two galaxies to the south (S1 and S2, $z=6.3592\pm0.0001$), two galaxies to the west (W1, $z=6.3550\pm0.0001$; W2, $z=6.3628\pm0.0001$), and two low-redshift interlopers (G1, $z=3.4806\pm0.0001$; G2, $z=2.00\pm0.01$). We present spectral fits and morpho-kinematic maps for each bright emission line (e.g., \oiiiB, \ha, \niiB) from the R2700 data for all sources except G2 (whose spectral lines fall outside the observed wavelengths of the R2700 data). From a line ratio analysis, the galaxies in component C are likely powered by star formation, while we cannot rule out or confirm the presence of AGN in the other high-redshift sources. We perform gravitational lens modelling, finding evidence for a two-source composition of the lensed central object and a comparable magnification factor ($\mu=2.1-2.4$) to previous work. The projected distances and velocity offsets of each galaxy suggest that they will merge within the next $\sim1$\,Gyr. Finally, we examine the dust extinction-corrected SFR$_{\rm H\alpha}$ of each $z>6$ source, finding that the total star formation ($510\pm140$\,M$_{\odot}$\,yr$^{-1}$, magnification-corrected) is distributed across the six $z\sim6.34-6.36$ objects over a region of diameter $\sim11$\,kpc. Altogether, this suggests that HFLS3 is not a single starburst galaxy, but instead is a merging system of star-forming galaxies in the Epoch of Reionisation.}


\maketitle

\section{Introduction}\label{intro}
Observations have revealed that the mode of galaxy evolution in the first $2$\,Gyr of the Universe ($z\gtrsim3$) was drastically different from that of following epochs. This can be seen in the similar evolution of the global star formation rate density (SFRD) and molecular gas mass, both of which increased with cosmic time during this epoch and decreased for $z<1$ (e.g., \citealt{bouw15,deca19}). At higher redshifts, the major merger rate is higher (e.g., \citealt{duca19}), as well as the mean star formation rate (SFR) and black hole accretion rate (BHAR) for a given stellar mass (e.g., \citealt{spea14,yang18,pope23}). Together, these results suggest that galaxies in the first 2\,Gyr formed rapidly, accreting gas via filaments and mergers, resulting in the buildup of stellar and black hole mass on shorter timescales than the following epochs.

The unique environment of the early Universe allowed galaxies to start forming stars quickly, resulting in high-redshift galaxies that have exhausted their gas supply and have already stopped forming stars by $z\gtrsim3$ (e.g., see recent JWST results of \citealt{loos23,stra23,carn23}). While this quenched appearance could be caused by a temporary minimum in a stochastic star formation history (e.g., \citealt{arat20,dome23}), there have also been a number of $z>4$ hyper-luminous infrared galaxies (HyLIRGs) detected with large infrared luminosities ($L_{IR}>10^{13}$\,L$_{\odot}$) and SFRs of $\gtrsim10^3$\,M$_{\odot}$\,yr$^{-1}$ (e.g., \citealt{wagg14,vene19,carn19,riec20,chen20}) that could deplete gas reservoirs rapidly. The elevated SFR of these sources was believed to be the result of ongoing hierarchical merging (e.g., \citealt{hopk06}), although the discovery of ordered rotation in some HyLIRGs (e.g., \citealt{tsuk21}) suggests that they could be fuelled by secular accretion.

One of the most extreme high-redshift HyLIRGs is 1HERMES S350 J170647.8+584623, or HFLS3. This source lies in the Spitzer First Look Survey (FLS; \citealt{fadd04}) extragalactic field (a four square degree field centred on 17h18m00s +59$^{\circ}30'00.0''$), and was included in the Herschel Multi-tiered Extragalactic Survey (HerMES; \citealt{oliv12}). Based on very red Herschel/SPIRE colours ($S_{500\mu m}>S_{350\mu m}>S_{250\mu m}$ and $S_{500\mu m}/S_{350\mu m}>1.3$), \citet{riec13} identified this source as a high-redshift dusty galaxy with a nearby low-redshift companion to the north (G1B, $z\sim2.092$ based on detection of CIV$\lambda1549$ with Keck/LRIS). A comprehensive suite of observations (e.g.; Plateau de Bure Interferometer, PdBI; the Very Large Array, VLA; Keck; William Herschel Telescope) enabled the creation of a NIR-radio spectral energy distribution (SED) and the measurement of a number of emission lines for HFLS3 ($z_{spec}=6.3369\pm0.0009$). Based on their analysis of the line and continuum emission, \citet{riec13} report a high mass of molecular gas ($\rm M_{H_2}=(1.04\pm0.09)\times10^{11}/\alpha_{CO}$\,M$_{\odot}$), dust ($\rm M_{dust}=1.31^{+0.32}_{-0.30}\times10^9$\,M$_{\odot}$), and stars ($\rm M_*\sim3.7\times10^{10}$\,M$_{\odot}$), as well as a high FIR luminosity ($\rm L_{FIR}$=$2.86^{+0.32}_{-0.31}\times10^{13}$\,L$_{\odot}$) and SFR$_{\rm FIR}\sim2900$\,M$_{\odot}$\,yr$^{-1}$ and evidence for a velocity gradient from PdBI [CII]158\,$\mu$m\footnote{The values reported by \citet{riec13} assume no gravitational lensing.}. Some of the spectral lines were asymmetric, which \citet{riec13} interpret as evidence for a possible close separation merger. However, some of the data were taken at low resolution ($>2''$) and thus did not allow for detailed source differentiation and characterisation.

Follow-up Keck/NIRC2 and Hubble Space Telescope (HST) WFC3 imaging by \citet{coor14} revealed the presence of three close-by companions: two to the north (G1 \& G2) and one to the south (R1). The northern sources were assumed to be at the same redshift ($z\sim2.1$), while a photometric redshift of $z_{phot}\sim6$ was found for R1. Gravitational lens modelling confirmed that HFLS3 may be composed of two sources, which are slightly magnified ($\mu\sim2.2\pm0.3$) by the foreground sources G1 \& G2. 

The idea that the starburst nature of HFLS3 could be caused by a history of major mergers in a dense environment inspired multiple searches for an overdensity of galaxies. \citet{robs14} used JCMT/SCUBA-2 to perform a low-resolution ($\sim14''$), wide-area (67\,arcmin$^2$) search for significant submm emission at two wavelengths ($450\,\mu$m and $850\,\mu$m), but found no overdensity. This was followed by a search at optical wavelengths with HST and GTC \citep{lapo15}, finding no large-scale overdensity but a possible small-scale ($\sim36$\,kpc) overdensity of faint objects.

These observations suggest that HFLS3 is a gas-rich, starbursting, dusty galaxy with a velocity gradient (suggesting merging activity) when the Universe was only $\sim850$\,Myr old. Because of these exceptional properties, it was chosen as a target for the JWST/NIRSpec GTO programme `Galaxy Assembly with NIRSpec IFS' (GA-NIFS; PI: S. Arribas \& R. Maiolino). This programme aims to observe (in cycles 1 and 3) a sample of 55\,galaxies at $z\sim3-11$ spanning a variety of types (e.g., QSOs, AGN, SFGs, strongly lensed galaxies, quenched sources, major mergers), to show the power of the integral field unit (IFU; \citealt{boke22}) on NIRSpec (\citealt{jako22}) for exploring resolved kinematics and gas properties. The survey is ongoing, and detailed studies of $z=3-7$ AGN and QSOs are now published (\citealt{mars23,uble23,pern23}).

Here, we present the JWST/NIRSpec IFU observations of HFLS3. The high resolution and sensitivity of these observations reveal a more complex system than implied by previous data, with six strongly detected sources at $z\simeq 6.3-6.5$ and two low-redshift interlopers ($z\sim2.0-3.5$) within a $3''\times3''$ field. We present the details of our dataset in Section \ref{ifudes}, and characterise the field in Section \ref{fieldchar}. Section \ref{analysis} contains further analysis (i.e., morpho-kinematic maps, gravitational lens modelling, line ratio-based excitation conditions, star formation rate derivation, and galaxy merger discussion), and we conclude in Section \ref{conc}. 

We use a standard concordance cosmology ($h_o$, $\Omega_m$, $\Omega_{\Lambda}$ = 0.7, 0.3, 0.7) throughout, where $1''$ corresponds to $\sim$5.5\,kpc at $z\sim6.34$, $\sim$7.3\,kpc at $z\sim3.48$, and $\sim$8.3\,kpc at $z\sim2.1$. To match the notation of other works (e.g., \citealt{curt20}), emission lines are named based on their air wavelength, while we use their vacuum wavelengths for analysis (e.g., $\lambda_{[OIII]\lambda5007}=5008.24\angstrom$). We adopt a Salpeter stellar initial mass function \citep[IMF;][]{salp55}.

\section{JWST/NIRSpec IFU data description}\label{ifudes}

HFLS3 was observed with the JWST/NIRSpec IFU using two disperser/filter combinations (G395H/290LP, hereafter R2700; and PRISM/CLEAR, hereafter R100) on 1 September 2022 as part of PID 1264 (PI: L. Colina; see Table \ref{JWST_spec}). Both settings used the improved reference sampling and subtraction pattern IRS$^2$ of \citet{raus17} and adopted a 4-point dither with a medium `cycling' pattern and a starting point of 1.

\begin{table*}
    \caption{JWST NIRSpec/IFU observation properties.}
    \centering
    \begin{tabular}{c|ccccc}
Grating/Filter & Readout      & Groups/Int & Ints/Exp  & Exposures  & Total Time\\
               & Pattern      &            &           &  & [s] \\ \hline
G395H/290LP    & NRSIRS2      & 25         & 1         & 4 & 7352.801 \\
PRISM/CLEAR    & NRSIRS2RAPID & 60         & 1         & 4 & 3559.689\\ \hline
    \end{tabular}
    \label{JWST_spec}
\end{table*}

We use v1.8.2 of the GTO pipeline with CRDS context 1105 to create R2700 and R100 cubes with drizzle weighting (\citealt{fruc02}). This context accounts for flux leakage from 68\,failed open shutters in the multi-shutter array (MSA; $\sim0.03\%$ of the full MSA) into the IFU aperture. A patch was included to correct some important bugs affecting this specific pipeline version (see details in \citealt{pern23}). We have corrected count-rate images for 1/f noise through a polynomial fit. During stage 2, we have removed all data in regions of known failed open MSA shutters. We have also masked pixels at the edge of the slices (one pixel wide) to conservatively exclude pixels with unreliable flat-field reference files for the spectrograph optics (i.e., SFLAT files), and implement the outlier rejection of \cite{DEugenio2023}.

A recent in-depth investigation of the NIRSpec IFU point spread function (PSF) by \citet{DEugenio2023} found that it may be non-circular. Using a standard star, they determine that the major axis FWHM is wavelength dependent (FWHM$_{\rm major}\sim0.12-0.20''$ between $\lambda_{\rm obs}=3.0-5.0$\,$\mu$m). The combination of a dither and drizzle weighting allowed us to sub-sample the detector pixels, resulting in cube spaxels of $0.05''$. The resolving power of the R2700 data increases from $\rm R_{2700}\sim2000-3500$ from $\lambda_{\rm obs}=3.0-5.0$\,$\mu$m, while the resolving power of the R100 data decreases from $\rm R_{100}\sim90-30$ over $\lambda_{\rm obs}=0.5-1.2$\,$\mu$m and increases from $\rm R_{100}\sim30-290$ over $\lambda_{\rm obs}=1.2-5.0$\,$\mu$m.

\subsection{Astrometric correction}\label{astromet}

A comparison of the JWST/NIRSpec IFU data with archival HST data\footnote{HST images from Programme 13045 (PI: A. R. Cooray) were retrieved from the MAST archive.} revealed a positional offset between images at comparable wavelengths. Astrometric errors like this are currently common in NIRSpec IFU data at the time of analysis (e.g., \citealt{wyle22,pern23}). Since we wish to estimate spatially resolved quantities, it is crucial to have each image aligned to a common reference frame. By shifting to the Gaia DR3 frame, we find an offset of $0.17\pm0.07''$ (see Appendix \ref{astrover} for details).

\subsection{Background subtraction}

Since these observations did not include a dedicated background exposure, no background subtraction was performed by the pipeline. To estimate the background emission, we extract a mean spectrum from the R100 and R2700 data cubes using a 25\,spaxel ($12.5''$) diameter aperture (MPDAF task \textit{aperture}; \citealt{baco16}) from a signal-free region to the southeast. 

We assume that the background emission is uniform across the field of view (FoV) and create a correction by subtracting this background spectrum from the spectrum of each spaxel. Note that for the R2700 spectrum, we only include channels that are not affected by chip gap/edge issues ($\lambda_{obs}\approx 2.87057-3.9778$, $4.2078-5.26643\,\mu$m).

\section{Source distribution}\label{fieldchar}
Previous analyses of the HFLS3 field found evidence for multiple sources (\citealt{riec13,coor14}): the primary lensed starburst (HFLS3, $z\sim6.34$), two low-redshift sources to the north (G1 and G2, reported to be $z\sim2.0$), and a fourth source only detected in rest-frame UV emission to the southeast (R1). Initial exploration of the IFU data cubes shows evidence for a complex distribution of flux with multiple spatial components. It is not clear \textit{a priori} if these are the same as previously detected, so we proceed with an uninformed search for emission. 

\subsection{Field characterisation}\label{fieldmor}
\begin{figure}
\centering
\includegraphics[width=0.49\textwidth]{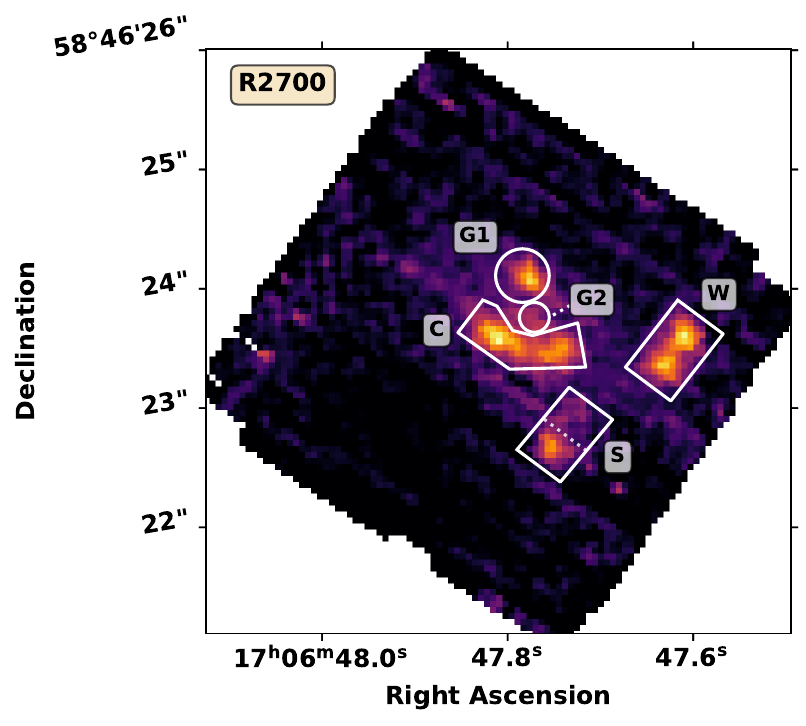}
\includegraphics[width=0.49\textwidth]{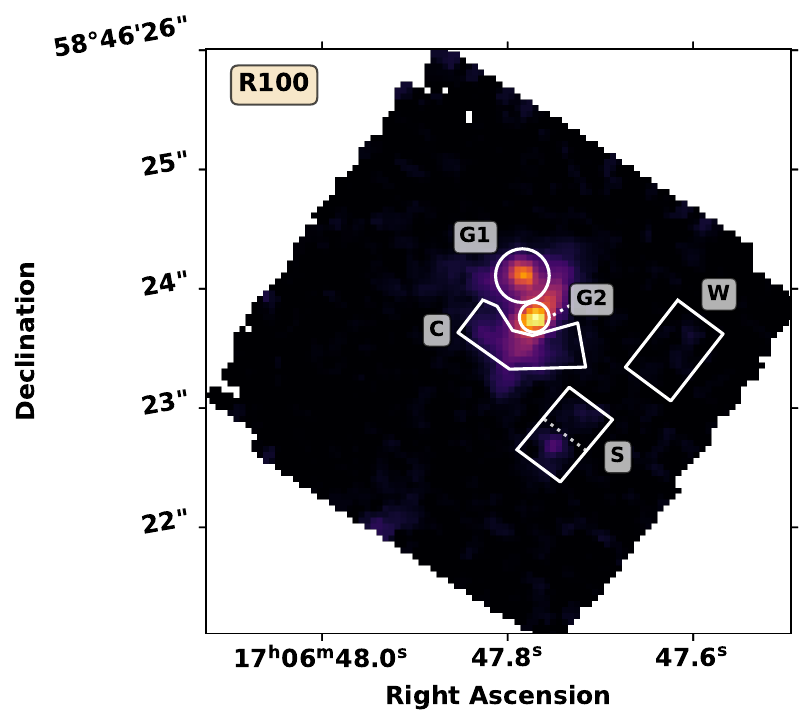}
\caption{Integrated emission of the HFLS3 field, using two illustrative wavelength ranges: redshifted \ha for $z\sim6.34$ for the R2700 cube ($\lambda_{obs}=4.79954\,-\,4.84467\,\mu$m, left panel) and the approximate wavelength range of HST/WFC3 F160W for the R100 cube ($\lambda_{obs}=1.4\,-\,1.6\,\mu$m; right panel). The adopted source masks are shown with white boundaries. North is up and east is to the left. The right panel is nearly identical to the observed HST/WFC3 F160W image, but at higher spatial resolution (see Fig. \ref{spatoff}).}
\label{white}
\end{figure}

As an example of the complex source distribution, we present an integrated map of the R2700 cube over the wavelength range that contains \ha redshifted to $z=6.34$ (i.e., $\lambda_{obs}=4.79954\,-\,4.84467\,\mu$m, or $z_{H\alpha}\sim6.31-6.38$; left panel in Fig. \ref{white}). There are four separate areas of emission: a bright source to the north, an east-west arc near the centre, a southern source, and a double-lobed source to the west. 

If we instead integrate the R100 cube over a wavelength range similar to the HST/WFC3 F160W transmission (i.e., $\lambda_{obs}=1.4\,-\,1.6\,\mu$m), the distribution is drastically changed (Fig. \ref{white}, right panel). Note that this distribution is identical to that seen in previous HST/WFC3 F160W observations (see Fig. \ref{spatoff}), but at higher spatial resolution. The northern source is still present, and there is an extended object with a bright core between this source and the central arc (as seen in the HST images; \citealt{coor14}). The other three sets of sources are much weaker.

From these maps, we define five regions (see Fig. \ref{white}): the bright northern component (`G1'), the core of the extended object in the R100 data (`G2'), the arc (`C'), the southern sources (`S'), and the western sources (`W'). The regions G1 and G2 are identical to those of \citet{coor14}, while the C region contains the HFLS3 source. The previously found source R1 is the brightest region within the `S' mask, but we use a larger area to encompass nearby emission. \citet{lapo15} identified the S and W sources as possible faint sources (ID2 and ID3, respectively) in HST maps. We also note that weaker line emission is detected in the field of view, which we explore in Appendix \ref{candsec}. 

\subsection{R2700 integrated spectra}\label{intspecsec}

To investigate the distribution of flux in this field, we extract spectra from the R2700 cube using each of the spatial masks (see top panel of Fig. \ref{white}) except `G2', which will be examined in the next subsection. These spectra are fit with multiple pseudo-Voigt profiles (one pseudo-Voigt profile per line). This profile is an adjustable combination of a Gaussian and Lorentzian, which represent line broadening due to pressure and Doppler (or thermal) motion, respectively (\citealt{arms67}; see Appendix \ref{morevoigt} for profile details). All lines are fit simultaneously using \lmfit with a least-squared minimisation, assuming a single redshift for most lines and a power law continuum normalised to the $4\,\mu$m value:
\begin{equation}\label{conteq}
\rm F(\lambda_{obs})=F_{4\mu m} \left(\frac{\lambda_{obs}}{4\mu m}\right)^{\alpha}
\end{equation}
The models are convolved with the line spread function (LSF)\footnote{As recorded in the JWST documentation; \url{https://jwst-docs.stsci.edu/jwst-near-infrared-spectrograph/nirspec-instrumentation/nirspec-dispersers-and-filters}} before comparison to the data. The velocity widths of line pairs (i.e., \siiAB, \niiAB, \oiiiAB, and each Balmer line) are fixed to be identical. We adopt the standard assumptions of \niiB/\niiA$=2.94$ (e.g., \citealt{dojc23}) and \oiiiB/\oiiiA$=2.98$ (e.g., \citealt{dimi07}). Following other NIRSpec IFU investigations (\citealt{uble23,pern23}), we rescale the error spectrum generated by the calibration pipeline (`ERR') using the standard deviation of the line-free regions in the observed spectrum. Uncertainties on each parameter are estimated using the standard error output from \lmfit. Only lines with $>1\sigma$ emission at the expected centroid wavelength are fit.

The best-fit spectra are shown in Fig. \ref{spec1}, while the redshifts and continuum fluxes for each source are presented in Table \ref{redtable} and the line properties are listed in Table \ref{specfittab}.

With a redshift of $z_{G1}=3.4805\pm0.0001$, the northern galaxy `G1' is well-detected in \ha, \niiAB, \siiAB, \siii, and \hei. Because \hei may be resonantly scattered (e.g., \citealt{rudy89,bell23}), we allow the centroid of this line to vary (resulting in an offset of $137\pm19$\,km\,s$^{-1}$). In addition, \siii may be blended with Pa-$\epsilon$ (e.g., \citealt{kehr06}), so we allow the model of this line to feature a velocity offset ($54\pm14$\,km\,s$^{-1}$). The best-fitting model shows minor ($\sim1\sigma$) residuals for the \ha complex, possibly suggesting the presence of an additional galaxy or outflow.

The central component `C' ($z_C=6.3425\pm0.0002$) features well-detected \ha, \hb, \oiiiB, and \niiB emission with broad profiles (FWHM$\sim500-700$\,km\,s$^{-1}$). We find evidence for weak \oiiiA, \niiA, \siiAB, with no \hg. The \siiAB emission is broad and low-level, so fits with a free FWHM returned unphysical values ($\sim1500$\,km\,s$^{-1}$). To better constrain these lines, we fixed the width of each to be identical to that of \ha. The $4\,\mu$m continuum is strong but features no significant slope in $\rm F_{\lambda}$. While the best-fit model spectrum does not show significant residuals (Fig. \ref{spec1}), we note that this component is found to feature two discrete sources with a velocity offset (see Section \ref{whatisc}) with a velocity separation smaller than our LSF. With this in mind, the large linewidth is due to the blending of the two galaxies.

The integrated spectra of the S and W components are quite similar to each other, with very strong detections in \hg, \hb, \oiiiAB, and \ha. 
But while the S emission is well-fit by a single redshift ($z_S=6.3592\pm0.0001$), the double-peaked appearance of the W component suggests two galaxies ($z_{W1}=6.3549\pm0.0001$ and $z_{W2}=6.3627\pm0.0001$, with a velocity difference of $368\pm7$\,km\,s$^{-1}$) that may each be characterised directly from the integrated spectrum. No significant \niiAB or \siiAB emission is detected in the integrated W spectrum.

While component S is well-fit by a single velocity component, we find that it is composed of two spatially separated sources at the same redshift (see Section \ref{mommap}). To investigate each component (S1 and S2), we create two sub-masks (see dashed line in Fig. \ref{white}) and fit the integrated spectra of each. S1 is brighter in line emission (with significant \niiAB emission), but the continuum value of each component is approximately equal.

\begin{figure*}
\centering
\begin{subfigure}{\textwidth}
\includegraphics[width=\textwidth]{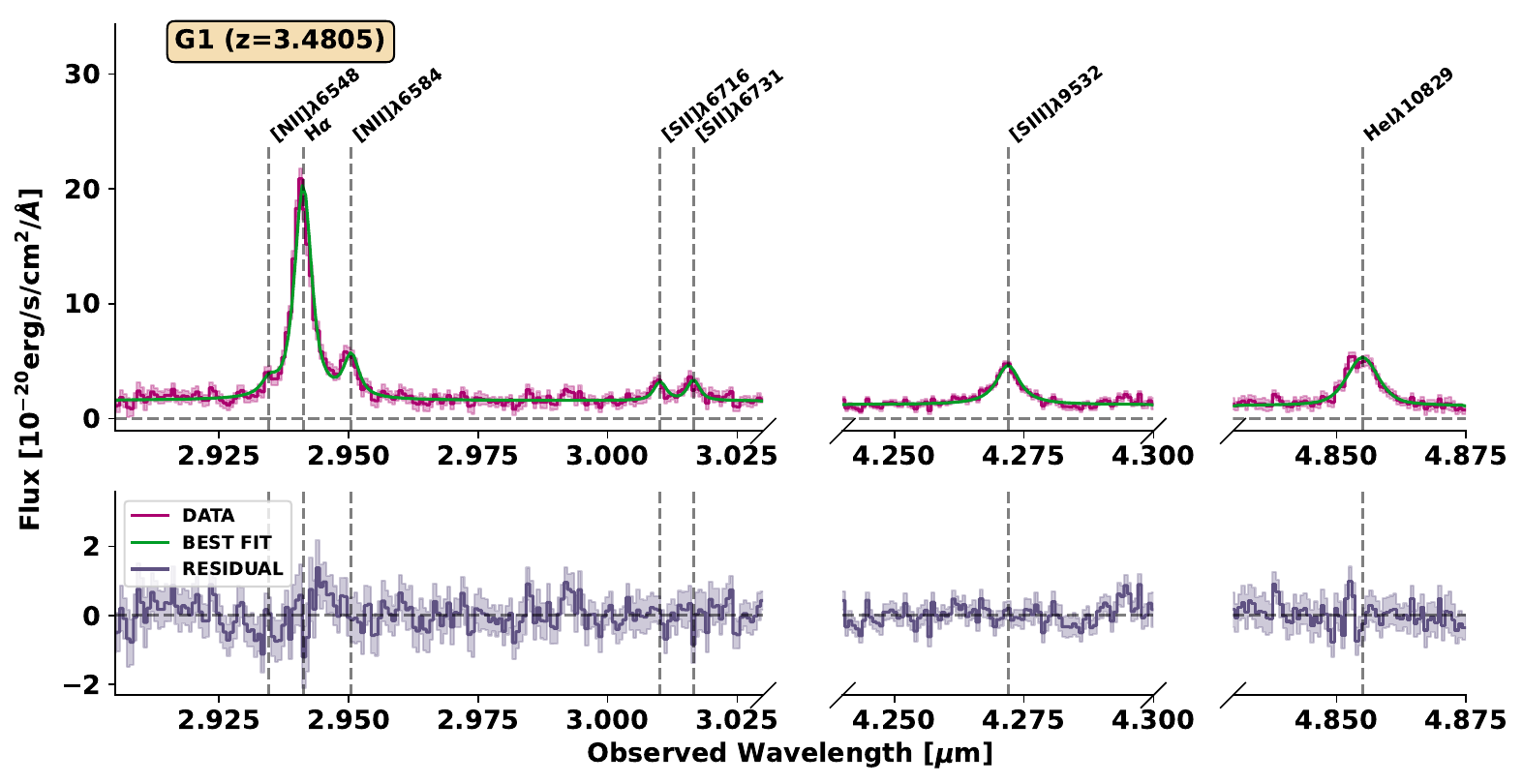}
\includegraphics[width=\textwidth]{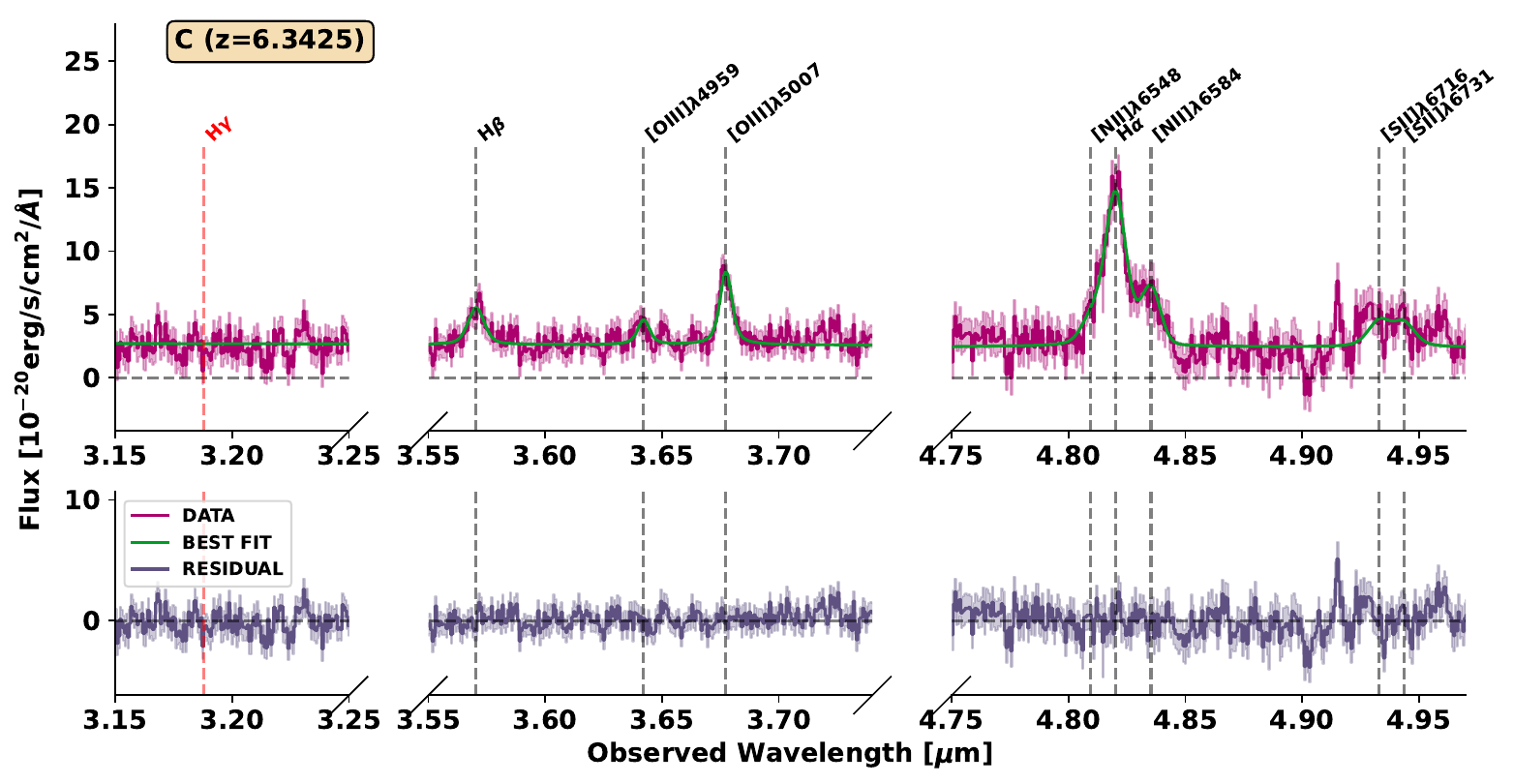}
\end{subfigure}
\caption{Integrated spectra of the R2700 cube using the masks of Fig. \ref{white} (magenta; excluding G2), with $1\sigma$ errors from the associated error spectrum shown as shaded region. Best-fit models (line emission and continuum) are shown by green lines. The centroids of each line are depicted by dashed lines, with red lines indicating that the spectral line was not fit. The lower panel shows the residual.}
\label{spec1}
\end{figure*}
\begin{figure*}\ContinuedFloat
\begin{subfigure}{\textwidth}
\includegraphics[width=\textwidth]{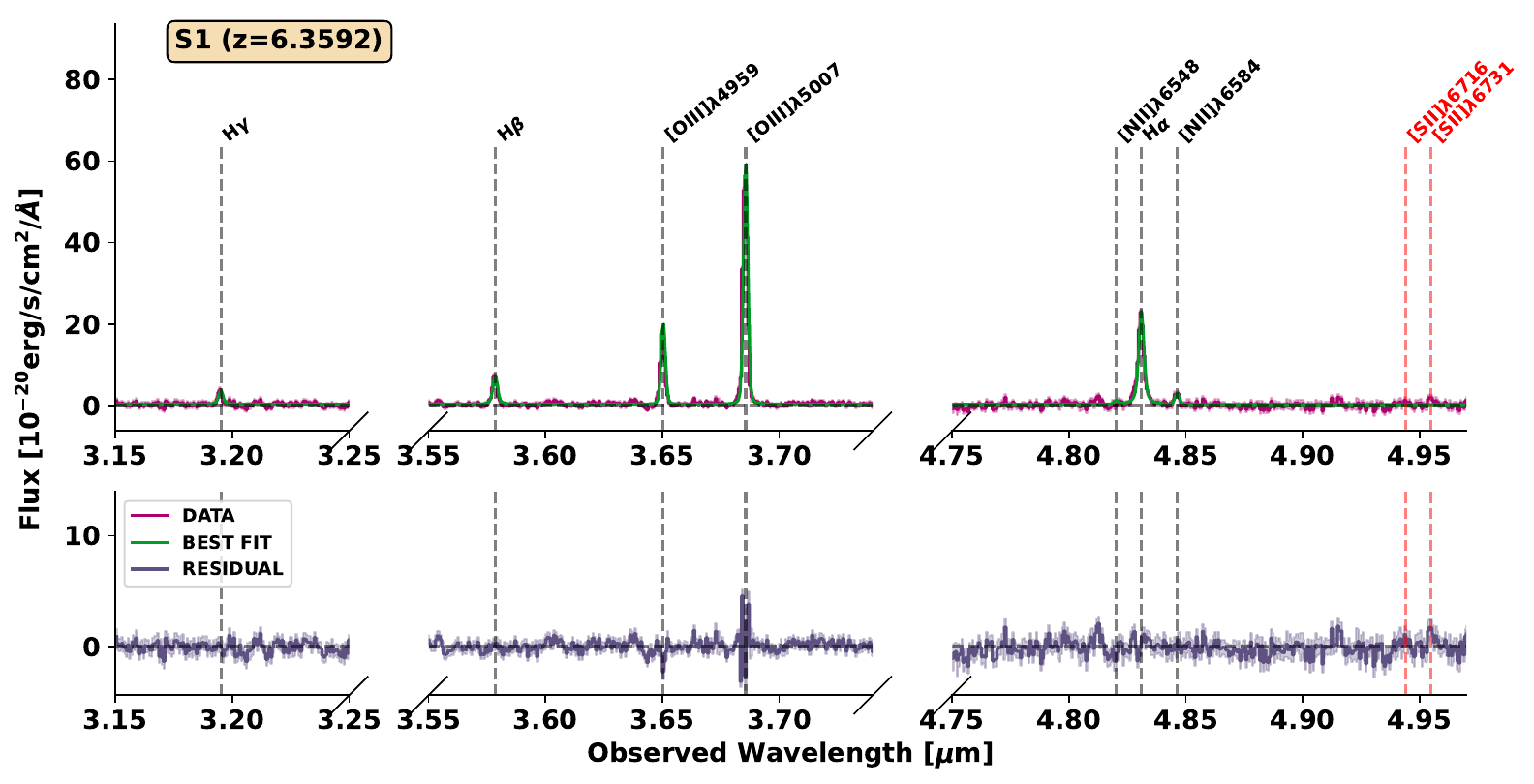}
\includegraphics[width=\textwidth]{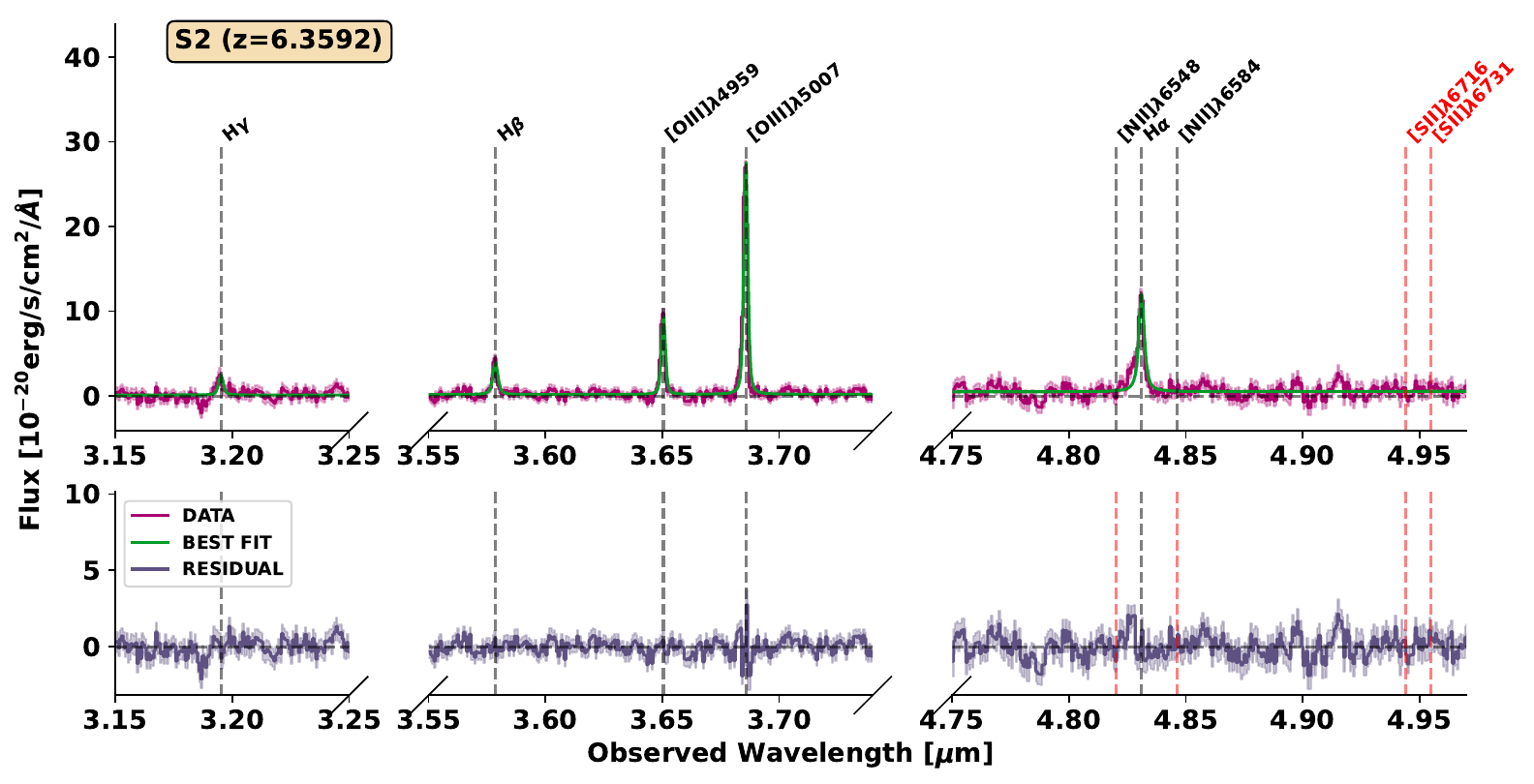}
\end{subfigure}
\caption{(Cont.)}
\end{figure*}
\begin{figure*}\ContinuedFloat
\begin{subfigure}{\textwidth}
\includegraphics[width=\textwidth]{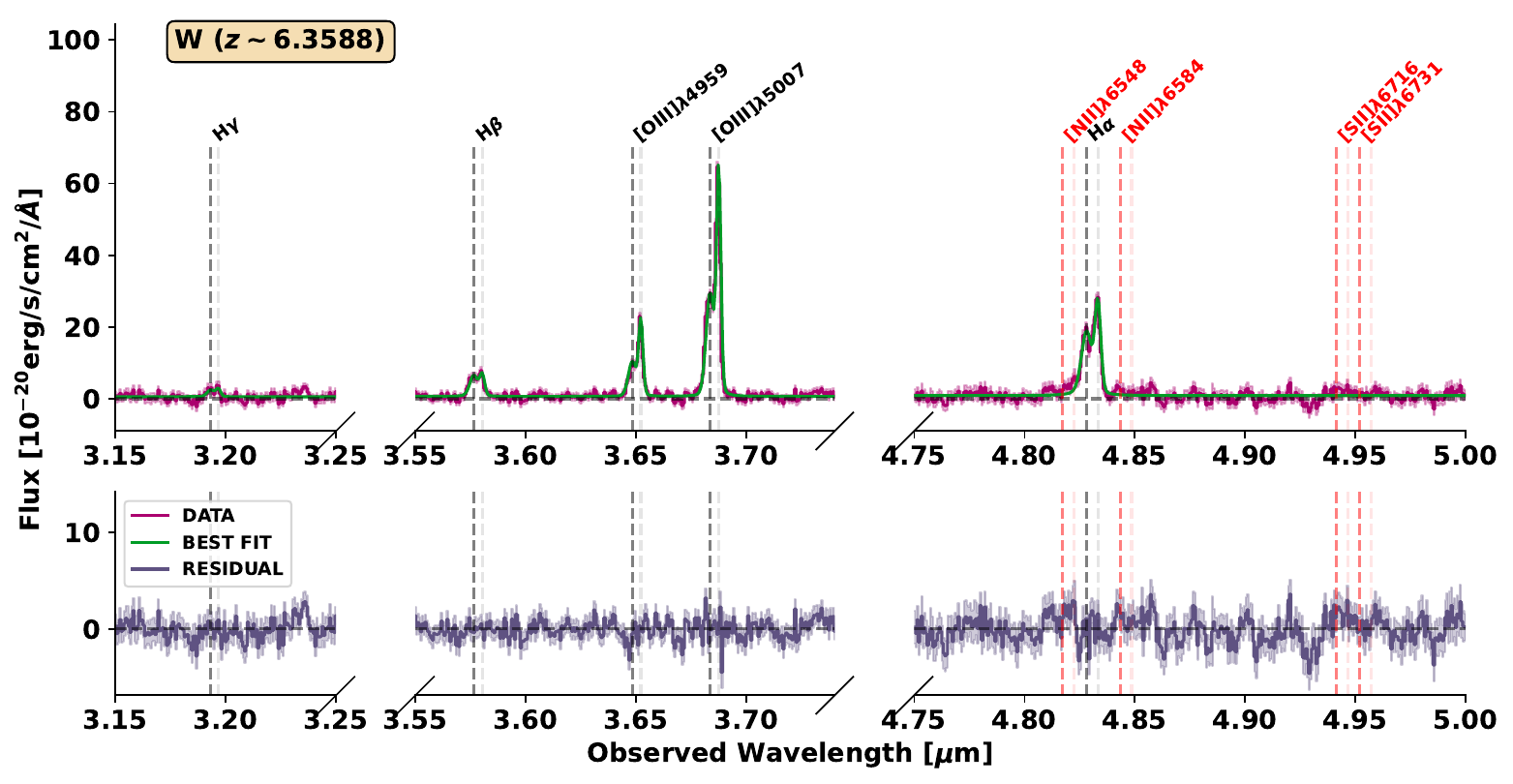}
\end{subfigure}
\caption{(Cont.)}
\end{figure*}

\begin{table}
\caption{Best-fit redshift and continuum flux levels for each of the HFLS3 field components.}
\centering
\begin{tabular}{c|ccc}
	&	Redshift	        &	$4\,\mu$m Continuum Flux & Spectral	\\ 
        &                       &   [$10^{-20}$\,erg\,s$^{-1}$\,cm$^{-2}$\,\angstrom$^{-1}$] & Slope \\\hline
G1	&	$3.4805\pm0.0001$	&	$1.26\pm0.02$	&	$-0.7\pm0.1$	\\ 
G2	&	$2.00\pm0.01$	&	-	&	-	\\
C	&	$6.3425\pm0.0002$	&	$2.54\pm0.06$	&	$-0.3\pm0.1$	\\
S1	&	$6.3592\pm0.0001$	&	$0.25\pm0.03$	&	$-0.9\pm0.8$	\\
S2	&	$6.3593\pm0.0001$	&	$0.25\pm0.02$	&	$3.8\pm0.5$	\\
W1	&	$6.3549\pm0.0001$	&	$0.69\pm0.05$	&	$1.1\pm0.4$	\\
W2	&	$6.3627\pm0.0001$	&	---	&	---	\\ \hline
\end{tabular}
\tablefoot{Fits derived through a multi-line and continuum simultaneous fit of R2700 spectra (except for G2, which used the R100 spectrum). Spectral slope given for $F_{\lambda}(\lambda_{obs})$ (see equation \ref{conteq}). Since the results of $\rm W1$ and $\rm W2$ come from the same fit, the best-fit continuum properties listed for $\rm W1$ describe the combined continuum.}
\label{redtable}
\end{table}

\clearpage
\onecolumn
\begin{landscape}

\begin{table*}
\caption{Best fit FWHM and integrated fluxes of observed spectral lines.}
\centering
\begin{tabular}{cc|cccccccccccc}
   &                                                        & H$\gamma$ 							   & H$\beta$ 			 & [OIII]$\lambda$4959 					  & [OIII]$\lambda$5007 				   & [NII]$\lambda$6548      			   & H$\alpha$              			  & [NII]$\lambda$6584      			  & [SII]$\lambda$6716      			   & [SII]$\lambda$6731      & [SIII]$\lambda$9532     & HeI$\lambda$10829\\ \hline

G1  & FWHM  & $-$   										& $-$	   							  & $-$				 		  	 & $-$				 				   		  & $282\pm85$  						& $333\pm53$  					   & $282\pm85$  						 & $194\pm70$   				  	   & $194\pm70$  						 & $352\pm18$  			   & $432\pm39$\\
  & I 	  & $-$   										& $-$	   							  & $-$				 		  	 & $-$				 				   		  & $63\pm13$ & $1035\pm65$	&$184\pm39$	  &$64\pm14$	  & $71\pm19$	  &$286\pm11$ &$402\pm34$\\ \hline

C  & FWHM  & (500)				 		& $679\pm43$  				  & $476\pm53$  						 & $476\pm53$  					  & $571\pm101$  						& $679\pm43$  				   & $571\pm101$  						 & $679\pm43$  				   & $679\pm43$  						 & $-$ 					   & $-$ \\
  & I 	  & $<172$  & $310\pm37$ 	  & $165\pm14$ 	 & $492\pm42$ 	  & $144\pm28$ 		& $1747\pm190$	&$423\pm82$ 	 &$290\pm70$	& $265\pm68$	 & $-$ 						   & $-$ \\ \hline


S1  & FWHM  & $116\pm15$   						& $116\pm15$  			  & $119\pm5$  						 & $119\pm5$  					  & $98\pm58$				 		& $116\pm15$  					   & $98\pm58$				 		 & (150)				 	   & (150)				 		 & $-$ 						   & $-$ \\
  & I 	  & $95\pm14$ 		& $216\pm10$ 	  & $498\pm9$ 	 & $1484\pm26$ 	  & $22\pm6$ & $802\pm29$	& $64\pm16$ & $<61$ & $<64$ & $-$ 								   & $-$ \\ \hline

S2  & FWHM  & $125\pm14$   						& $125\pm14$  			  & $65\pm5$  						 & $65\pm5$  					  & (150)  					  & $125\pm14$  					   &  (150)  					  & (150)				 	   & (150)				 		 & $-$ 							   & $-$ \\
  & I 	  & $75\pm18$ 		& $130\pm10$ 	  & $235\pm7$ 	 & $701\pm20$ 	  & $<53$ 	  & $458\pm30$& $<44$	& $<55$ & $<64$ & $-$ 									   & $-$ \\ \hline

W1   & FWHM & $332\pm27$   						& $332\pm27$  			  & $324\pm14$  						 & $324\pm14$  					  & (300)				 		& $332\pm27$  					   & (300)				 		 & (300)				 	   & (300)				 		 & $-$ 						   & $-$ \\
   & I	& $92\pm29$ 		& $310\pm40$ 	  & $452\pm24$ 	 & $1348\pm73$ 	  & $<205$ & $1223\pm158$	& $<227$ & $<236$ & $<199$ & $-$ 								   & $-$ \\ \hline
W2   & FWHM & $172\pm16$  						& $172\pm16$ 			  & $135\pm8$ 						 & $135\pm8$ 					  & (200)				 		& $172\pm16$ 					   & (200)				 		 & (200)				 	   & (200)				 		 & $-$ 						   & $-$ \\
   & I	& $59\pm19$	& $169\pm26$ & $555\pm21$   & $1655\pm64$ & $<136$ & $891\pm120$ & $<151$ & $<157$ & $<133$ & $-$ 									   & $-$ \\ \hline \hline

X1  & FWHM & (150)				 		& $270\pm69$  					  & $131\pm28$  						 & $131\pm28$  					  & (150)  					  & $270\pm69$  					   &  (150)  					  & (150)				 	   & (150)				 		 & $-$ 							   & $-$ \\
  & I 	  & $<27$  & $39\pm12$ 	  & $30\pm3$ 	 & $88\pm9$ 	  & 54 	  & $324\pm53$	& $<51$	& $<59$ & $<55$ & $-$ 									   & $-$ \\ \hline

\end{tabular}
\tablefoot{FWHM and integrated fluxes given in units of [km\,s$^{-1}$[] and [$10^{-20}$erg\,s$^{-1}$\,cm$^{-2}$], respectively. Values found using simultaneous fit of a power law continuum and 1-D Gaussian for each line. For undetected lines, we list the $3\sigma$ upper limit on the integrated flux based on the error spectrum and an assumed FWHM (see value in parentheses). No correction for dust reddening or gravitational magnification has been applied. We include the best-fit line fluxes for a weaker candidate galaxy (X1, see Appendix \ref{candsec}).}
\label{specfittab}
\end{table*}

\end{landscape}
\clearpage
\twocolumn

\subsection{R100 integrated spectra}\label{intspecsec100}
Due to the coarse LSF of the R100 data ($\sim10^4$\,km\,s$^{-1}$; \citealt{boke23}), lines are frequently blended. In order to fit models to these data, one must create a high-resolution model and convolve it with the line spread function (e.g., \citealt{hein23,jone23,umed23}). While the R100 data is therefore not useful for kinematics or precise redshift derivations, we may examine the spectrum of `G2' to find a general redshift.

\begin{figure}
\centering
\includegraphics[width=0.5\textwidth]{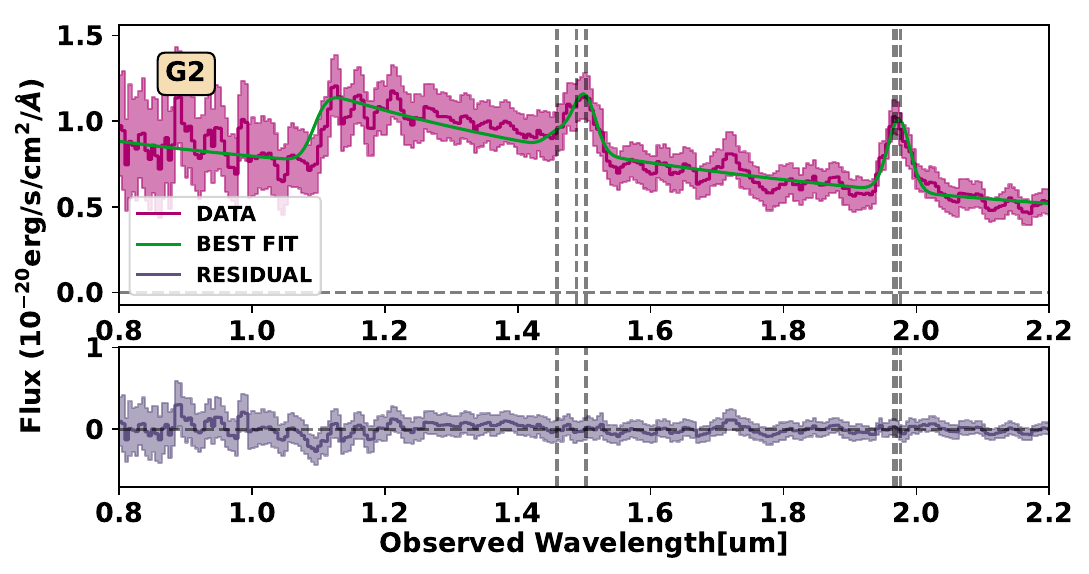}
\caption{Integrated spectrum extracted from the R100 cube using the `G2' mask of Fig. \ref{white} with illustrative $1\sigma$ errors shown as shaded region. A basic fit to the data is shown by the green line. The lower panel shows the residual, with the centroids of each line depicted by dashed lines. From left to right, the lines are \hb, \oiiiAB, \niiA, \ha, and \niiB. The best-fit redshift is $z_{G2}=2.00\pm0.01$}
\label{r100g2}
\end{figure}

`G2' features two clear emission lines in a wavelength regime ($\lambda_{\rm obs}=1.4-2.0\,\mu$m) that is not covered by the R2700 data. To fit this, we first create a higher-resolution model spectrum ($\delta \lambda=0.001\,\mu$m) that is populated with the strongest expected lines (i.e., \hb, \oiiiAB, \niiAB, and \ha)\footnote{We find no evidence for significant [OII]$\lambda\lambda3726,3729$ emission in G1.}. Since the lines are unresolved, we adopt Gaussian profiles instead of the more detailed pseudo-Voigt profiles used in the R2700 analysis. We find evidence for a spectral break at $\lambda_{\rm obs}\sim1.1\,\mu$m, which we tentatively interpret as a Balmer break. The continuum is thus modelled as a power law red-wards of this break ($\lambda_{\rm rest}=364.5$\,nm) and a separate power law bluewards of this value. The combined line + continuum model is then convolved with the resolving power to account for the LSF, and is rebinned to the R100 spectral bins. We use \lmfit with a least-square minimisation to find the best-fit continuum and line parameters and redshift, resulting in the fit of Fig. \ref{r100g2}. The error spectrum was rescaled to match the standard deviation of the residual spectrum.

Due to the heavy blending of the lines, we do not present best-fit widths or integrated fluxes. However, we find a best-fit redshift of $z=2.00\pm0.01$, which is comparable to the CIV-based redshift of \citet{riec13} ($z\sim2.092$). The redshift difference may be ascribed to a different extraction aperture or the simplicity of our current model. Indeed, the best-fit model features residuals around the putative Balmer break and around a possible line at $\lambda_{obs}\sim1.72$\,$\mu$m. While the line amplitude is quite weak ($\sim1\sigma$), the centroid wavelength of this line is in agreement with the faint auroral line \niiMAYBE.

We note that the rich R100 spectrum of each source (which contain more emission lines) will be modelled with advanced codes in future works, and our current model is solely used to confirm the approximate redshift of the `G2' object. 

\section{Analysis}\label{analysis}

\subsection{Morpho-kinematic maps}\label{mommap}

Next, we investigate the morpho-kinematics of these components using the R2700 data. This is commonly done using non-parametric measures (e.g., cumulative velocity distributions or moment maps), which are well-suited to sources with isolated, asymmetric lines. Since the emission lines of our sources feature overlapping lines in some spectra (e.g., \ha and \niiAB) we instead use a parametric model. The standard assumption of a Gaussian model resulted in some low-level residuals, so we instead assume that each line may be modelled by a pseudo-Voigt profile (see details in Appendix \ref{momsec}).

\begin{figure*}
\centering
\includegraphics[width=0.8\textwidth]{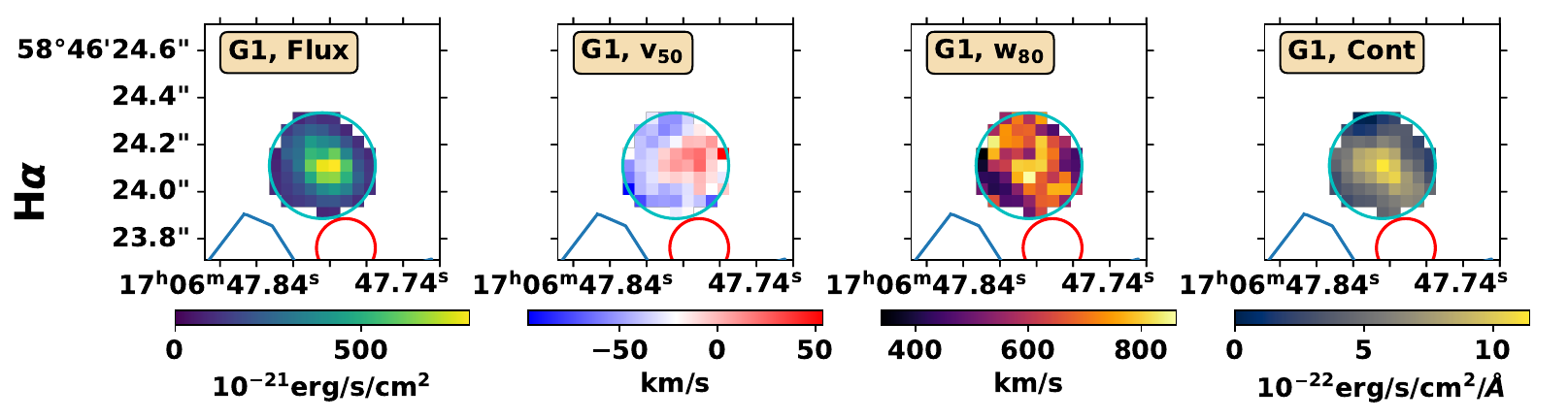}
\includegraphics[width=0.8\textwidth]{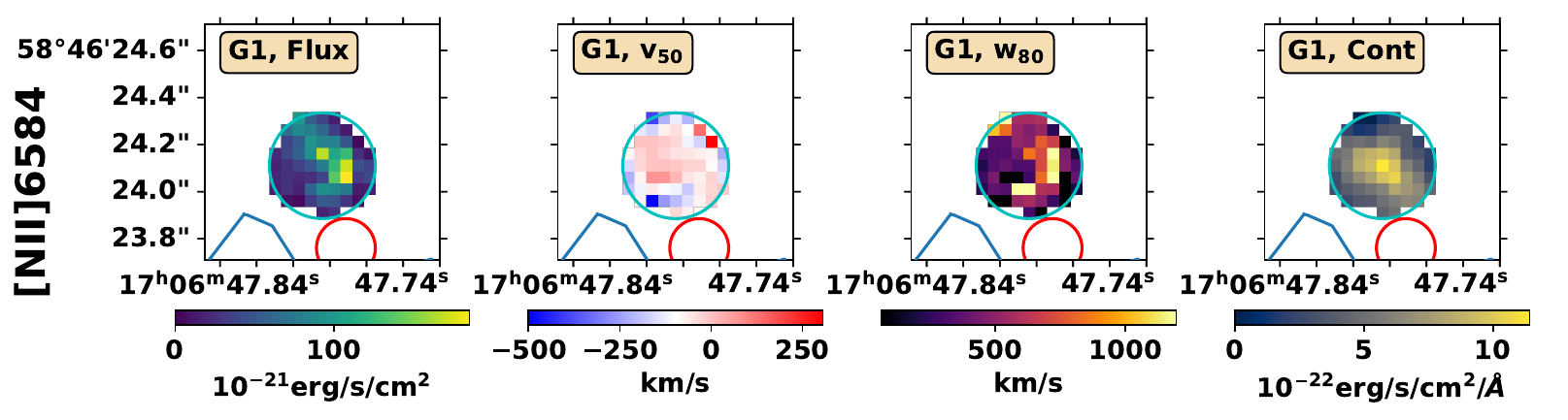}
\includegraphics[width=0.8\textwidth]{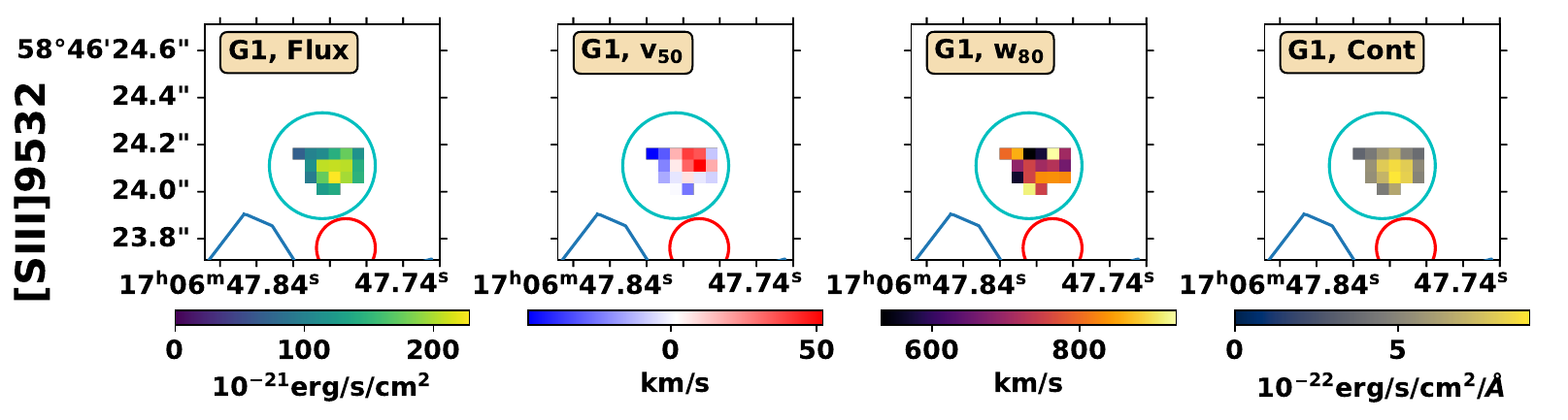}
\includegraphics[width=0.8\textwidth]{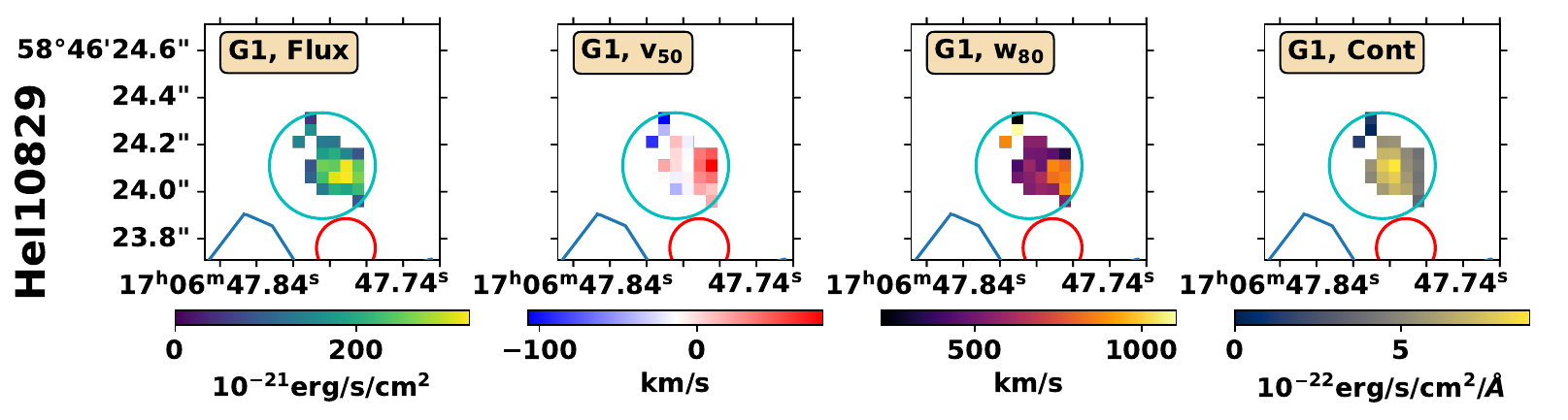}
\caption{Morpho-kinematic maps of the strong emission lines detected in the component G1 ($z=3.4806$). Integrated intensities are given per spaxel. The corresponding maps of \siiAB are not presented due to low S/N. Due to the construction of our model, \niiA has the same integrated velocity field, velocity dispersion map, and continuum map as \niiB, but an integrated intensity map that is a factor 2.98 lower. For \siii and \hei, $v_{50}$ values are given with respect to their best-fit redshifts ($z=3.4810$ and 3.4821, respectively).}
\label{NM}
\end{figure*}

\begin{figure*}
\centering
\includegraphics[width=0.8\textwidth]{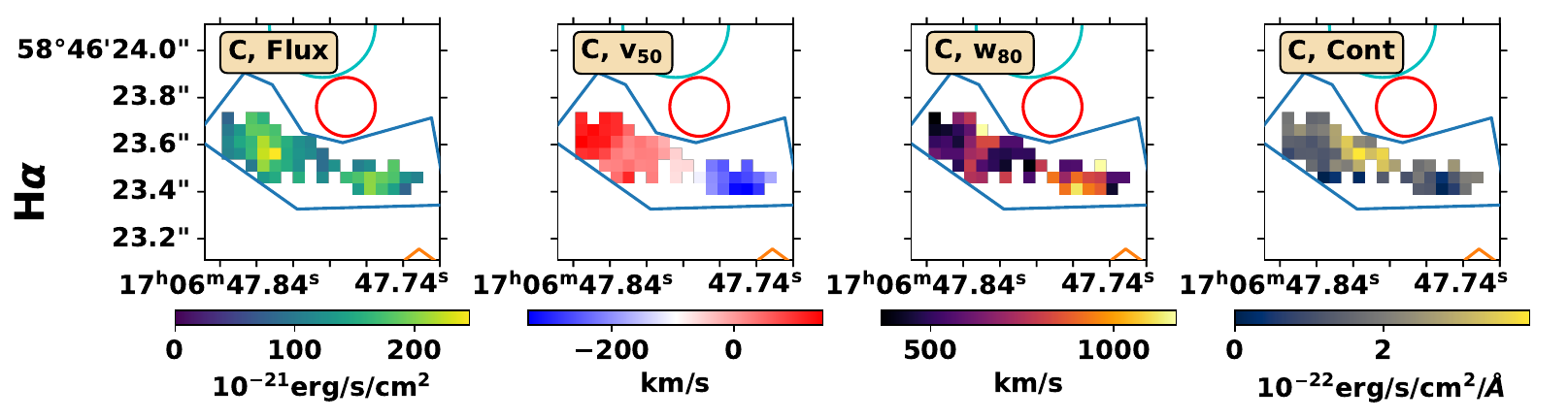}
\includegraphics[width=0.8\textwidth]{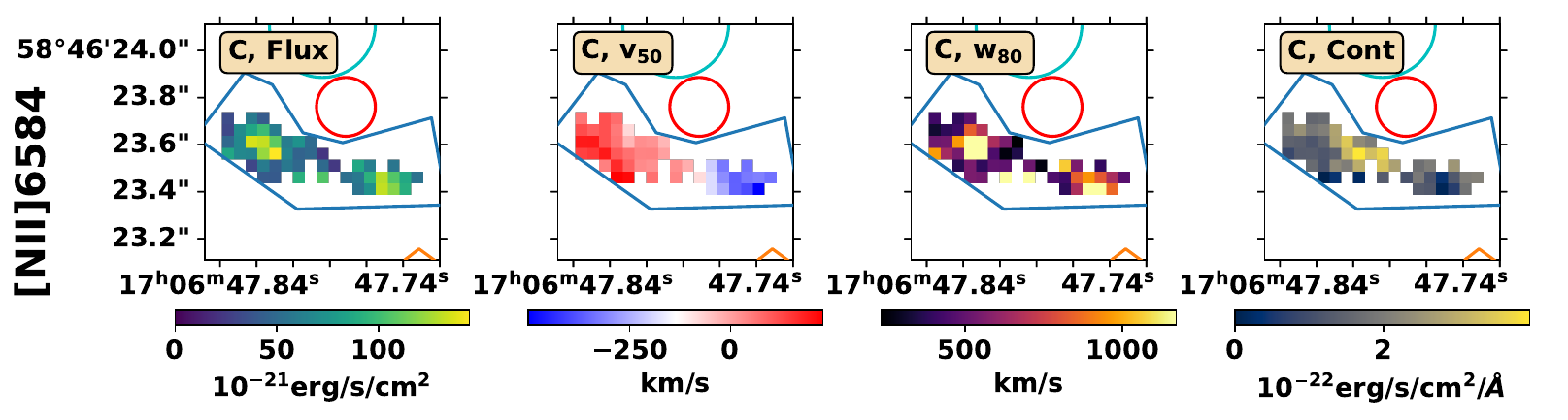}
\caption{Morpho-kinematic maps of the strong emission lines detected in the C component ($z=6.3425$). Integrated intensities are given per spaxel. The corresponding maps of \hb, \oiiiAB, and \siiAB are not presented due to low S/N. Due to the construction of our model, \niiA has the same integrated velocity field, velocity dispersion map, and continuum map as \niiB, but an integrated intensity map that is a factor 2.98 lower.}
\label{CM}
\end{figure*}

\begin{figure*}
\centering
\includegraphics[width=0.8\textwidth]{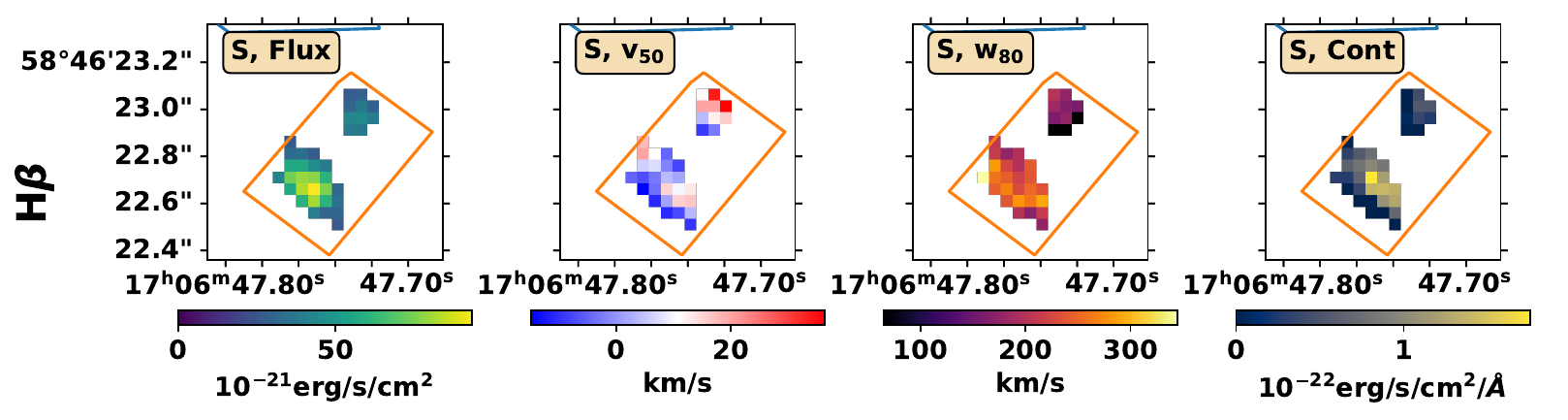}
\includegraphics[width=0.8\textwidth]{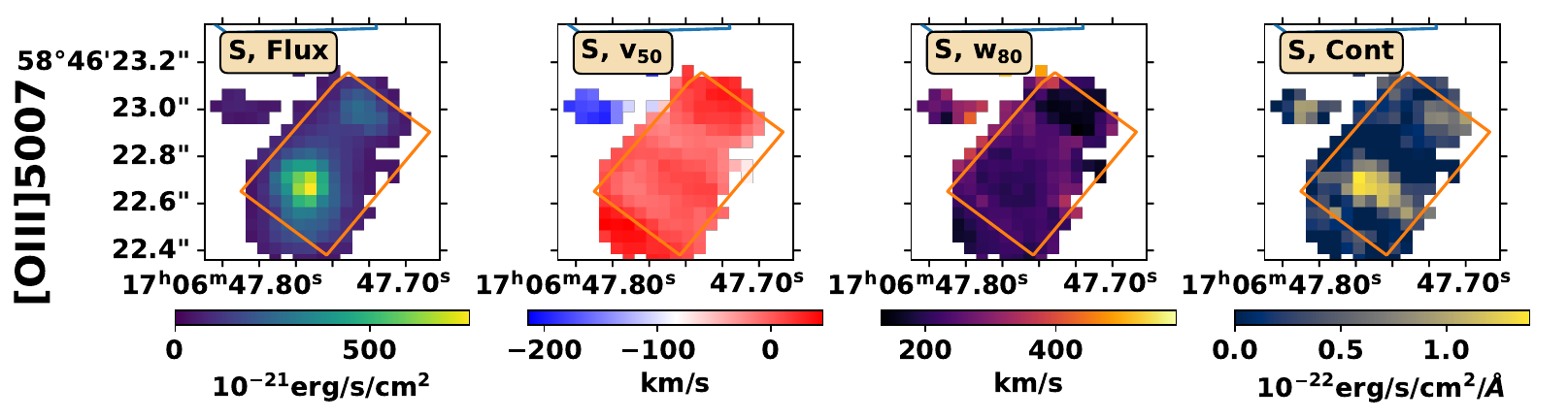}
\includegraphics[width=0.8\textwidth]{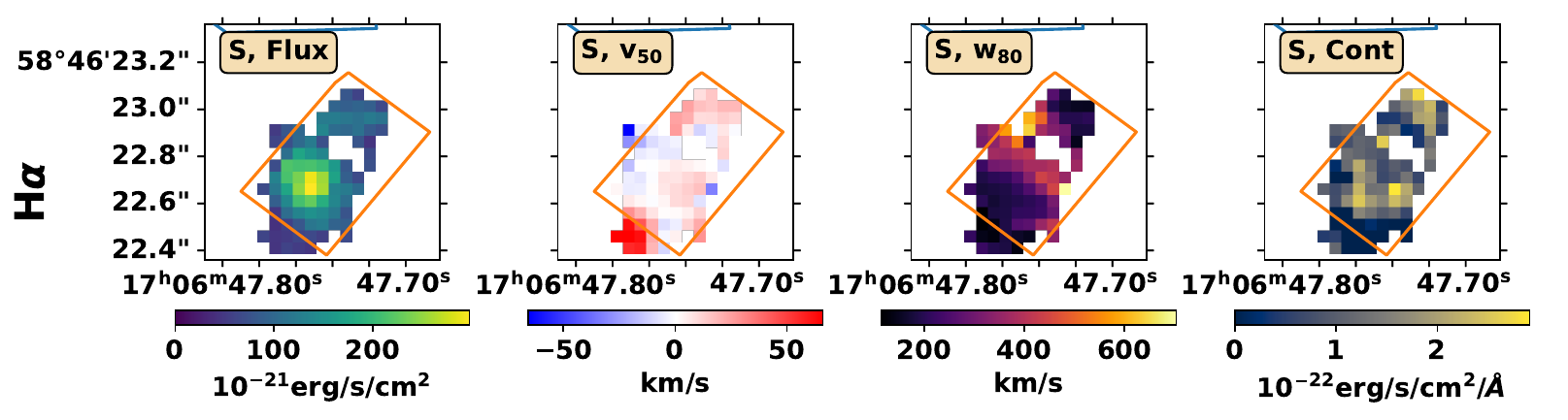}
\caption{Morpho-kinematic maps of the strong emission lines detected in the S components ($z\sim6.3592$). Integrated intensities are given per spaxel. The corresponding maps of \hg are not presented due to low S/N. Due to the construction of our model, \oiiiA has the same velocity field, velocity dispersion map, and continuum map as \oiiiB, but an integrated intensity map that is a factor 2.94 lower.}
\label{SM}
\end{figure*}

\begin{figure*}
\centering
\includegraphics[width=0.8\textwidth]{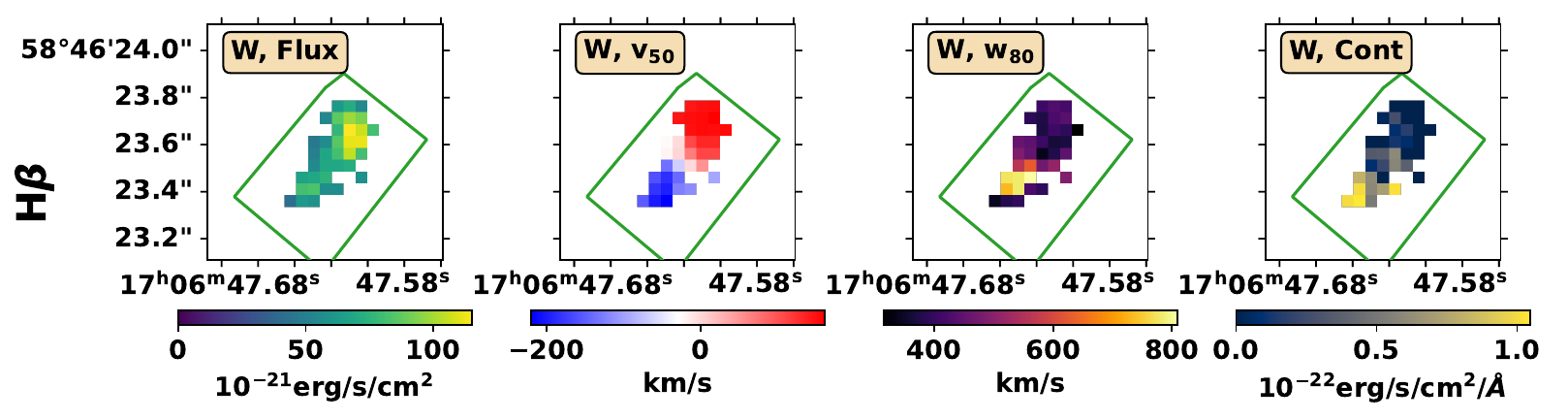}
\includegraphics[width=0.8\textwidth]{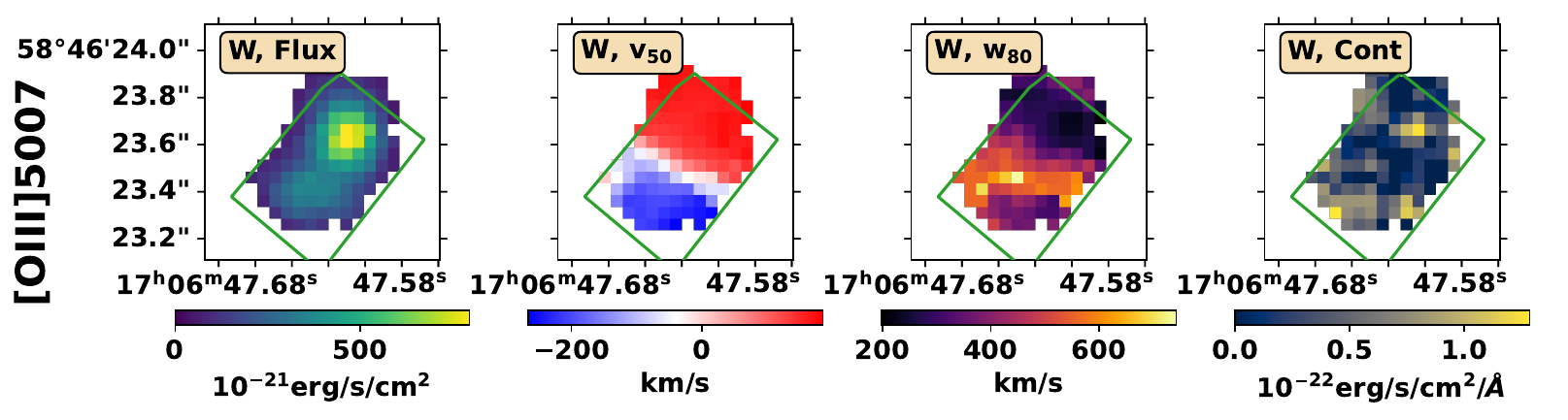}
\includegraphics[width=0.8\textwidth]{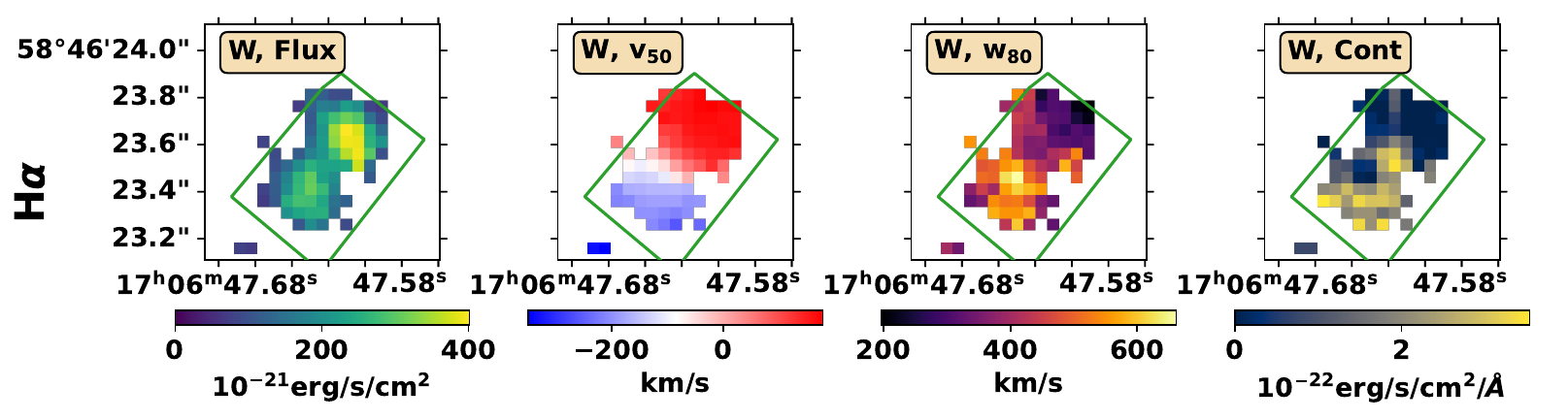}
\caption{Morpho-kinematic maps of the strong emission lines detected in the W components ($z_{W1}=6.3550$, $z_{W2}=6.3628$). Integrated intensities are given per spaxel. The corresponding maps of \hg are not presented due to low S/N. Due to the construction of our model, \oiiiA has the same velocity field, velocity dispersion map, and continuum map as \oiiiB, but an integrated intensity map that is a factor 2.94 lower.}
\label{WM}
\end{figure*}

The resulting integrated intensity, velocity offset, velocity dispersion, and continuum maps are shown in Fig. \ref{NM}, \ref{CM}, \ref{SM}, and \ref{WM}. Some weaker lines (e.g., \siiB in G1) are detected in the integrated spectrum, but lack the S/N to be significantly detected in the spectrum of multiple individual spaxels.

The maps for component G1 are shown in Fig. \ref{NM}. For the two strongest lines (\ha and \niiB), we see that the emission is circular with a slight concentration in the northeast-southwest diagonal. There is no strong velocity gradient, and both the velocity dispersion and continuum emission feature a central peak. The two lower-S/N lines (\siii and \hei) are detected in the core, with a similar distribution to the core of the bright lines. This elongation may be influenced by a non-circular PSF (e.g., \citealt{DEugenio2023}). These combined morpho-kinematics may be interpreted either as a face-on disk or a dispersion-dominated galaxy.

From the maps of component C (Fig. \ref{CM}), an east-west velocity gradient is apparent. This has previously been seen in PdBI [CII] observations (\citealt{riec13}), and the higher spatial resolution of our data allows for a more in-depth investigation. We find that while the continuum emission peaks in the centre (possibly with a contribution from G2, see Figure \ref{white}), the line intensity peaks in the east and west sides. In addition, the velocity dispersion does not feature a central maximum. This argues for the presence of two separate galaxies, which we will investigate further in Section \ref{gravlenssec}.

The S component is clearly composed of two spatially separated galaxies at a similar redshift (i.e., low velocity dispersion, no apparent velocity gradient; Fig. \ref{SM}) with continuum and \ha intensity minima between the galaxies. The brighter galaxy (to the southeast; S1) was originally identified as `R1' by \citet{coor14}.

The W component features a strong velocity gradient (Figure \ref{WM}). However, the spatially offset integrated intensity peak (to the north) and higher velocity dispersion in the south argues against rotation. Instead, this appears to be two galaxies that are spatially and spectrally separated. This is supported by the asymmetric double-peak nature seen in the integrated spectrum of this component (Fig. \ref{spec1}) and the two spatial peaks seen in Fig. \ref{white}.
\vspace{1mm}

Altogether, our analysis suggests that the HFLS3 field is composed of several components: two low-redshift sources to the north (G1, $z\sim3.481$; G2, $z\sim2.01$), a lensed source with complex kinematics suggesting two components (C1 and C2, $z\sim6.342$), two galaxies to the south (S1 and S2, $z\sim6.359$), and two galaxies to the west (W1, $z\sim6.355$; W2, $z\sim6.363$).

\subsection{Gravitational lens modelling}\label{gravlenssec}
The gravitational lens model of the HFLS3 field was previously analysed by \citet{coor14}, who assumed that G1 and G2 were at the same redshift ($z\sim2.1$) and used the marginally resolved PdBI [CII]158$\mu$m image of HFLS3 to derive a low magnification factor ($\mu\sim2.2$). With the precise spectroscopic redshifts of G1 and G2 and a resolved map of HFLS3, we refine this model using the public lens modelling software \textlcsc{\PAL} \footnote{https://github.com/Jammy2211/PyAutoLens} \citep{nigh15, nigh18,nigh21}.

\subsubsection{Methods}
Using \PAL, we may derive the intrinsic (source-plane) mass and light profiles for each source in the field of view. Each light profile is given by a S\'{e}rsic profile \citep{sers63}, while we assume elliptical isothermal mass profiles. The mass and light profiles of each object are assumed to have shared centres, axis ratios, and position angles. We model the PSF as a circular Gaussian with FWHM$\sim0.1''$. The fitting process results in the best-fit centroid, intrinsic axis ratio ($q\equiv a/b$), and position angle ($\phi$) of the mass and light profiles, the effective radius ($\rm r_{eff}$)and S\'{e}rsic index ($n$) of the light profile, and the Einstein radius ($\rm r_{Ein}$) of the mass profile. By dividing the total fluxes of the best-fit image-plane and source-plane models for each component, we calculate a total magnification factor.

\begin{figure*}
\centering
\includegraphics[width=0.3\textwidth]{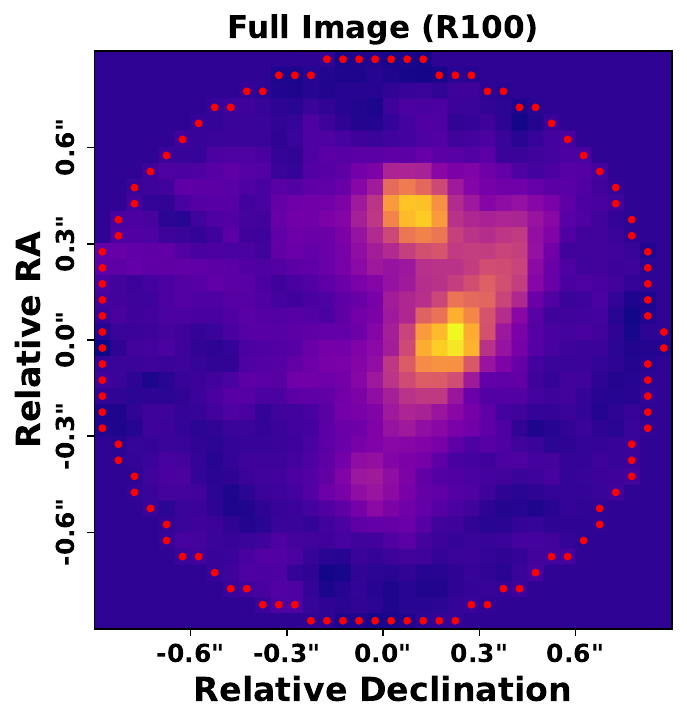}
\includegraphics[width=0.3\textwidth]{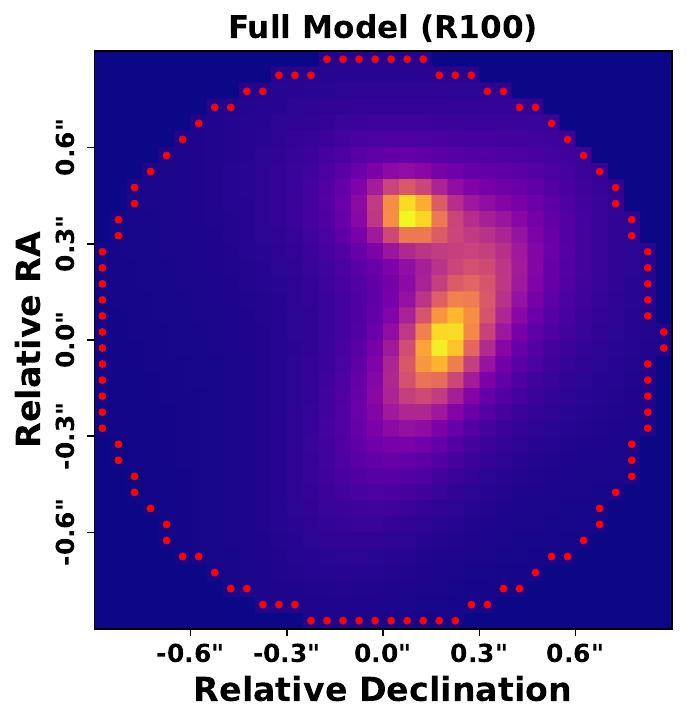}
\includegraphics[width=0.3\textwidth]{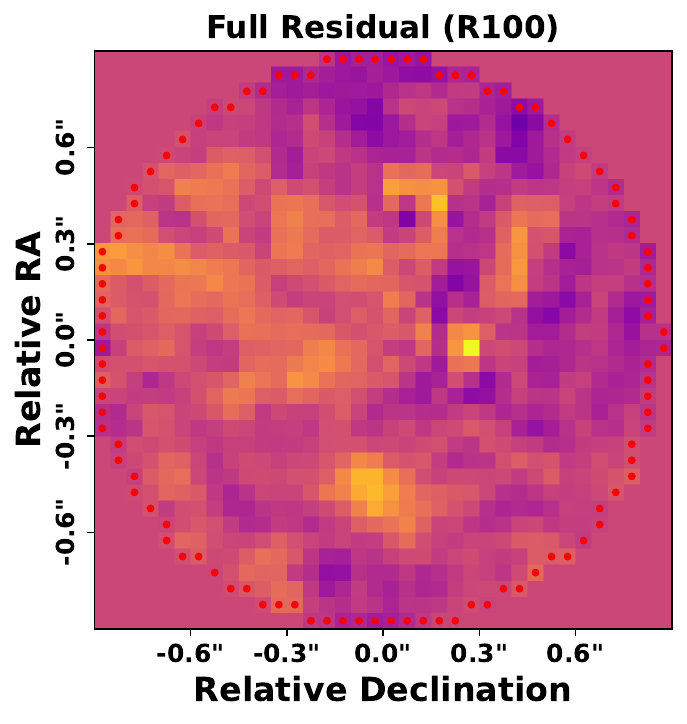}
\includegraphics[width=0.3\textwidth]{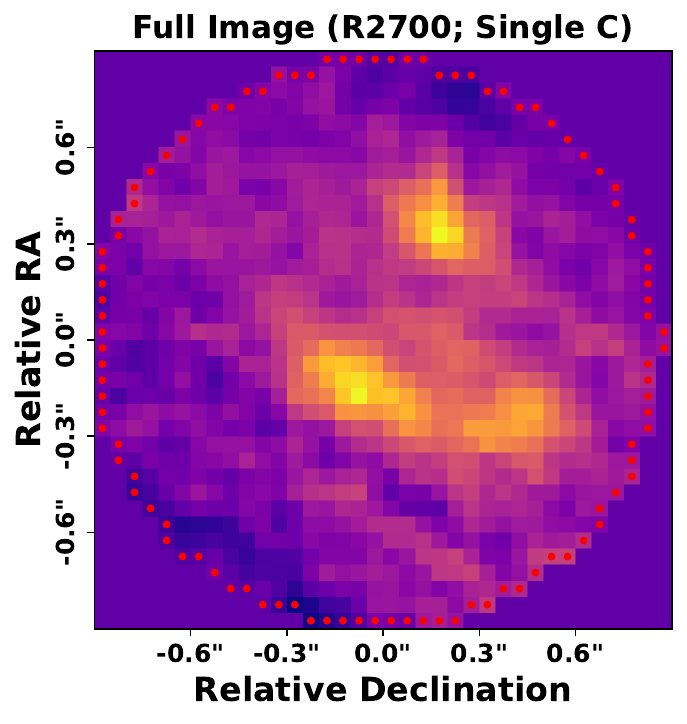}
\includegraphics[width=0.3\textwidth]{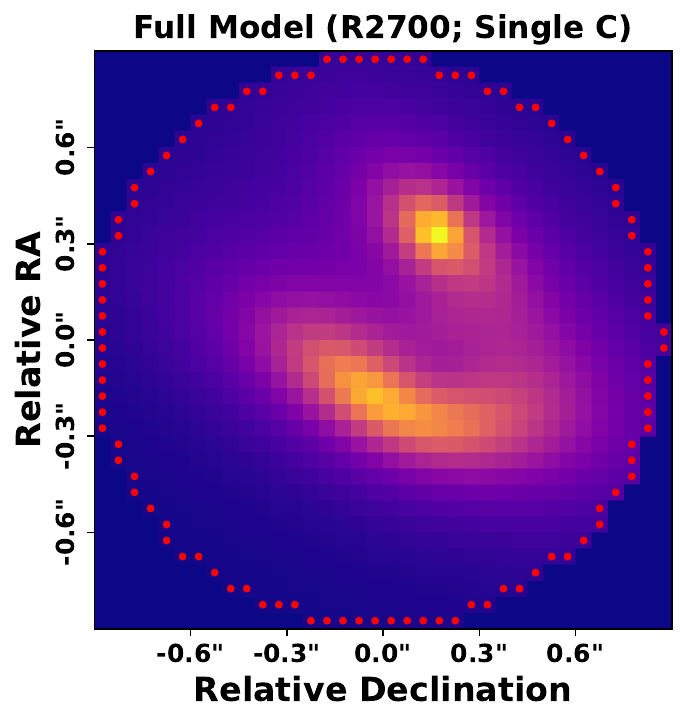}
\includegraphics[width=0.3\textwidth]{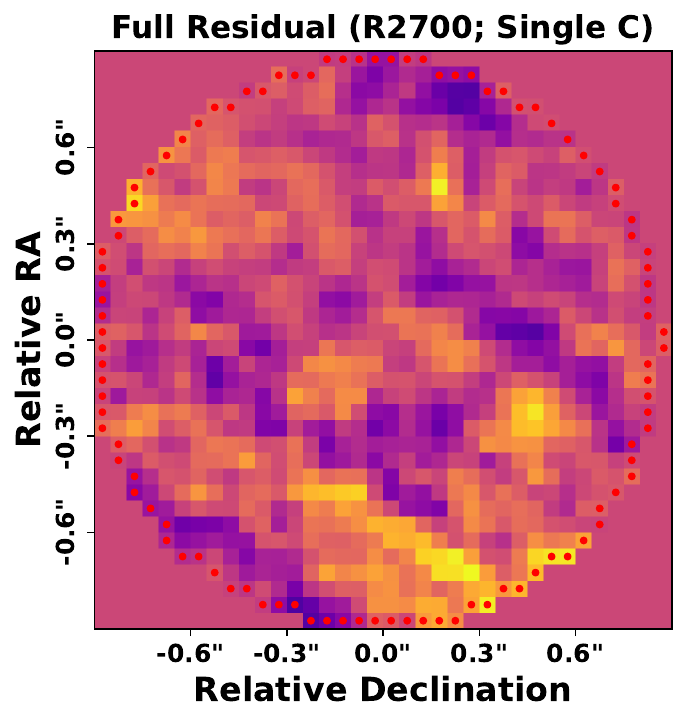}
\includegraphics[width=0.3\textwidth]{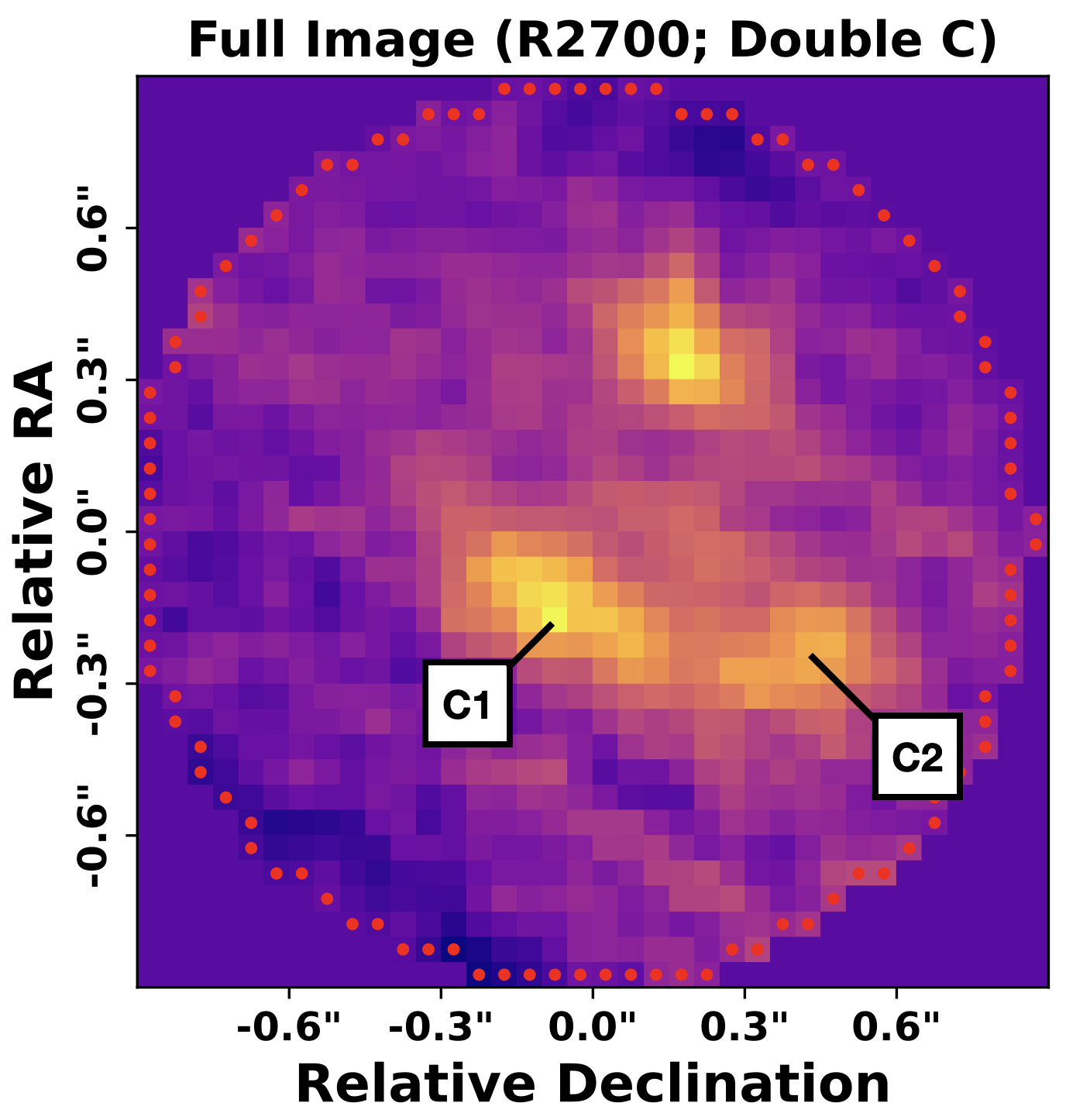}
\includegraphics[width=0.3\textwidth]{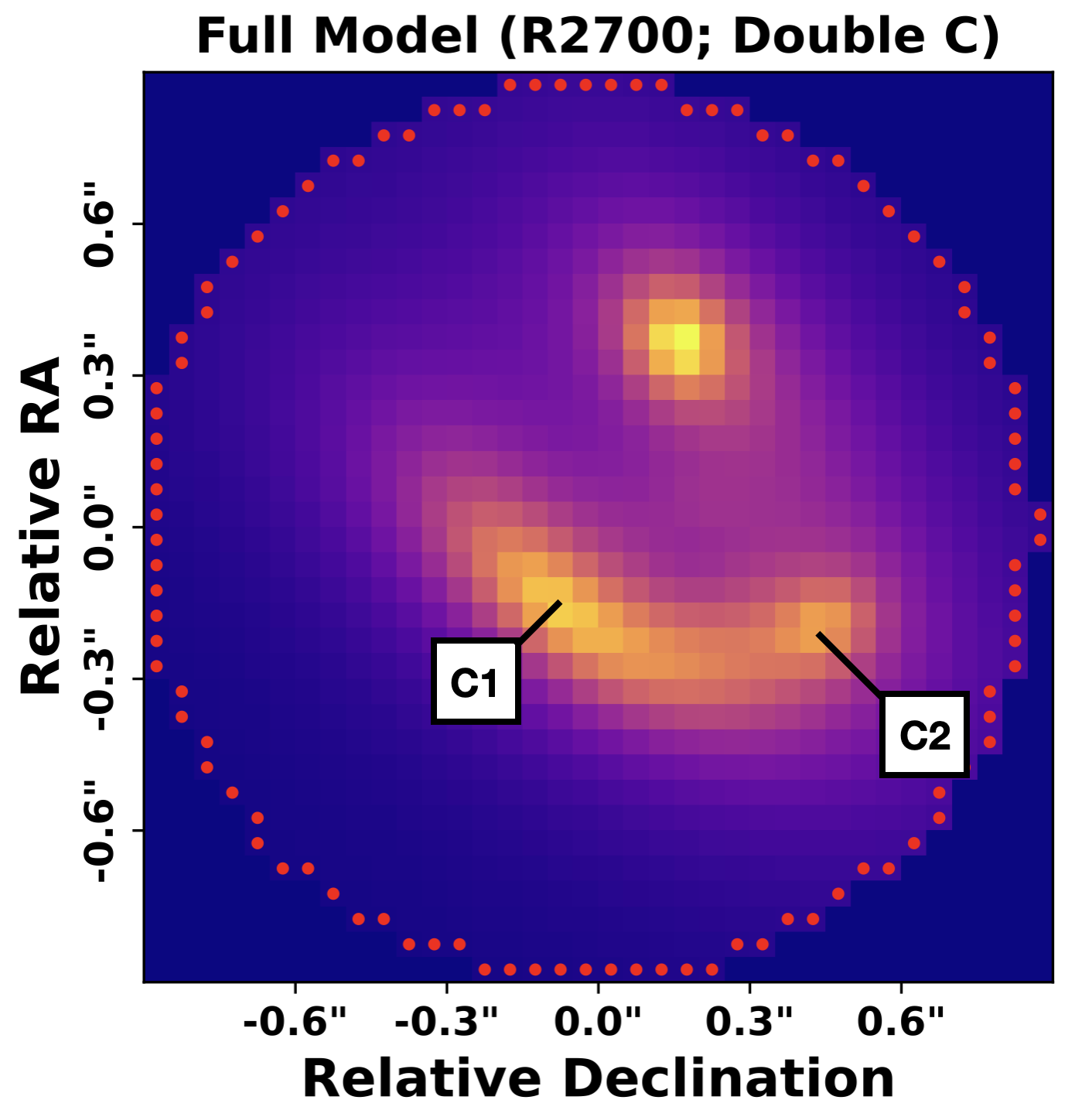}
\includegraphics[width=0.3\textwidth]{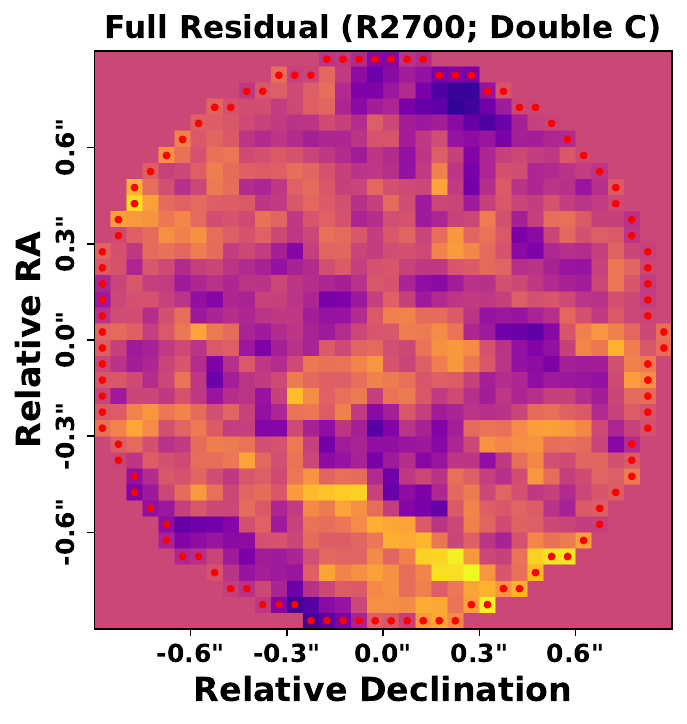}
\caption{Results of \PAL lensing analysis of collapsed images of the R100 cube ($\lambda_{\rm obs}=0.8-1.1\,\mu$m; top row) and the R2700 cube ($\lambda_{\rm obs}=4.80-4.85\,\mu$m) assuming that the component C is composed of a single source (middle row) or two sources (C1 to the west and C2 to the east; lower row). The observed data, best-fit model, and residual (all in the image plane) are presented from left to right. Mass and light profiles are assumed to be S\'{e}rsic and isothermal ellipse profiles, respectively. The outline of the spatial mask is shown by red markers.}
\label{r100_lens}
\end{figure*}

Due to the complexity of the field, we begin by examining the two low-redshift sources (G1 and G2). These are isolated by collapsing the R100 data cube over $\lambda_{\rm obs}=0.8-1.1\,\mu$m (which is not covered by the R2700 cube). This collapsed image (top left panel of Fig. \ref{r100_lens}) has little contribution from the $z>6$ components. Because of this, we may examine how G2 ($z\sim2.0$, the northernmost source in the top row of Fig. \ref{r100_lens}) lenses G1 ($z\sim3.5$, the elongated central source in the top row of Fig. \ref{r100_lens}). Our model contains the mass and light profile of G2 and the light profile of G1. The resulting fit is shown in the top row of Fig. \ref{r100_lens}, with parameters listed in Table \ref{lenstable} and source-plane models of G1 and G2 given in Appendix \ref{BFLMC}.

We next turn to a wavelength range in the R2700 cube containing \ha at $z\sim6.34$ ($\lambda_{obs}=4.80-4.85\,\mu$m) to examine whether component C is better modelled by a single or double source. This wavelength range contains strong \ha emission from component C, moderate continuum from G1, and no strong emission from G2. We model the light and mass profile of G1, assuming the best-fit mass model for G2 from the R100 collapsed image but with a variable Einstein radius. Component C is fit using either a single S\'{e}rsic profile or two spatially offset S\'{e}rsic profiles. The resulting fits for a single and double component for C are shown in the middle and lower row of Fig. \ref{r100_lens}, with parameters listed in Table \ref{lenstable}. 

\clearpage
\onecolumn
\begin{landscape}

\begin{table*}
\centering
\caption{Best-fit parameters from \textlcsc{PyAutoLens} analysis.}
\begin{tabular}{lc|ccccccc}

     &    & ($\delta$x,$\delta$y) & $q$ & $\phi$      & $\rm r_{eff}$ & $n$ & $\mu$ & $\rm r_{Ein}$\\
     &    & [$"$,$"$]             &     & [$^{\circ}$] & [$"$]         &     &        & [$"$]        \\ \hline
     
R100 & G1& ($0.14\pm0.004,0.301\pm0.009$) & $1.43\pm0.01$ & $57.02\pm1.51$ & $1.028\pm0.018$ & $5.06\pm0.16$ & 1.2 & - \\
 & G2 & ($0.194\pm0.002,0.004\pm0.002$) & $2.39\pm0.05$ & $-30.52\pm0.95$ & $0.251\pm0.007$ & $1.22\pm0.06$ & - & $0.12\pm0.011$\\ \hline

R2700 (Single C) & G1 & ($0.181\pm0.002,0.268\pm0.004$) & $1.94\pm0.08$ & $36.13\pm0.85$ & $1.072\pm0.041$ & $3.19\pm0.1$ & 1.12 & $0.145\pm0.008$\\
 & G2 & - & - & - & - & - & - & $0.11\pm0.01$ \\
 & C & ($0.13\pm0.005,-0.017\pm0.007$) & $2.26\pm0.04$ & $69.65\pm0.77$ & $0.298\pm0.002$ & $1.19\pm0.04$ & 1.89 & - \\ \hline

R2700 (Double C) & G1 & ($0.176\pm0.003,0.298\pm0.009$) & $1.77\pm0.09$ & $29.2\pm0.42$ & $2.673\pm0.26$ & $4.73\pm0.23$ & 1.1 & $0.169\pm0.006$\\
 & G2 & - & - & - & - & - & - & $0.12\pm0.01$ \\
 & C1 & ($0.102\pm0.005,0.012\pm0.004$) & $2.00\pm0.05$ & $57.1\pm1.2$ & $0.293\pm0.006$ & $1.34\pm0.07$ & 2.09 & - \\
 & C2 & ($0.293\pm0.009,0.004\pm0.002$) & $1.49\pm0.04$ & $46.22\pm4.7$ & $0.211\pm0.019$ & $3.17\pm0.26$ & 2.43 & - \\ \hline

\end{tabular}
\tablefoot{Parameters include the offset from central position (17h06m47.799s, $+58^{\circ}46'23.729''$) in arcseconds, axis ratio ($q\equiv a/b$), position angle ($\phi$), effective radius ($\rm r_{eff}$), S\'{e}rsic index ($n$), lensing magnification ($\mu$), and Einstein radius ($\rm r_{Ein}$). We present the lensing analysis results for a collapsed region of the R100 cube that contains mainly emission from G1 and G2 ($\lambda_{\rm obs}=0.8-1.1\,\mu$m, top section) and for a collapsed region of the R2700 cube that primarily contains \ha from C and continuum emission from G1 ($\lambda_{obs}=4.80-4.85\,\mu$m). The middle section shows the results assuming C is composed of one source, while the lower section assumes two components. For each R2700 result, we assume that the mass profile of G2 is identical to the best-fit light profile of the R100 image, but with a variable Einstein radius.}
\label{lenstable}
\end{table*}

\end{landscape}
\clearpage
\twocolumn

\subsubsection{Results}
From this analysis, we find that the lowest-redshift source (G2) is elongated ($q\sim2.4$), compact ($r_{\rm eff}\sim0.25''$, or $\sim2.1$\,kpc at $z=2.00$), and not centrally peaked ($n\sim1.2$). On the other hand, the source-plane morphology of G1 is characterised as elongated ($q\sim1.4-1.9$), extended ($r_{\rm eff}\sim1.0-2.7''$, or $\sim7.3-19.8$\,kpc at $z=3.48$), centrally peaked ($n\sim3-5$), and only moderately magnified by G2 ($\mu\sim1.1-1.2$). 

Component C in the collapsed R2700 image appears as a curved arc with two bright regions. We first examine whether this may be explained as a single lensed source, yielding a best-fit source-plane morphology that is elongated ($q\sim2.3$), compact ($r_{\rm eff}\sim0.3''$, or $\sim1.6$\,kpc at $z=6.34$), with an exponential profile ($n\sim1.2$), and a low magnification factor ($\mu=1.89$). As seen in the middle row of Fig. \ref{r100_lens}, the image-plane morphology of this source is indeed arced, but has a central peak rather than two spatially separate clumps. This results in a significant residual at the location of the western spot in the observed arc.

Next, we assume that component C is composed of two spatially separate S\'{e}rsic profiles (C1 to the west and C2 to the east). The best-fit model returns one source (C1) that is similar to the similar to the best-fit single-component model, with $q\sim2.0$, $r_{\rm eff}\sim0.3''$, $n\sim1.2$, and a higher magnification factor ($\mu=2.09$). However, this model contains an additional component to the west that is more circular ($q\sim1.5$), smaller ($r_{\rm eff}\sim0.2''$), more centrally peaked ($n\sim3$), and is more magnified ($\mu=2.43$). The resulting image-plane morphology of this best-fit model is double-peaked, resulting in a better representation (i.e., a lower $\chi^2$). The two best-fit components have a small source-plane projected distance ($0.2''$, or $\sim1.1$\,kpc at $z=6.34$) that is comparable to their effective radius. This is discussed in the next subsection.

We note that this analysis suggests that none of the emission peaks in the observed fields represent multiply-imaged objects. The $z<6$ sources (G1 and G2) feature significantly different redshifts and are quite compact, implying that they are individual galaxies. The C emission is better fit by two separate components, rather than a single galaxy imaged twice. The effects of the gravitational lens are quite small at larger spatial separations from G2, so it is unlikely that the S and W components are magnified or multiply imaged (as supported by the different morpho-kinematics of each sub-component). If the PSF is truly larger than $0.1''$ (see Section \ref{ifudes}), then our general conclusions are unchanged, although C1 and C2 feature smaller best-fit source-plane effective radii with higher magnification.

\subsubsection{Nature of component C}\label{whatisc}
This lensing analysis of the \ha emission in the central component suggests that it is composed of two spatially separate ($\sim1$\,kpc projected offset) galaxies. By examining the best-fit morphologies (Fig. \ref{C_source}), it is clear that their proximity to each other results in overlapping emission in the image-plane. So while the far east and west sides of the central arc in the image plane are dominated by the emission from each source, the space between the two apparent galaxies contains emission from both sources. 

Returning to the morpho-kinematic maps (Fig. \ref{CM}), we see that each line shows a weak ($\delta v\sim100$\,km\,s$^{-1}$) east-west velocity gradient (as seen in previous [CII] observations; \citealt{riec13}). Since none of the integrated intensity or velocity dispersion maps show a central peak, as expected for a single rotating disk (see the best-fit single component model in Fig. \ref{r100_lens}), these maps instead imply multiple galaxies separated in velocity.

This is supported by the previous morpho-kinematic investigation of FIR lines by \citet{riec13}. An east-west velocity gradient was found in [CII], but with a high velocity dispersion which peaked in the southeast. In addition, the integrated spectrum exhibited two velocity components in multiple FIR lines ($z=6.3335$ and $z=6.3427$, resulting in a weighted average of $z=6.3369\pm0.0009$). From our best-fit redshift ($z_C=6.3425\pm0.0003$), this suggests that the rest-frame optical emission is dominated by the higher redshift component.  Due to the overlapping image-plane distributions of the two components (see bottom-centre and bottom-right panels of Figure \ref{C_source}), they are not easily separable without lens modelling. 

To summarise, the lensing and morpho-kinematic analyses of our new high-resolution data agree with previous results (\citealt{riec13,coor14}, who claimed that the component C of HFLS3 is composed of two components. The best-fit magnification of our model is also similar ($\mu\sim2.1-2.4$, compared to $\mu=2.2\pm0.3$). 

\subsection{Excitation conditions}\label{linerat}
\begin{figure}
\centering
\includegraphics[width=0.5\textwidth]{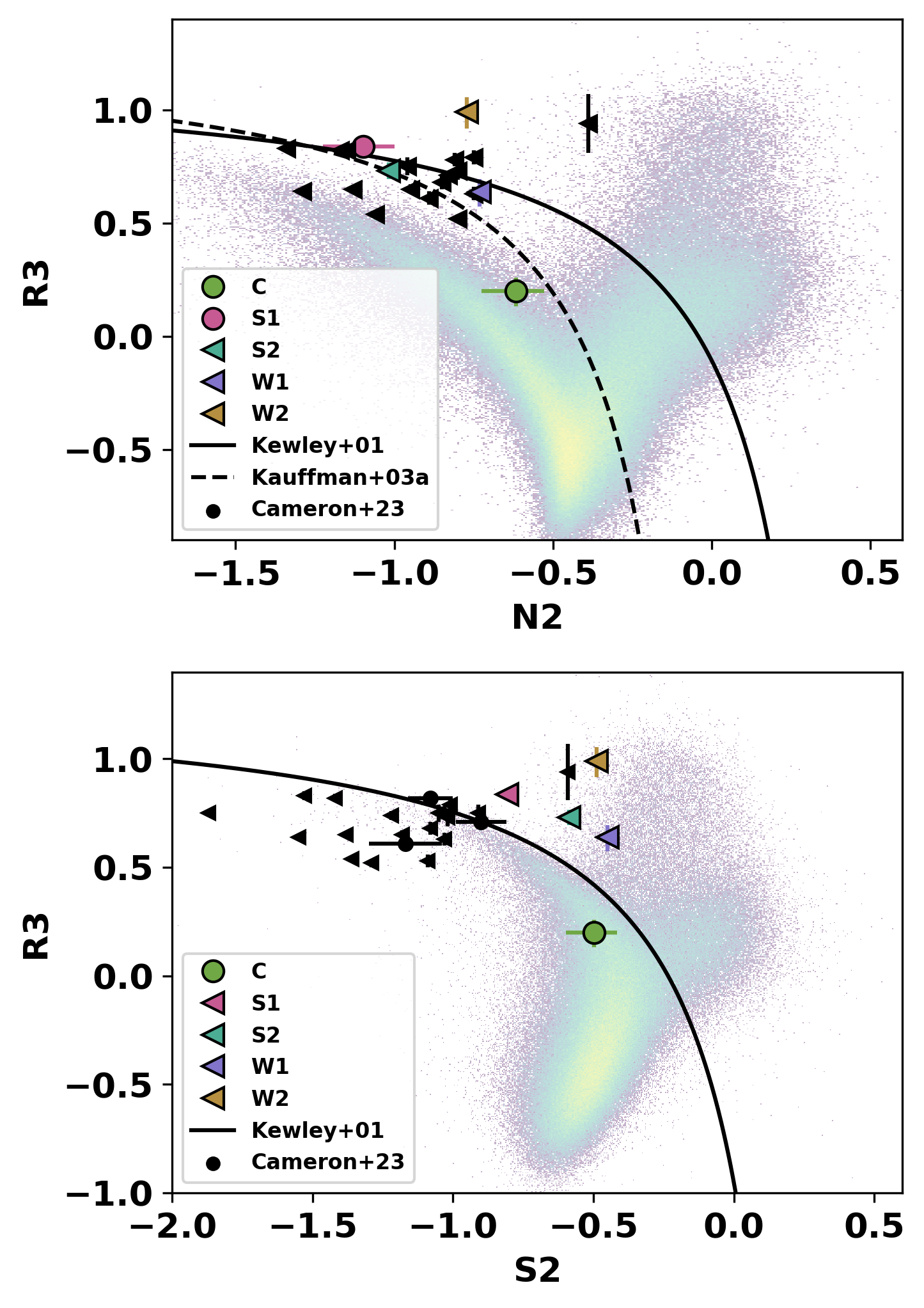}
\caption{[SII]-VO87 (top) and [NII]-BPT (lower) plots created using best-fit line fluxes for each source (see values in Table \ref{specfittab}). Distribution of low-redshift galaxies from SDSS (MPA-JHU DR8 catalogue; \citealt{kauf03sdss,brin04}) are shown by background points. Left-facing arrows represent $3\sigma$ upper limits on N2 and S2. We include the demarcation lines for $z\sim0$ galaxies of \citet[solid lines]{kewl01} and \citet[dashed line]{kauf03}. We compare our results with values from $z\sim5.5-7.0$ galaxies
observed with JWST/NIRSpec MSA as part of the JADES survey (\citealt{came23}; black symbols).}
\label{BPT}
\end{figure}

Using the well-detected lines for our sample, we may place constraints on the excitation conditions of each source using line ratio diagnostic diagrams. To do this, we first calculate three line ratios:
\begin{equation}
N2=log_{10}([NII]\lambda6584/H\alpha)
\end{equation}
\begin{equation}
S2=log_{10}([SII]\lambda\lambda6716,6731/H\alpha)
\end{equation}
\begin{equation}
R3=log_{10}([OIII]\lambda5007/H\beta)
\end{equation}
These are used to create the [NII]-BPT (R3 vs N2; \citealt{bald81}) and [SII]-VO87 (R3 vs S2; \citealt{veil87}) plots (see Fig. \ref{BPT}) for each high-redshift ($z>6$) source in the field. We examine the line ratios of both galaxies in component S (S1 and S2) and W (W1 and W2) separately. On the other hand, the galaxies in C are blended in the image plane (see best-fit source-plane images in Figure \ref{C_source}), so we present the combined line ratios of each galaxy in this close pair. Since we are examining ratios of nearby emission line fluxes, we do not apply corrections for dust reddening or gravitational lensing.

For the [NII]-BPT, sources that lie above the solid \citet{kewl01} line are believed to be dominated by AGN excitation, while those that lie beneath the dashed \citet{kauf03} line are mainly star-forming, and those that lie between the lines are a combination of the two excitation sources (i.e., `composite'). Similarly, sources below (above) the solid \citet{kauf03} line in the [SII]-VO87 are thought to be star-forming (AGN-dominated). 

We find that while the sources in S and W have high R3 ratios (comparable to the R3 ratios of $z\sim5.5-7.0$ galaxies observed with JWST/NIRSpec MSA as part of the JADES survey; \citealt{came23}), the high upper limits on N2 and S2 do not allow us to rule out or confirm the presence of AGN for most components. However, the detection of \niiB in S1 shows that this point lies on the demarcation line for AGN activity. Component C lies in the star-forming regime, but with large errors that may place it in the composite region. 

We note that these demarcation lines were derived for $z\sim0$ galaxies with approximately solar metallicity. Recent results suggest that these lines may not separate SF- and AGN-driven ionisation for high-redshift, low-metallicity sources (e.g., \citealt{kewl13,naka22,uble23}). This interpretation is further complicated by the fact that the C and S components are composed of two sub-components each (see Section \ref{mommap} and \ref{gravlenssec}), which may exhibit different ionisation sources and/or metallicities. These line ratios may also be affected by shocks within galaxies (e.g., \citealt{alle08}), which we explore further in Appendix \ref{shocksec}.

These results suggest that the line ratios of components S and W do not allow us to robustly claim the presence or absence of AGN, while component C is likely composed of star-forming galaxies but may include an AGN.

\subsection{Distribution of star formation}

HFLS3 was originally found to be an extreme starburst, with an SFR$_{\rm FIR}=2900$\,M$_{\odot}$\,yr$^{-1}$ \citep{riec13}. Further studies found that the central galaxy was lensed, resulting in a lower intrinsic $\rm SFR\sim1300$\,M$_{\odot}$\,yr$^{-1}$ or a 100-Myr averaged $\rm SFR\sim660$\,M$_{\odot}$\,yr$^{-1}$ \citep{coor14}. However, both of these SFRs are based on fits to SEDs that contain low-resolution observations (e.g., Herschel/SPIRE with $>20''$, Spitzer/IRAC with $\sim2''$) which would capture emission from multiple components of the HFLS3 system (i.e., the lensed emission in C, both galaxies in S, and both galaxies in W). This raises the question of how the star formation activity is distributed in the HFLS3 system: a single dominant starburst or multiple star-forming galaxies within an area of projected diameter $\sim11$\,kpc? This may be explored using the \ha luminosity (which mainly traces star formation on timescales of $\sim10-20$\,Myr; e.g., \citealt{kenn98,glaz99}), corrected for dust attenuation.

To investigate this, we first derive an estimate of the B-V colour excess from the Balmer decrement (e.g., \citealt{domi13}): 
\begin{equation}
E(B-V)_{BD}=\frac{2.5}{k(\lambda_{H\beta})-k(\lambda_{H\alpha})}log_{10}\left(\frac{F_{H\alpha,obs}/F_{H\beta,obs}}{2.86}\right)
\end{equation}
where $k(\lambda)$ is the assumed dust attenuation curve \citep{calz00}, $F_{H\alpha,obs}/F_{H\beta,obs}$ is the observed line flux ratio, and we have assumed an intrinsic line ratio of 2.86 \citep{oste89}. We then use this value to estimate the intrinsic line flux of \ha, assuming case B recombination and $T_e=10^4$\,K:
\begin{equation}
F_{H\alpha,int}=F_{H\alpha,obs}10^{k(\lambda_{H\alpha})E(B-V)_{BD}/2.5}
\end{equation}
Finally, this dust-corrected flux is used to estimate the SFR assuming a Salpeter IMF \citep{kenn98}:
\begin{equation}
SFR_{H\alpha}=\frac{4\pi D_L^2 F_{H\alpha,int}}{1.26\times10^{41}\mathrm{erg\,s^{-1}}}
\end{equation}
where $D_L$ is the luminosity distance. The resulting values are presented in Table \ref{sfr_dust}. The \citet{calz00} dust attenuation curve was chosen to ease comparison to previous results, but other curves could be adopted. For example, the quadratic law of \citealt{redd20}) results in slightly higher ($\lesssim1\sigma$ difference) values of SFR, while the high-mass, high-redshift analogue law of \citet{sali18} returns slightly lower ($\lesssim2\sigma$ difference) values of SFR. 

From this analysis, we see that the central component indeed features a high SFR. But as found in Section \ref{gravlenssec}, this emission is lensed, resulting in a dust- and magnification-corrected SFR of $160\pm80$\,M$_{\odot}$\,yr$^{-1}$ (assuming an average $\mu\sim2.26\pm0.17$). The combined SFR of the $z\sim6.3$ galaxies is $510\pm140$\,M$_{\odot}$\,yr$^{-1}$, which is comparable to the previous 100\,Myr-averaged SFR of \citet{coor14}, or $654_{-90}^{104}$\,M$_{\odot}$\,yr$^{-1}$. But if S1, S2, W1, and/or W2 contain AGN (which is not ruled out by their line ratios, see Section \ref{linerat}), then their SFR$_{H\alpha}$ values may be contaminated (e.g., \citealt{garn10}). On the other hand, if these sources contain star formation in optically thick regions, then the true SFR may be higher.

In summary, the star formation (as traced by H$\alpha$) is not concentrated in a single source but is distributed in multiple objects: a central lensed arc that has two components ($\sim$32$\%$ of the SFR), two galaxies to the south ($\sim$17$\%$ of the SFR), and two galaxies to the west ($\sim$51$\%$ of the SFR). While the absolute SFR of this system is dependent on the presence of dust and AGN (which require further high-resolution radio and submm observations to be confirmed), the current data imply that HFLS3 is not a single starbursting galaxy, as previously reported.

\begin{table}
\centering
\caption{Observed B-V colour excess derived from Balmer-decrement, as well as integrated \ha flux and resulting SFR corrected for dust reddening and gravitational magnification.}
\begin{tabular}{c|c|cc}
Component & E(B-V)$_{BD}$ & $F_{H\alpha,intrinsic}$ & SFR$_{H\alpha}$ \\ \hline
C & $0.58\pm0.14$ & $4.6\pm2.3$ & $160\pm80$\\
S1 & $0.22\pm0.05$ & $1.6\pm0.3$ & $58\pm10$\\
S2 & $0.18\pm0.09$ & $0.8\pm0.2$ & $29\pm9$\\
W1 & $0.27\pm0.16$ & $2.8\pm1.6$ & $100\pm60$\\
W2 & $0.52\pm0.17$ & $4.4\pm2.8$ & $160\pm100$\\
\end{tabular}
\tablefoot{Flux and SFR values are in units of [$10^{-17}$\,erg\,s$^{-1}$\,cm$^{-2}$] and [M$_{\odot}$\,yr$^{-1}$], respectively. We assume a \citet{salp55} IMF and \citet{calz00} extinction law.}
\label{sfr_dust}
\end{table}

\subsection{HFLS3: A galaxy group}\label{groupdisc}
In the previous Sections, we have found that the HFLS3 field contains at least six strongly detected sources at $z\approx6.35$. Here, we discuss the implications of this apparent high density of galaxies.

Previous large-scale optical searches around HFLS3 found little evidence for an overdensity on the scale of $\gtrsim100$\,kpc (\citealt{robs14,lapo15}). However, \citet{lapo15} detected three faint optical companions to HFLS3 on small scales ($\sim36$\,kpc), based on HST imaging with comparable spatial resolution to the current work. Two of these sources are detected in our data (S and W), while the third source falls outside the IFU FoV. Since we have found that each of these sources and the central source are composed of two galaxies, the true density is likely higher. Whether this represents a true galaxy overdensity requires knowledge of the H$\alpha$ emitter space density at $z>6$, which is not yet well predicted (e.g., \citealt{pozz16}). Alternatively, we may compare the observed galaxy distribution to a cosmological simulation that includes H$\alpha$ emission and dust extinction (e.g., \citealt{hash23}). This will be deferred to a future work.

At this time, we may use the spatial and velocity offsets of each galaxy to see if they are likely gravitationally interacting. Since these values are projected, they represent lower limits on the true three-dimensional offset and relative velocity of galaxy pairs, but may be used as a first test of close association. Studies of galaxy pairs usually adopt criteria of $\Delta r \lesssim20$\,kpc and $\Delta$v$\lesssim500$\,km\,s$^{-1}$ for `close' pairs (e.g., \citealt{dunc19,vent19,roma21}). The six $z>6$ galaxies detected here are all located within a circle of diameter $\sim2''$ ($\sim11$\,kpc at $z=6.34$) with a maximum redshift difference of $\delta z\sim0.02$ (i.e., a line of sight velocity difference of $\sim800$\,km\,s$^{-1}$). So while each galaxy in this field firmly meets the distance criterion, the total velocity difference is high.

However, each galaxy is a member of a pair with smaller spatial offsets and redshift offsets: C1 and C2 ($\delta z\sim0.0092$, using the redshifts of \citealt{riec13}), S1 and S2 ($\delta z\sim0.0001$), and W1 and W2 ($\delta z\sim0.0078$). Since the distance and velocity criteria are met for each pair, they are likely interacting. In addition, the galaxy pairs in the S and W regions are also likely interacting, while C features a slightly higher velocity offset ($\Delta$v$=591\pm11$\,km\,s$^{-1}$ with respect to $z_{W1}$). Based on these small velocity and spatial offsets, it is likely that these galaxies represent an interacting system that will merge within $\sim1$\,Gyr (e.g., \citealt{cons06}).

\section{Conclusion}\label{conc}

In this work, we present JWST/NIRSpec IFU observations of a field containing the $z=6.34$ source HFLS3, as part of the GA-NIFS programme. By exploring both the low ($R\sim100$) and high spectral resolution ($R\sim2700$) data, we find a crowded field, with two low-redshift sources (G1, $z\sim3.4806$; G2, $z\sim2.00$), a central gravitationally lensed arc that is composed of two sources (C, $z\sim6.3425$), two close galaxies at the same redshift to the south (S1 \& S2, $z\sim6.3592$), and two close galaxies with a velocity offset to the west (W1, $z\sim6.3550$; W2, $z\sim6.3628$). All of the $z>6$ galaxies are located within an area of $\sim2''$, or $\sim11\,$kpc at $z\sim6.34$.

The spectral fits and morpho-kinematic map analysis of these data reveal a variety of kinematic features. G1 has no strong velocity gradient and is likely a dispersion-dominated source. However, C features a strong velocity gradient across the length of its arc, hinting at possible merging activity. The galaxies in component S are distinct, but are at similar redshifts with no velocity gradient. This is different from the galaxies in component W, which are closely associated but have a strong velocity gradient as seen in the double-peak profile of each line. The red component ($W2$) is brighter but has a lower velocity dispersion. Because of the asymmetry in the integrated intensity and velocity dispersion maps, this likely represents a merger between two galaxies, rather than a single rotating disk.

Next, we use our high-quality IFU data and updated source redshifts to examine the gravitational lensing of the HFLS3 field. Our best-fit models show that G1 is moderately magnified by G2 ($\mu\sim1.1-1.2$), while the \ha emission of C is more strongly magnified ($\mu\sim2.1-2.4$, comparable to the value of $\sim2.2$ from \citealt{coor14}) and is composed of two closely separated components.

The integrated line fluxes are plotted on [NII]-BPT and [SII]-VO87 plots, showing that component C is likely powered by star formation, while we may not rule out or confirm the presence of AGN in the S and W components. 

The observed Balmer decrements are then used to derive extinction-corrected SFR$_{\rm H\alpha}$ values for each source. This shows that the star formation is distributed across the field ($\rm SFR_{\rm H\alpha}=510\pm140$\,M$_{\odot}$\,yr$^{-1}$, corrected for lensing and extinction), with the largest contribution ($\sim51\%$) from the W galaxies. However, the presence of AGN in some sources may inflate these SFRs.

We use the projected spatial offsets and relative line of sight velocities of each $z>6$ galaxy to investigate whether they likely represent a closed, merging group. Using the standard `close pair' criteria of $\delta r<$20\,kpc and $\delta$v$<500$\,km\,s$^{-1}$, we find that the field contains multiple likely mergers: C1/C2, S1/S2, and W1/W2. In addition, the combined W and S components meet these criteria and will likely merge. While the C components feature a slightly higher velocity offset ($\sim600$\,km\,s$^{-1}$), the HFLS3 group contains multiple close pairs.

Taken together, our results require a drastic reinterpretation of the HFLS3 field. It is composed of at least six distinct sources within $z\sim6.34-6.36$ that are lensed by two foreground galaxies at $z\sim2.1$ and $z\sim3.5$. All of the $z>6$ galaxies either feature strong velocity gradients and/or are closely associated with another galaxy, implying ongoing interaction. This behaviour has been seen at high redshift (e.g., \citealt{gino22}), including four galaxies at $z\sim7.9$ detected with the NIRSpec IFU (\citealt{hash23}). Thus, HFLS3 is likely not an extreme starburst, but instead represents one of the densest groups of interacting star-forming galaxies within the first 1\,Gyr of the Universe. Recent and ongoing high-resolution observations with JWST/MIRI, NOEMA, and the JVLA (as well as a future in-depth study of the R100 data cube briefly explored here) will help to further characterise this unique field.

\section*{Acknowledgements}
GCJ, AJB, AJC, JC, AS acknowledges funding from the ``FirstGalaxies'' Advanced Grant from the European Research Council (ERC) under the European Union's Horizon 2020 research and innovation programme (Grant agreement No. 789056).
H{\"U} gratefully acknowledges support by the Isaac Newton Trust and by the Kavli Foundation through a Newton-Kavli Junior Fellowship.
MP, SA, and BRdP acknowledge grant PID2021-127718NB-I00 funded by the Spanish Ministry of Science and Innovation/State Agency of Research (MICIN/AEI/ 10.13039/501100011033).
MP also acknowledges Programa Atracci\'on de Talento de la Comunidad de Madrid via grant 2018-T2/TIC-11715.
SC and GV acknowledge support from the European Union (ERC, WINGS,101040227).
RM, FDE, JS, and JW acknowledge support by the Science and Technology Facilities Council (STFC), from the ERC Advanced Grant 695671 ``QUENCH'', and by the UKRI Frontier Research grant RISEandFALL.
RM also acknowledges funding from a research professorship from the Royal Society.
RB acknowledges support from an STFC Ernest Rutherford Fellowship [grant number ST/T003596/1].
GC acknowledges the support of the INAF Large Grant 2022 ``The metal circle: a new sharp view of the baryon cycle up to Cosmic Dawn with the latest generation IFU facilities''.
This work has made use of data from the European Space Agency (ESA) mission {\it Gaia} (\url{https://www.cosmos.esa.int/gaia}), processed by the {\it Gaia} Data Processing and Analysis Consortium (DPAC, \url{https://www.cosmos.esa.int/web/gaia/dpac/consortium}). Funding for the DPAC has been provided by national institutions, in particular the institutions participating in the {\it Gaia} Multilateral Agreement.

\bibliographystyle{aa}
\bibliography{references}

\begin{thebibliography}{83}
\expandafter\ifx\csname natexlab\endcsname\relax\def\natexlab#1{#1}\fi

\bibitem[{{Allen} {et~al.}(2008){Allen}, {Groves}, {Dopita}, {Sutherland}, \& {Kewley}}]{alle08}
{Allen}, M.~G., {Groves}, B.~A., {Dopita}, M.~A., {Sutherland}, R.~S., \& {Kewley}, L.~J. 2008, \apjs, 178, 20

\bibitem[{{Arata} {et~al.}(2020){Arata}, {Yajima}, {Nagamine}, {Abe}, \& {Khochfar}}]{arat20}
{Arata}, S., {Yajima}, H., {Nagamine}, K., {Abe}, M., \& {Khochfar}, S. 2020, \mnras, 498, 5541

\bibitem[{{Armstrong}(1967)}]{arms67}
{Armstrong}, B. 1967, \jqsrt, 7, 61

\bibitem[{{Bacon} {et~al.}(2016){Bacon}, {Piqueras}, {Conseil}, {Richard}, \& {Shepherd}}]{baco16}
{Bacon}, R., {Piqueras}, L., {Conseil}, S., {Richard}, J., \& {Shepherd}, M. 2016, {MPDAF: MUSE Python Data Analysis Framework}, Astrophysics Source Code Library, record ascl:1611.003

\bibitem[{{Baldwin} {et~al.}(1981){Baldwin}, {Phillips}, \& {Terlevich}}]{bald81}
{Baldwin}, J.~A., {Phillips}, M.~M., \& {Terlevich}, R. 1981, \pasp, 93, 5

\bibitem[{{Belli} {et~al.}(2023){Belli}, {Park}, {Davies}, {Mendel}, {Johnson}, {Conroy}, {Benton}, {Bugiani}, {Emami}, {Leja}, {Li}, {Maheson}, {Mathews}, {Naidu}, {Nelson}, {Tacchella}, {Terrazas}, \& {Weinberger}}]{bell23}
{Belli}, S., {Park}, M., {Davies}, R.~L., {et~al.} 2023, arXiv e-prints, arXiv:2308.05795

\bibitem[{{B{\"o}ker} {et~al.}(2022){B{\"o}ker}, {Arribas}, {L{\"u}tzgendorf}, {Alves de Oliveira}, {Beck}, {Birkmann}, {Bunker}, {Charlot}, {de Marchi}, {Ferruit}, {Giardino}, {Jakobsen}, {Kumari}, {L{\'o}pez-Caniego}, {Maiolino}, {Manjavacas}, {Marston}, {Moseley}, {Muzerolle}, {Ogle}, {Pirzkal}, {Rauscher}, {Rawle}, {Rix}, {Sabbi}, {Sargent}, {Sirianni}, {te Plate}, {Valenti}, {Willott}, \& {Zeidler}}]{boke22}
{B{\"o}ker}, T., {Arribas}, S., {L{\"u}tzgendorf}, N., {et~al.} 2022, \aap, 661, A82

\bibitem[{{B{\"o}ker} {et~al.}(2023){B{\"o}ker}, {Beck}, {Birkmann}, {Giardino}, {Keyes}, {Kumari}, {Muzerolle}, {Rawle}, {Zeidler}, {Abul-Huda}, {Alves de Oliveira}, {Arribas}, {Bechtold}, {Bhatawdekar}, {Bonaventura}, {Bunker}, {Cameron}, {Carniani}, {Charlot}, {Curti}, {Espinoza}, {Ferruit}, {Franx}, {Jakobsen}, {Karakla}, {L{\'o}pez-Caniego}, {L{\"u}tzgendorf}, {Maiolino}, {Manjavacas}, {Marston}, {Moseley}, {Ogle}, {Perna}, {Pe{\~n}a-Guerrero}, {Pirzkal}, {Plesha}, {Proffitt}, {Rauscher}, {Rix}, {Rodr{\'\i}guez del Pino}, {Rustamkulov}, {Sabbi}, {Sing}, {Sirianni}, {te Plate}, {{\'U}beda}, {Wahlgren}, {Wislowski}, {Wu}, \& {Willott}}]{boke23}
{B{\"o}ker}, T., {Beck}, T.~L., {Birkmann}, S.~M., {et~al.} 2023, \pasp, 135, 038001

\bibitem[{{Bouwens} {et~al.}(2015){Bouwens}, {Illingworth}, {Oesch}, {Trenti}, {Labb{\'e}}, {Bradley}, {Carollo}, {van Dokkum}, {Gonzalez}, {Holwerda}, {Franx}, {Spitler}, {Smit}, \& {Magee}}]{bouw15}
{Bouwens}, R.~J., {Illingworth}, G.~D., {Oesch}, P.~A., {et~al.} 2015, \apj, 803, 34

\bibitem[{{Brinchmann} {et~al.}(2004){Brinchmann}, {Charlot}, {White}, {Tremonti}, {Kauffmann}, {Heckman}, \& {Brinkmann}}]{brin04}
{Brinchmann}, J., {Charlot}, S., {White}, S.~D.~M., {et~al.} 2004, \mnras, 351, 1151

\bibitem[{{Calzetti} {et~al.}(2000){Calzetti}, {Armus}, {Bohlin}, {Kinney}, {Koornneef}, \& {Storchi-Bergmann}}]{calz00}
{Calzetti}, D., {Armus}, L., {Bohlin}, R.~C., {et~al.} 2000, \apj, 533, 682

\bibitem[{{Cameron} {et~al.}(2023){Cameron}, {Saxena}, {Bunker}, {D'Eugenio}, {Carniani}, {Maiolino}, {Curtis-Lake}, {Ferruit}, {Jakobsen}, {Arribas}, {Bonaventura}, {Charlot}, {Chevallard}, {Curti}, {Looser}, {Maseda}, {Rawle}, {Rodr{\'\i}guez Del Pino}, {Smit}, {{\"U}bler}, {Willott}, {Witstok}, {Egami}, {Eisenstein}, {Johnson}, {Hainline}, {Rieke}, {Robertson}, {Stark}, {Tacchella}, {Williams}, {Willmer}, {Bhatawdekar}, {Bowler}, {Boyett}, {Circosta}, {Helton}, {Jones}, {Kumari}, {Ji}, {Nelson}, {Parlanti}, {Sandles}, {Scholtz}, \& {Sun}}]{came23}
{Cameron}, A.~J., {Saxena}, A., {Bunker}, A.~J., {et~al.} 2023, \aap, 677, A115

\bibitem[{{Carnall} {et~al.}(2023){Carnall}, {McLeod}, {McLure}, {Dunlop}, {Begley}, {Cullen}, {Donnan}, {Hamadouche}, {Jewell}, {Jones}, {Pollock}, \& {Wild}}]{carn23}
{Carnall}, A.~C., {McLeod}, D.~J., {McLure}, R.~J., {et~al.} 2023, \mnras, 520, 3974

\bibitem[{{Carniani} {et~al.}(2019){Carniani}, {Gallerani}, {Vallini}, {Pallottini}, {Tazzari}, {Ferrara}, {Maiolino}, {Cicone}, {Feruglio}, {Neri}, {D'Odorico}, {Wang}, \& {Li}}]{carn19}
{Carniani}, S., {Gallerani}, S., {Vallini}, L., {et~al.} 2019, \mnras, 489, 3939

\bibitem[{{Cheng} {et~al.}(2020){Cheng}, {Cao}, {Lu}, {Li}, {Yang}, {Rigopoulou}, {Charmandaris}, {Gao}, {Xu}, {van der Werf}, {Diaz Santos}, {Privon}, {Zhao}, {Cao}, {Dai}, {Huang}, {Sanders}, {Wang}, {Wang}, \& {Zhu}}]{chen20}
{Cheng}, C., {Cao}, X., {Lu}, N., {et~al.} 2020, \apj, 898, 33

\bibitem[{{Conselice}(2006)}]{cons06}
{Conselice}, C.~J. 2006, \apj, 638, 686

\bibitem[{{Cooray} {et~al.}(2014){Cooray}, {Calanog}, {Wardlow}, {Bock}, {Bridge}, {Burgarella}, {Bussmann}, {Casey}, {Clements}, {Conley}, {Farrah}, {Fu}, {Gavazzi}, {Ivison}, {La Porte}, {Lo Faro}, {Ma}, {Magdis}, {Oliver}, {Osage}, {P{\'e}rez-Fournon}, {Riechers}, {Rigopoulou}, {Scott}, {Viero}, \& {Watson}}]{coor14}
{Cooray}, A., {Calanog}, J., {Wardlow}, J.~L., {et~al.} 2014, \apj, 790, 40

\bibitem[{{Curti} {et~al.}(2020){Curti}, {Mannucci}, {Cresci}, \& {Maiolino}}]{curt20}
{Curti}, M., {Mannucci}, F., {Cresci}, G., \& {Maiolino}, R. 2020, \mnras, 491, 944

\bibitem[{{Decarli} {et~al.}(2019){Decarli}, {Walter}, {G{\'o}nzalez-L{\'o}pez}, {Aravena}, {Boogaard}, {Carilli}, {Cox}, {Daddi}, {Popping}, {Riechers}, {Uzgil}, {Weiss}, {Assef}, {Bacon}, {Bauer}, {Bertoldi}, {Bouwens}, {Contini}, {Cortes}, {da Cunha}, {D{\'\i}az-Santos}, {Elbaz}, {Inami}, {Hodge}, {Ivison}, {Le F{\`e}vre}, {Magnelli}, {Novak}, {Oesch}, {Rix}, {Sargent}, {Smail}, {Swinbank}, {Somerville}, {van der Werf}, {Wagg}, \& {Wisotzki}}]{deca19}
{Decarli}, R., {Walter}, F., {G{\'o}nzalez-L{\'o}pez}, J., {et~al.} 2019, \apj, 882, 138

\bibitem[{{D'Eugenio} {et~al.}(2023){D'Eugenio}, {Perez-Gonzalez}, {Maiolino}, {Scholtz}, {Perna}, {Circosta}, {Uebler}, {Arribas}, {Boeker}, {Bunker}, {Carniani}, {Charlot}, {Chevallard}, {Cresci}, {Curtis-Lake}, {Jones}, {Kumari}, {Lamperti}, {Looser}, {Parlanti}, {Rix}, {Robertson}, {Rodriguez Del Pino}, {Tacchella}, {Venturi}, \& {Willott}}]{DEugenio2023}
{D'Eugenio}, F., {Perez-Gonzalez}, P., {Maiolino}, R., {et~al.} 2023, arXiv e-prints, arXiv:2308.06317

\bibitem[{{Dimitrijevi{\'c}} {et~al.}(2007){Dimitrijevi{\'c}}, {Popovi{\'c}}, {Kova{\v{c}}evi{\'c}}, {Da{\v{c}}i{\'c}}, \& {Ili{\'c}}}]{dimi07}
{Dimitrijevi{\'c}}, M.~S., {Popovi{\'c}}, L.~{\v{C}}., {Kova{\v{c}}evi{\'c}}, J., {Da{\v{c}}i{\'c}}, M., \& {Ili{\'c}}, D. 2007, \mnras, 374, 1181

\bibitem[{{Doj{\v{c}}inovi{\'c}} {et~al.}(2023){Doj{\v{c}}inovi{\'c}}, {Kova{\v{c}}evi{\'c}-Doj{\v{c}}inovi{\'c}}, \& {Popovi{\'c}}}]{dojc23}
{Doj{\v{c}}inovi{\'c}}, I., {Kova{\v{c}}evi{\'c}-Doj{\v{c}}inovi{\'c}}, J., \& {Popovi{\'c}}, L.~{\v{C}}. 2023, Advances in Space Research, 71, 1219

\bibitem[{{Dome} {et~al.}(2024){Dome}, {Tacchella}, {Fialkov}, {Ceverino}, {Dekel}, {Ginzburg}, {Lapiner}, \& {Looser}}]{dome23}
{Dome}, T., {Tacchella}, S., {Fialkov}, A., {et~al.} 2024, \mnras, 527, 2139

\bibitem[{{Dom{\'\i}nguez} {et~al.}(2013){Dom{\'\i}nguez}, {Siana}, {Henry}, {Scarlata}, {Bedregal}, {Malkan}, {Atek}, {Ross}, {Colbert}, {Teplitz}, {Rafelski}, {McCarthy}, {Bunker}, {Hathi}, {Dressler}, {Martin}, \& {Masters}}]{domi13}
{Dom{\'\i}nguez}, A., {Siana}, B., {Henry}, A.~L., {et~al.} 2013, \apj, 763, 145

\bibitem[{{Duarte Puertas} {et~al.}(2021){Duarte Puertas}, {Vilchez}, {Iglesias-P{\'a}ramo}, {Drissen}, {Kehrig}, {Martin}, {P{\'e}rez-Montero}, \& {Arroyo-Polonio}}]{duar21}
{Duarte Puertas}, S., {Vilchez}, J.~M., {Iglesias-P{\'a}ramo}, J., {et~al.} 2021, \aap, 645, A57

\bibitem[{{Duncan} {et~al.}(2019{\natexlab{a}}){Duncan}, {Conselice}, {Mundy}, {Bell}, {Donley}, {Galametz}, {Guo}, {Grogin}, {Hathi}, {Kartaltepe}, {Kocevski}, {Koekemoer}, {P{\'e}rez-Gonz{\'a}lez}, {Mantha}, {Snyder}, \& {Stefanon}}]{duca19}
{Duncan}, K., {Conselice}, C.~J., {Mundy}, C., {et~al.} 2019{\natexlab{a}}, \apj, 876, 110

\bibitem[{{Duncan} {et~al.}(2019{\natexlab{b}}){Duncan}, {Conselice}, {Mundy}, {Bell}, {Donley}, {Galametz}, {Guo}, {Grogin}, {Hathi}, {Kartaltepe}, {Kocevski}, {Koekemoer}, {P{\'e}rez-Gonz{\'a}lez}, {Mantha}, {Snyder}, \& {Stefanon}}]{dunc19}
{Duncan}, K., {Conselice}, C.~J., {Mundy}, C., {et~al.} 2019{\natexlab{b}}, \apj, 876, 110

\bibitem[{{Fadda} {et~al.}(2004){Fadda}, {Jannuzi}, {Ford}, \& {Storrie-Lombardi}}]{fadd04}
{Fadda}, D., {Jannuzi}, B.~T., {Ford}, A., \& {Storrie-Lombardi}, L.~J. 2004, \aj, 128, 1

\bibitem[{{Fruchter} \& {Hook}(2002)}]{fruc02}
{Fruchter}, A.~S. \& {Hook}, R.~N. 2002, \pasp, 114, 144

\bibitem[{{Gaia Collaboration} {et~al.}(2021){Gaia Collaboration}, {Brown}, {Vallenari}, {Prusti}, {de Bruijne}, {Babusiaux}, {Biermann}, {Creevey}, {Evans}, {Eyer}, {Hutton}, {Jansen}, {Jordi}, {Klioner}, {Lammers}, {Lindegren}, {Luri}, {Mignard}, {Panem}, {Pourbaix}, {Randich}, {Sartoretti}, {Soubiran}, {Walton}, {Arenou}, {Bailer-Jones}, {Bastian}, {Cropper}, {Drimmel}, {Katz}, {Lattanzi}, {van Leeuwen}, {Bakker}, {Cacciari}, {Casta{\~n}eda}, {De Angeli}, {Ducourant}, {Fabricius}, {Fouesneau}, {Fr{\'e}mat}, {Guerra}, {Guerrier}, {Guiraud}, {Jean-Antoine Piccolo}, {Masana}, {Messineo}, {Mowlavi}, {Nicolas}, {Nienartowicz}, {Pailler}, {Panuzzo}, {Riclet}, {Roux}, {Seabroke}, {Sordo}, {Tanga}, {Th{\'e}venin}, {Gracia-Abril}, {Portell}, {Teyssier}, {Altmann}, {Andrae}, {Bellas-Velidis}, {Benson}, {Berthier}, {Blomme}, {Brugaletta}, {Burgess}, {Busso}, {Carry}, {Cellino}, {Cheek}, {Clementini}, {Damerdji}, {Davidson}, {Delchambre}, {Dell'Oro}, {Fern{\'a}ndez-Hern{\'a}ndez}, {Galluccio}, {Garc{\'\i}a-Lario},
  {Garcia-Reinaldos}, {Gonz{\'a}lez-N{\'u}{\~n}ez}, {Gosset}, {Haigron}, {Halbwachs}, {Hambly}, {Harrison}, {Hatzidimitriou}, {Heiter}, {Hern{\'a}ndez}, {Hestroffer}, {Hodgkin}, {Holl}, {Jan{\ss}en}, {Jevardat de Fombelle}, {Jordan}, {Krone-Martins}, {Lanzafame}, {L{\"o}ffler}, {Lorca}, {Manteiga}, {Marchal}, {Marrese}, {Moitinho}, {Mora}, {Muinonen}, {Osborne}, {Pancino}, {Pauwels}, {Petit}, {Recio-Blanco}, {Richards}, {Riello}, {Rimoldini}, {Robin}, {Roegiers}, {Rybizki}, {Sarro}, {Siopis}, {Smith}, {Sozzetti}, {Ulla}, {Utrilla}, {van Leeuwen}, {van Reeven}, {Abbas}, {Abreu Aramburu}, {Accart}, {Aerts}, {Aguado}, {Ajaj}, {Altavilla}, {{\'A}lvarez}, {{\'A}lvarez Cid-Fuentes}, {Alves}, {Anderson}, {Anglada Varela}, {Antoja}, {Audard}, {Baines}, {Baker}, {Balaguer-N{\'u}{\~n}ez}, {Balbinot}, {Balog}, {Barache}, {Barbato}, {Barros}, {Barstow}, {Bartolom{\'e}}, {Bassilana}, {Bauchet}, {Baudesson-Stella}, {Becciani}, {Bellazzini}, {Bernet}, {Bertone}, {Bianchi}, {Blanco-Cuaresma}, {Boch}, {Bombrun}, {Bossini},
  {Bouquillon}, {Bragaglia}, {Bramante}, {Breedt}, {Bressan}, {Brouillet}, {Bucciarelli}, {Burlacu}, {Busonero}, {Butkevich}, {Buzzi}, {Caffau}, {Cancelliere}, {C{\'a}novas}, {Cantat-Gaudin}, {Carballo}, {Carlucci}, {Carnerero}, {Carrasco}, {Casamiquela}, {Castellani}, {Castro-Ginard}, {Castro Sampol}, {Chaoul}, {Charlot}, {Chemin}, {Chiavassa}, {Cioni}, {Comoretto}, {Cooper}, {Cornez}, {Cowell}, {Crifo}, {Crosta}, {Crowley}, {Dafonte}, {Dapergolas}, {David}, {David}, {de Laverny}, {De Luise}, {De March}, {De Ridder}, {de Souza}, {de Teodoro}, {de Torres}, {del Peloso}, {del Pozo}, {Delbo}, {Delgado}, {Delgado}, {Delisle}, {Di Matteo}, {Diakite}, {Diener}, {Distefano}, {Dolding}, {Eappachen}, {Edvardsson}, {Enke}, {Esquej}, {Fabre}, {Fabrizio}, {Faigler}, {Fedorets}, {Fernique}, {Fienga}, {Figueras}, {Fouron}, {Fragkoudi}, {Fraile}, {Franke}, {Gai}, {Garabato}, {Garcia-Gutierrez}, {Garc{\'\i}a-Torres}, {Garofalo}, {Gavras}, {Gerlach}, {Geyer}, {Giacobbe}, {Gilmore}, {Girona}, {Giuffrida}, {Gomel}, {Gomez},
  {Gonzalez-Santamaria}, {Gonz{\'a}lez-Vidal}, {Granvik}, {Guti{\'e}rrez-S{\'a}nchez}, {Guy}, {Hauser}, {Haywood}, {Helmi}, {Hidalgo}, {Hilger}, {H{\l}adczuk}, {Hobbs}, {Holland}, {Huckle}, {Jasniewicz}, {Jonker}, {Juaristi Campillo}, {Julbe}, {Karbevska}, {Kervella}, {Khanna}, {Kochoska}, {Kontizas}, {Kordopatis}, {Korn}, {Kostrzewa-Rutkowska}, {Kruszy{\'n}ska}, {Lambert}, {Lanza}, {Lasne}, {Le Campion}, {Le Fustec}, {Lebreton}, {Lebzelter}, {Leccia}, {Leclerc}, {Lecoeur-Taibi}, {Liao}, {Licata}, {Lindstr{\o}m}, {Lister}, {Livanou}, {Lobel}, {Madrero Pardo}, {Managau}, {Mann}, {Marchant}, {Marconi}, {Marcos Santos}, {Marinoni}, {Marocco}, {Marshall}, {Martin Polo}, {Mart{\'\i}n-Fleitas}, {Masip}, {Massari}, {Mastrobuono-Battisti}, {Mazeh}, {McMillan}, {Messina}, {Michalik}, {Millar}, {Mints}, {Molina}, {Molinaro}, {Moln{\'a}r}, {Montegriffo}, {Mor}, {Morbidelli}, {Morel}, {Morris}, {Mulone}, {Munoz}, {Muraveva}, {Murphy}, {Musella}, {Noval}, {Ord{\'e}novic}, {Orr{\`u}}, {Osinde}, {Pagani}, {Pagano},
  {Palaversa}, {Palicio}, {Panahi}, {Pawlak}, {Pe{\~n}alosa Esteller}, {Penttil{\"a}}, {Piersimoni}, {Pineau}, {Plachy}, {Plum}, {Poggio}, {Poretti}, {Poujoulet}, {Pr{\v{s}}a}, {Pulone}, {Racero}, {Ragaini}, {Rainer}, {Raiteri}, {Rambaux}, {Ramos}, {Ramos-Lerate}, {Re Fiorentin}, {Regibo}, {Reyl{\'e}}, {Ripepi}, {Riva}, {Rixon}, {Robichon}, {Robin}, {Roelens}, {Rohrbasser}, {Romero-G{\'o}mez}, {Rowell}, {Royer}, {Rybicki}, {Sadowski}, {Sagrist{\`a} Sell{\'e}s}, {Sahlmann}, {Salgado}, {Salguero}, {Samaras}, {Sanchez Gimenez}, {Sanna}, {Santove{\~n}a}, {Sarasso}, {Schultheis}, {Sciacca}, {Segol}, {Segovia}, {S{\'e}gransan}, {Semeux}, {Shahaf}, {Siddiqui}, {Siebert}, {Siltala}, {Slezak}, {Smart}, {Solano}, {Solitro}, {Souami}, {Souchay}, {Spagna}, {Spoto}, {Steele}, {Steidelm{\"u}ller}, {Stephenson}, {S{\"u}veges}, {Szabados}, {Szegedi-Elek}, {Taris}, {Tauran}, {Taylor}, {Teixeira}, {Thuillot}, {Tonello}, {Torra}, {Torra}, {Turon}, {Unger}, {Vaillant}, {van Dillen}, {Vanel}, {Vecchiato}, {Viala}, {Vicente},
  {Voutsinas}, {Weiler}, {Wevers}, {Wyrzykowski}, {Yoldas}, {Yvard}, {Zhao}, {Zorec}, {Zucker}, {Zurbach}, \& {Zwitter}}]{gaia21}
{Gaia Collaboration}, {Brown}, A.~G.~A., {Vallenari}, A., {et~al.} 2021, \aap, 649, A1

\bibitem[{{Gaia Collaboration} {et~al.}(2016){Gaia Collaboration}, {Prusti}, {de Bruijne}, {Brown}, {Vallenari}, {Babusiaux}, {Bailer-Jones}, {Bastian}, {Biermann}, {Evans}, {Eyer}, {Jansen}, {Jordi}, {Klioner}, {Lammers}, {Lindegren}, {Luri}, {Mignard}, {Milligan}, {Panem}, {Poinsignon}, {Pourbaix}, {Randich}, {Sarri}, {Sartoretti}, {Siddiqui}, {Soubiran}, {Valette}, {van Leeuwen}, {Walton}, {Aerts}, {Arenou}, {Cropper}, {Drimmel}, {H{\o}g}, {Katz}, {Lattanzi}, {O'Mullane}, {Grebel}, {Holland}, {Huc}, {Passot}, {Bramante}, {Cacciari}, {Casta{\~n}eda}, {Chaoul}, {Cheek}, {De Angeli}, {Fabricius}, {Guerra}, {Hern{\'a}ndez}, {Jean-Antoine-Piccolo}, {Masana}, {Messineo}, {Mowlavi}, {Nienartowicz}, {Ord{\'o}{\~n}ez-Blanco}, {Panuzzo}, {Portell}, {Richards}, {Riello}, {Seabroke}, {Tanga}, {Th{\'e}venin}, {Torra}, {Els}, {Gracia-Abril}, {Comoretto}, {Garcia-Reinaldos}, {Lock}, {Mercier}, {Altmann}, {Andrae}, {Astraatmadja}, {Bellas-Velidis}, {Benson}, {Berthier}, {Blomme}, {Busso}, {Carry}, {Cellino}, {Clementini},
  {Cowell}, {Creevey}, {Cuypers}, {Davidson}, {De Ridder}, {de Torres}, {Delchambre}, {Dell'Oro}, {Ducourant}, {Fr{\'e}mat}, {Garc{\'\i}a-Torres}, {Gosset}, {Halbwachs}, {Hambly}, {Harrison}, {Hauser}, {Hestroffer}, {Hodgkin}, {Huckle}, {Hutton}, {Jasniewicz}, {Jordan}, {Kontizas}, {Korn}, {Lanzafame}, {Manteiga}, {Moitinho}, {Muinonen}, {Osinde}, {Pancino}, {Pauwels}, {Petit}, {Recio-Blanco}, {Robin}, {Sarro}, {Siopis}, {Smith}, {Smith}, {Sozzetti}, {Thuillot}, {van Reeven}, {Viala}, {Abbas}, {Abreu Aramburu}, {Accart}, {Aguado}, {Allan}, {Allasia}, {Altavilla}, {{\'A}lvarez}, {Alves}, {Anderson}, {Andrei}, {Anglada Varela}, {Antiche}, {Antoja}, {Ant{\'o}n}, {Arcay}, {Atzei}, {Ayache}, {Bach}, {Baker}, {Balaguer-N{\'u}{\~n}ez}, {Barache}, {Barata}, {Barbier}, {Barblan}, {Baroni}, {Barrado y Navascu{\'e}s}, {Barros}, {Barstow}, {Becciani}, {Bellazzini}, {Bellei}, {Bello Garc{\'\i}a}, {Belokurov}, {Bendjoya}, {Berihuete}, {Bianchi}, {Bienaym{\'e}}, {Billebaud}, {Blagorodnova}, {Blanco-Cuaresma}, {Boch},
  {Bombrun}, {Borrachero}, {Bouquillon}, {Bourda}, {Bouy}, {Bragaglia}, {Breddels}, {Brouillet}, {Br{\"u}semeister}, {Bucciarelli}, {Budnik}, {Burgess}, {Burgon}, {Burlacu}, {Busonero}, {Buzzi}, {Caffau}, {Cambras}, {Campbell}, {Cancelliere}, {Cantat-Gaudin}, {Carlucci}, {Carrasco}, {Castellani}, {Charlot}, {Charnas}, {Charvet}, {Chassat}, {Chiavassa}, {Clotet}, {Cocozza}, {Collins}, {Collins}, {Costigan}, {Crifo}, {Cross}, {Crosta}, {Crowley}, {Dafonte}, {Damerdji}, {Dapergolas}, {David}, {David}, {De Cat}, {de Felice}, {de Laverny}, {De Luise}, {De March}, {de Martino}, {de Souza}, {Debosscher}, {del Pozo}, {Delbo}, {Delgado}, {Delgado}, {di Marco}, {Di Matteo}, {Diakite}, {Distefano}, {Dolding}, {Dos Anjos}, {Drazinos}, {Dur{\'a}n}, {Dzigan}, {Ecale}, {Edvardsson}, {Enke}, {Erdmann}, {Escolar}, {Espina}, {Evans}, {Eynard Bontemps}, {Fabre}, {Fabrizio}, {Faigler}, {Falc{\~a}o}, {Farr{\`a}s Casas}, {Faye}, {Federici}, {Fedorets}, {Fern{\'a}ndez-Hern{\'a}ndez}, {Fernique}, {Fienga}, {Figueras}, {Filippi},
  {Findeisen}, {Fonti}, {Fouesneau}, {Fraile}, {Fraser}, {Fuchs}, {Furnell}, {Gai}, {Galleti}, {Galluccio}, {Garabato}, {Garc{\'\i}a-Sedano}, {Gar{\'e}}, {Garofalo}, {Garralda}, {Gavras}, {Gerssen}, {Geyer}, {Gilmore}, {Girona}, {Giuffrida}, {Gomes}, {Gonz{\'a}lez-Marcos}, {Gonz{\'a}lez-N{\'u}{\~n}ez}, {Gonz{\'a}lez-Vidal}, {Granvik}, {Guerrier}, {Guillout}, {Guiraud}, {G{\'u}rpide}, {Guti{\'e}rrez-S{\'a}nchez}, {Guy}, {Haigron}, {Hatzidimitriou}, {Haywood}, {Heiter}, {Helmi}, {Hobbs}, {Hofmann}, {Holl}, {Holland}, {Hunt}, {Hypki}, {Icardi}, {Irwin}, {Jevardat de Fombelle}, {Jofr{\'e}}, {Jonker}, {Jorissen}, {Julbe}, {Karampelas}, {Kochoska}, {Kohley}, {Kolenberg}, {Kontizas}, {Koposov}, {Kordopatis}, {Koubsky}, {Kowalczyk}, {Krone-Martins}, {Kudryashova}, {Kull}, {Bachchan}, {Lacoste-Seris}, {Lanza}, {Lavigne}, {Le Poncin-Lafitte}, {Lebreton}, {Lebzelter}, {Leccia}, {Leclerc}, {Lecoeur-Taibi}, {Lemaitre}, {Lenhardt}, {Leroux}, {Liao}, {Licata}, {Lindstr{\o}m}, {Lister}, {Livanou}, {Lobel}, {L{\"o}ffler},
  {L{\'o}pez}, {Lopez-Lozano}, {Lorenz}, {Loureiro}, {MacDonald}, {Magalh{\~a}es Fernandes}, {Managau}, {Mann}, {Mantelet}, {Marchal}, {Marchant}, {Marconi}, {Marie}, {Marinoni}, {Marrese}, {Marschalk{\'o}}, {Marshall}, {Mart{\'\i}n-Fleitas}, {Martino}, {Mary}, {Matijevi{\v{c}}}, {Mazeh}, {McMillan}, {Messina}, {Mestre}, {Michalik}, {Millar}, {Miranda}, {Molina}, {Molinaro}, {Molinaro}, {Moln{\'a}r}, {Moniez}, {Montegriffo}, {Monteiro}, {Mor}, {Mora}, {Morbidelli}, {Morel}, {Morgenthaler}, {Morley}, {Morris}, {Mulone}, {Muraveva}, {Musella}, {Narbonne}, {Nelemans}, {Nicastro}, {Noval}, {Ord{\'e}novic}, {Ordieres-Mer{\'e}}, {Osborne}, {Pagani}, {Pagano}, {Pailler}, {Palacin}, {Palaversa}, {Parsons}, {Paulsen}, {Pecoraro}, {Pedrosa}, {Pentik{\"a}inen}, {Pereira}, {Pichon}, {Piersimoni}, {Pineau}, {Plachy}, {Plum}, {Poujoulet}, {Pr{\v{s}}a}, {Pulone}, {Ragaini}, {Rago}, {Rambaux}, {Ramos-Lerate}, {Ranalli}, {Rauw}, {Read}, {Regibo}, {Renk}, {Reyl{\'e}}, {Ribeiro}, {Rimoldini}, {Ripepi}, {Riva}, {Rixon},
  {Roelens}, {Romero-G{\'o}mez}, {Rowell}, {Royer}, {Rudolph}, {Ruiz-Dern}, {Sadowski}, {Sagrist{\`a} Sell{\'e}s}, {Sahlmann}, {Salgado}, {Salguero}, {Sarasso}, {Savietto}, {Schnorhk}, {Schultheis}, {Sciacca}, {Segol}, {Segovia}, {Segransan}, {Serpell}, {Shih}, {Smareglia}, {Smart}, {Smith}, {Solano}, {Solitro}, {Sordo}, {Soria Nieto}, {Souchay}, {Spagna}, {Spoto}, {Stampa}, {Steele}, {Steidelm{\"u}ller}, {Stephenson}, {Stoev}, {Suess}, {S{\"u}veges}, {Surdej}, {Szabados}, {Szegedi-Elek}, {Tapiador}, {Taris}, {Tauran}, {Taylor}, {Teixeira}, {Terrett}, {Tingley}, {Trager}, {Turon}, {Ulla}, {Utrilla}, {Valentini}, {van Elteren}, {Van Hemelryck}, {van Leeuwen}, {Varadi}, {Vecchiato}, {Veljanoski}, {Via}, {Vicente}, {Vogt}, {Voss}, {Votruba}, {Voutsinas}, {Walmsley}, {Weiler}, {Weingrill}, {Werner}, {Wevers}, {Whitehead}, {Wyrzykowski}, {Yoldas}, {{\v{Z}}erjal}, {Zucker}, {Zurbach}, {Zwitter}, {Alecu}, {Allen}, {Allende Prieto}, {Amorim}, {Anglada-Escud{\'e}}, {Arsenijevic}, {Azaz}, {Balm}, {Beck}, {Bernstein},
  {Bigot}, {Bijaoui}, {Blasco}, {Bonfigli}, {Bono}, {Boudreault}, {Bressan}, {Brown}, {Brunet}, {Bunclark}, {Buonanno}, {Butkevich}, {Carret}, {Carrion}, {Chemin}, {Ch{\'e}reau}, {Corcione}, {Darmigny}, {de Boer}, {de Teodoro}, {de Zeeuw}, {Delle Luche}, {Domingues}, {Dubath}, {Fodor}, {Fr{\'e}zouls}, {Fries}, {Fustes}, {Fyfe}, {Gallardo}, {Gallegos}, {Gardiol}, {Gebran}, {Gomboc}, {G{\'o}mez}, {Grux}, {Gueguen}, {Heyrovsky}, {Hoar}, {Iannicola}, {Isasi Parache}, {Janotto}, {Joliet}, {Jonckheere}, {Keil}, {Kim}, {Klagyivik}, {Klar}, {Knude}, {Kochukhov}, {Kolka}, {Kos}, {Kutka}, {Lainey}, {LeBouquin}, {Liu}, {Loreggia}, {Makarov}, {Marseille}, {Martayan}, {Martinez-Rubi}, {Massart}, {Meynadier}, {Mignot}, {Munari}, {Nguyen}, {Nordlander}, {Ocvirk}, {O'Flaherty}, {Olias Sanz}, {Ortiz}, {Osorio}, {Oszkiewicz}, {Ouzounis}, {Palmer}, {Park}, {Pasquato}, {Peltzer}, {Peralta}, {P{\'e}turaud}, {Pieniluoma}, {Pigozzi}, {Poels}, {Prat}, {Prod'homme}, {Raison}, {Rebordao}, {Risquez}, {Rocca-Volmerange}, {Rosen},
  {Ruiz-Fuertes}, {Russo}, {Sembay}, {Serraller Vizcaino}, {Short}, {Siebert}, {Silva}, {Sinachopoulos}, {Slezak}, {Soffel}, {Sosnowska}, {Strai{\v{z}}ys}, {ter Linden}, {Terrell}, {Theil}, {Tiede}, {Troisi}, {Tsalmantza}, {Tur}, {Vaccari}, {Vachier}, {Valles}, {Van Hamme}, {Veltz}, {Virtanen}, {Wallut}, {Wichmann}, {Wilkinson}, {Ziaeepour}, \& {Zschocke}}]{gaia16}
{Gaia Collaboration}, {Prusti}, T., {de Bruijne}, J.~H.~J., {et~al.} 2016, \aap, 595, A1

\bibitem[{{Garn} {et~al.}(2010){Garn}, {Sobral}, {Best}, {Geach}, {Smail}, {Cirasuolo}, {Dalton}, {Dunlop}, {McLure}, \& {Farrah}}]{garn10}
{Garn}, T., {Sobral}, D., {Best}, P.~N., {et~al.} 2010, \mnras, 402, 2017

\bibitem[{{Ginolfi} {et~al.}(2022){Ginolfi}, {Piconcelli}, {Zappacosta}, {Jones}, {Pentericci}, {Maiolino}, {Travascio}, {Menci}, {Carniani}, {Rizzo}, {Arrigoni Battaia}, {Cantalupo}, {De Breuck}, {Graziani}, {Knudsen}, {Laursen}, {Mainieri}, {Schneider}, {Stanley}, {Valiante}, \& {Verhamme}}]{gino22}
{Ginolfi}, M., {Piconcelli}, E., {Zappacosta}, L., {et~al.} 2022, Nature Communications, 13, 4574

\bibitem[{{Glazebrook} {et~al.}(1999){Glazebrook}, {Blake}, {Economou}, {Lilly}, \& {Colless}}]{glaz99}
{Glazebrook}, K., {Blake}, C., {Economou}, F., {Lilly}, S., \& {Colless}, M. 1999, \mnras, 306, 843

\bibitem[{{Hashimoto} {et~al.}(2023){Hashimoto}, {{\'A}lvarez-M{\'a}rquez}, {Fudamoto}, {Colina}, {Inoue}, {Nakazato}, {Ceverino}, {Yoshida}, {Costantin}, {Sugahara}, {G{\'o}mez}, {Blanco-Prieto}, {Mawatari}, {Arribas}, {Marques-Chaves}, {Pereira-Santaella}, {Bakx}, {Hagimoto}, {Hashigaya}, {Matsuo}, {Tamura}, {Usui}, \& {Ren}}]{hash23}
{Hashimoto}, T., {{\'A}lvarez-M{\'a}rquez}, J., {Fudamoto}, Y., {et~al.} 2023, \apjl, 955, L2

\bibitem[{{Heintz} {et~al.}(2023){Heintz}, {Watson}, {Brammer}, {Vejlgaard}, {Hutter}, {Strait}, {Matthee}, {Oesch}, {Jakobsson}, {Tanvir}, {Laursen}, {Naidu}, {Mason}, {Killi}, {Jung}, {Hsiao}, {Abdurro'uf}, {Coe}, {Arrabal Haro}, {Finkelstein}, \& {Toft}}]{hein23}
{Heintz}, K.~E., {Watson}, D., {Brammer}, G., {et~al.} 2023, arXiv e-prints, arXiv:2306.00647

\bibitem[{{Hopkins} {et~al.}(2006){Hopkins}, {Hernquist}, {Cox}, {Di Matteo}, {Robertson}, \& {Springel}}]{hopk06}
{Hopkins}, P.~F., {Hernquist}, L., {Cox}, T.~J., {et~al.} 2006, \apjs, 163, 1

\bibitem[{{Jakobsen} {et~al.}(2022){Jakobsen}, {Ferruit}, {Alves de Oliveira}, {Arribas}, {Bagnasco}, {Barho}, {Beck}, {Birkmann}, {B{\"o}ker}, {Bunker}, {Charlot}, {de Jong}, {de Marchi}, {Ehrenwinkler}, {Falcolini}, {Fels}, {Franx}, {Franz}, {Funke}, {Giardino}, {Gnata}, {Holota}, {Honnen}, {Jensen}, {Jentsch}, {Johnson}, {Jollet}, {Karl}, {Kling}, {K{\"o}hler}, {Kolm}, {Kumari}, {Lander}, {Lemke}, {L{\'o}pez-Caniego}, {L{\"u}tzgendorf}, {Maiolino}, {Manjavacas}, {Marston}, {Maschmann}, {Maurer}, {Messerschmidt}, {Moseley}, {Mosner}, {Mott}, {Muzerolle}, {Pirzkal}, {Pittet}, {Plitzke}, {Posselt}, {Rapp}, {Rauscher}, {Rawle}, {Rix}, {R{\"o}del}, {Rumler}, {Sabbi}, {Salvignol}, {Schmid}, {Sirianni}, {Smith}, {Strada}, {te Plate}, {Valenti}, {Wettemann}, {Wiehe}, {Wiesmayer}, {Willott}, {Wright}, {Zeidler}, \& {Zincke}}]{jako22}
{Jakobsen}, P., {Ferruit}, P., {Alves de Oliveira}, C., {et~al.} 2022, \aap, 661, A80

\bibitem[{{Jones} {et~al.}(2023){Jones}, {Bunker}, {Saxena}, {Witstok}, {Stark}, {Arribas}, {Baker}, {Bhatawdekar}, {Bowler}, {Boyett}, {Cameron}, {Carniani}, {Charlot}, {Chevallard}, {Curti}, {Curtis-Lake}, {Eisenstein}, {Hainline}, {Hausen}, {Ji}, {Johnson}, {Kumari}, {Looser}, {Maiolino}, {Maseda}, {Parlanti}, {Rix}, {Robertson}, {Sandles}, {Scholtz}, {Smit}, {Tacchella}, {Ubler}, {Williams}, \& {Willott}}]{jone23}
{Jones}, G.~C., {Bunker}, A.~J., {Saxena}, A., {et~al.} 2023, arXiv e-prints, arXiv:2306.02471

\bibitem[{{Kauffmann} {et~al.}(2003{\natexlab{a}}){Kauffmann}, {Heckman}, {Tremonti}, {Brinchmann}, {Charlot}, {White}, {Ridgway}, {Brinkmann}, {Fukugita}, {Hall}, {Ivezi{\'c}}, {Richards}, \& {Schneider}}]{kauf03}
{Kauffmann}, G., {Heckman}, T.~M., {Tremonti}, C., {et~al.} 2003{\natexlab{a}}, \mnras, 346, 1055

\bibitem[{{Kauffmann} {et~al.}(2003{\natexlab{b}}){Kauffmann}, {Heckman}, {White}, {Charlot}, {Tremonti}, {Brinchmann}, {Bruzual}, {Peng}, {Seibert}, {Bernardi}, {Blanton}, {Brinkmann}, {Castander}, {Cs{\'a}bai}, {Fukugita}, {Ivezic}, {Munn}, {Nichol}, {Padmanabhan}, {Thakar}, {Weinberg}, \& {York}}]{kauf03sdss}
{Kauffmann}, G., {Heckman}, T.~M., {White}, S. D.~M., {et~al.} 2003{\natexlab{b}}, \mnras, 341, 33

\bibitem[{{Kehrig} {et~al.}(2006){Kehrig}, {V{\'\i}lchez}, {Telles}, {Cuisinier}, \& {P{\'e}rez-Montero}}]{kehr06}
{Kehrig}, C., {V{\'\i}lchez}, J.~M., {Telles}, E., {Cuisinier}, F., \& {P{\'e}rez-Montero}, E. 2006, \aap, 457, 477

\bibitem[{{Kennicutt}(1998)}]{kenn98}
{Kennicutt}, Robert~C., J. 1998, \apj, 498, 541

\bibitem[{{Kewley} {et~al.}(2013){Kewley}, {Dopita}, {Leitherer}, {Dav{\'e}}, {Yuan}, {Allen}, {Groves}, \& {Sutherland}}]{kewl13}
{Kewley}, L.~J., {Dopita}, M.~A., {Leitherer}, C., {et~al.} 2013, \apj, 774, 100

\bibitem[{{Kewley} {et~al.}(2001){Kewley}, {Dopita}, {Sutherland}, {Heisler}, \& {Trevena}}]{kewl01}
{Kewley}, L.~J., {Dopita}, M.~A., {Sutherland}, R.~S., {Heisler}, C.~A., \& {Trevena}, J. 2001, \apj, 556, 121

\bibitem[{{Laporte} {et~al.}(2015){Laporte}, {P{\'e}rez-Fournon}, {Calanog}, {Cooray}, {Wardlow}, {Bock}, {Bridge}, {Burgarella}, {Bussmann}, {Cabrera-Lavers}, {Casey}, {Clements}, {Conley}, {Dannerbauer}, {Farrah}, {Fu}, {Gavazzi}, {Gonz{\'a}lez-Solares}, {Ivison}, {Lo Faro}, {Ma}, {Magdis}, {Marques-Chaves}, {Mart{\'\i}nez-Navajas}, {Oliver}, {Osage}, {Riechers}, {Rigopoulou}, {Scott}, {Streblyanska}, \& {Vieira}}]{lapo15}
{Laporte}, N., {P{\'e}rez-Fournon}, I., {Calanog}, J.~A., {et~al.} 2015, \apj, 810, 130

\bibitem[{{Looser} {et~al.}(2023){Looser}, {D'Eugenio}, {Maiolino}, {Witstok}, {Sandles}, {Curtis-Lake}, {Chevallard}, {Tacchella}, {Johnson}, {Baker}, {Suess}, {Carniani}, {Ferruit}, {Arribas}, {Bonaventura}, {Bunker}, {Cameron}, {Charlot}, {Curti}, {de Graaff}, {Maseda}, {Rawle}, {Rix}, {Rodriguez Del Pino}, {Smit}, {{\"U}bler}, {Willott}, {Alberts}, {Egami}, {Eisenstein}, {Endsley}, {Hausen}, {Rieke}, {Robertson}, {Shivaei}, {Williams}, {Boyett}, {Chen}, {Ji}, {Jones}, {Kumari}, {Nelson}, {Perna}, {Saxena}, \& {Scholtz}}]{loos23}
{Looser}, T.~J., {D'Eugenio}, F., {Maiolino}, R., {et~al.} 2023, arXiv e-prints, arXiv:2302.14155

\bibitem[{{Marshall} {et~al.}(2023){Marshall}, {Perna}, {Willott}, {Maiolino}, {Scholtz}, {{\"U}bler}, {Carniani}, {Arribas}, {L{\"u}tzgendorf}, {Bunker}, {Charlot}, {Ferruit}, {Jakobsen}, {Rix}, {Rodr{\'\i}guez Del Pino}, {B{\"o}ker}, {Cameron}, {Cresci}, {Curtis-Lake}, {Jones}, {Kumari}, {P{\'e}rez-Gonz{\'a}lez}, \& {Reed}}]{mars23}
{Marshall}, M.~A., {Perna}, M., {Willott}, C.~J., {et~al.} 2023, \aap, 678, A191

\bibitem[{{Nakajima} \& {Maiolino}(2022)}]{naka22}
{Nakajima}, K. \& {Maiolino}, R. 2022, \mnras, 513, 5134

\bibitem[{Nightingale \& Dye(2015)}]{nigh15}
Nightingale, J.~W. \& Dye, S. 2015, MNRAS, 452, 2940

\bibitem[{Nightingale {et~al.}(2018)Nightingale, Dye, \& Massey}]{nigh18}
Nightingale, J.~W., Dye, S., \& Massey, R.~J. 2018, MNRAS, 478, 4738

\bibitem[{Nightingale {et~al.}(2021)Nightingale, Hayes, Kelly, Amvrosiadis, Etherington, He, Li, Cao, Frawley, Cole, Enia, Frenk, Harvey, Li, Massey, Negrello, \& Robertson}]{nigh21}
Nightingale, J.~W., Hayes, R.~G., Kelly, A., {et~al.} 2021, J. Open Source Softw., 6, 2825

\bibitem[{{Oliver} {et~al.}(2012){Oliver}, {Bock}, {Altieri}, {Amblard}, {Arumugam}, {Aussel}, {Babbedge}, {Beelen}, {B{\'e}thermin}, {Blain}, {Boselli}, {Bridge}, {Brisbin}, {Buat}, {Burgarella}, {Castro-Rodr{\'\i}guez}, {Cava}, {Chanial}, {Cirasuolo}, {Clements}, {Conley}, {Conversi}, {Cooray}, {Dowell}, {Dubois}, {Dwek}, {Dye}, {Eales}, {Elbaz}, {Farrah}, {Feltre}, {Ferrero}, {Fiolet}, {Fox}, {Franceschini}, {Gear}, {Giovannoli}, {Glenn}, {Gong}, {Gonz{\'a}lez Solares}, {Griffin}, {Halpern}, {Harwit}, {Hatziminaoglou}, {Heinis}, {Hurley}, {Hwang}, {Hyde}, {Ibar}, {Ilbert}, {Isaak}, {Ivison}, {Lagache}, {Le Floc'h}, {Levenson}, {Faro}, {Lu}, {Madden}, {Maffei}, {Magdis}, {Mainetti}, {Marchetti}, {Marsden}, {Marshall}, {Mortier}, {Nguyen}, {O'Halloran}, {Omont}, {Page}, {Panuzzo}, {Papageorgiou}, {Patel}, {Pearson}, {P{\'e}rez-Fournon}, {Pohlen}, {Rawlings}, {Raymond}, {Rigopoulou}, {Riguccini}, {Rizzo}, {Rodighiero}, {Roseboom}, {Rowan-Robinson}, {S{\'a}nchez Portal}, {Schulz}, {Scott}, {Seymour}, {Shupe},
  {Smith}, {Stevens}, {Symeonidis}, {Trichas}, {Tugwell}, {Vaccari}, {Valtchanov}, {Vieira}, {Viero}, {Vigroux}, {Wang}, {Ward}, {Wardlow}, {Wright}, {Xu}, \& {Zemcov}}]{oliv12}
{Oliver}, S.~J., {Bock}, J., {Altieri}, B., {et~al.} 2012, \mnras, 424, 1614

\bibitem[{{Osterbrock}(1989)}]{oste89}
{Osterbrock}, D.~E. 1989, {Astrophysics of gaseous nebulae and active galactic nuclei}

\bibitem[{{Pallottini} {et~al.}(2019){Pallottini}, {Ferrara}, {Decataldo}, {Gallerani}, {Vallini}, {Carniani}, {Behrens}, {Kohandel}, \& {Salvadori}}]{pall19}
{Pallottini}, A., {Ferrara}, A., {Decataldo}, D., {et~al.} 2019, \mnras, 487, 1689

\bibitem[{{Perna} {et~al.}(2023){Perna}, {Arribas}, {Marshall}, {D'Eugenio}, {{\"U}bler}, {Bunker}, {Charlot}, {Carniani}, {Jakobsen}, {Maiolino}, {Rodr{\'\i}guez Del Pino}, {Willott}, {B{\"o}ker}, {Circosta}, {Cresci}, {Curti}, {Husemann}, {Kumari}, {Lamperti}, {P{\'e}rez-Gonz{\'a}lez}, \& {Scholtz}}]{pern23}
{Perna}, M., {Arribas}, S., {Marshall}, M., {et~al.} 2023, \aap, 679, A89

\bibitem[{{Popesso} {et~al.}(2023){Popesso}, {Concas}, {Cresci}, {Belli}, {Rodighiero}, {Inami}, {Dickinson}, {Ilbert}, {Pannella}, \& {Elbaz}}]{pope23}
{Popesso}, P., {Concas}, A., {Cresci}, G., {et~al.} 2023, \mnras, 519, 1526

\bibitem[{{Pozzetti} {et~al.}(2016){Pozzetti}, {Hirata}, {Geach}, {Cimatti}, {Baugh}, {Cucciati}, {Merson}, {Norberg}, \& {Shi}}]{pozz16}
{Pozzetti}, L., {Hirata}, C.~M., {Geach}, J.~E., {et~al.} 2016, \aap, 590, A3

\bibitem[{{Rauscher} {et~al.}(2017){Rauscher}, {Arendt}, {Fixsen}, {Greenhouse}, {Lander}, {Lindler}, {Loose}, {Moseley}, {Mott}, {Wen}, {Wilson}, \& {Xenophontos}}]{raus17}
{Rauscher}, B.~J., {Arendt}, R.~G., {Fixsen}, D.~J., {et~al.} 2017, \pasp, 129, 105003

\bibitem[{{Reddy} {et~al.}(2020){Reddy}, {Shapley}, {Kriek}, {Steidel}, {Shivaei}, {Sanders}, {Mobasher}, {Coil}, {Siana}, {Freeman}, {Azadi}, {Fetherolf}, {Leung}, {Price}, \& {Zick}}]{redd20}
{Reddy}, N.~A., {Shapley}, A.~E., {Kriek}, M., {et~al.} 2020, \apj, 902, 123

\bibitem[{{Riechers} {et~al.}(2013){Riechers}, {Bradford}, {Clements}, {Dowell}, {P{\'e}rez-Fournon}, {Ivison}, {Bridge}, {Conley}, {Fu}, {Vieira}, {Wardlow}, {Calanog}, {Cooray}, {Hurley}, {Neri}, {Kamenetzky}, {Aguirre}, {Altieri}, {Arumugam}, {Benford}, {B{\'e}thermin}, {Bock}, {Burgarella}, {Cabrera-Lavers}, {Chapman}, {Cox}, {Dunlop}, {Earle}, {Farrah}, {Ferrero}, {Franceschini}, {Gavazzi}, {Glenn}, {Solares}, {Gurwell}, {Halpern}, {Hatziminaoglou}, {Hyde}, {Ibar}, {Kov{\'a}cs}, {Krips}, {Lupu}, {Maloney}, {Martinez-Navajas}, {Matsuhara}, {Murphy}, {Naylor}, {Nguyen}, {Oliver}, {Omont}, {Page}, {Petitpas}, {Rangwala}, {Roseboom}, {Scott}, {Smith}, {Staguhn}, {Streblyanska}, {Thomson}, {Valtchanov}, {Viero}, {Wang}, {Zemcov}, \& {Zmuidzinas}}]{riec13}
{Riechers}, D.~A., {Bradford}, C.~M., {Clements}, D.~L., {et~al.} 2013, \nat, 496, 329

\bibitem[{{Riechers} {et~al.}(2020){Riechers}, {Hodge}, {Pavesi}, {Daddi}, {Decarli}, {Ivison}, {Sharon}, {Smail}, {Walter}, {Aravena}, {Capak}, {Carilli}, {Cox}, {Cunha}, {Dannerbauer}, {Dickinson}, {Neri}, \& {Wagg}}]{riec20}
{Riechers}, D.~A., {Hodge}, J.~A., {Pavesi}, R., {et~al.} 2020, \apj, 895, 81

\bibitem[{{Rigby} {et~al.}(2023){Rigby}, {Perrin}, {McElwain}, {Kimble}, {Friedman}, {Lallo}, {Doyon}, {Feinberg}, {Ferruit}, {Glasse}, {Rieke}, {Rieke}, {Wright}, {Willott}, {Colon}, {Milam}, {Neff}, {Stark}, {Valenti}, {Abell}, {Abney}, {Abul-Huda}, {Acton}, {Adams}, {Adler}, {Aguilar}, {Ahmed}, {Albert}, {Alberts}, {Aldridge}, {Allen}, {Altenburg}, {{\'A}lvarez-M{\'a}rquez}, {Alves de Oliveira}, {Andersen}, {Anderson}, {Anderson}, {Argyriou}, {Armstrong}, {Arribas}, {Artigau}, {Arvai}, {Atkinson}, {Bacon}, {Bair}, {Banks}, {Barrientes}, {Barringer}, {Bartosik}, {Bast}, {Baudoz}, {Beatty}, {Bechtold}, {Beck}, {Bergeron}, {Bergkoetter}, {Bhatawdekar}, {Birkmann}, {Blazek}, {Blome}, {Boccaletti}, {B{\"o}ker}, {Boia}, {Bonaventura}, {Bond}, {Bosley}, {Boucarut}, {Bourque}, {Bouwman}, {Bower}, {Bowers}, {Boyer}, {Bradley}, {Brady}, {Braun}, {Breda}, {Bresnahan}, {Bright}, {Britt}, {Bromenschenkel}, {Brooks}, {Brooks}, {Brown}, {Brown}, {Brown}, {Bunker}, {Burger}, {Bushouse}, {Cale}, {Cameron}, {Cameron},
  {Canipe}, {Caplinger}, {Caputo}, {Cara}, {Carey}, {Carniani}, {Carrasquilla}, {Carruthers}, {Case}, {Catherine}, {Chance}, {Chapman}, {Charlot}, {Charlow}, {Chayer}, {Chen}, {Cherinka}, {Chichester}, {Chilton}, {Chonis}, {Clampin}, {Clark}, {Clark}, {Coe}, {Coleman}, {Comber}, {Comeau}, {Connolly}, {Cooper}, {Cooper}, {Coppock}, {Correnti}, {Cossou}, {Coulais}, {Coyle}, {Cracraft}, {Curti}, {Cuturic}, {Davis}, {Davis}, {Dean}, {DeLisa}, {deMeester}, {Dencheva}, {Dencheva}, {DePasquale}, {Deschenes}, {Hunor Detre}, {Diaz}, {Dicken}, {DiFelice}, {Dillman}, {Dixon}, {Doggett}, {Donaldson}, {Douglas}, {DuPrie}, {Dupuis}, {Durning}, {Easmin}, {Eck}, {Edeani}, {Egami}, {Ehrenwinkler}, {Eisenhamer}, {Eisenhower}, {Elie}, {Elliott}, {Elliott}, {Ellis}, {Engesser}, {Espinoza}, {Etienne}, {Etxaluze}, {Falini}, {Feeney}, {Ferry}, {Filippazzo}, {Fincham}, {Fix}, {Flagey}, {Florian}, {Flynn}, {Fontanella}, {Ford}, {Forshay}, {Fox}, {Franz}, {Fu}, {Fullerton}, {Galkin}, {Galyer}, {Garc{\'\i}a Mar{\'\i}n}, {Gardner},
  {Gardner}, {Garland}, {Garrett}, {Gasman}, {Gaspar}, {Gaudreau}, {Gauthier}, {Geers}, {Geithner}, {Gennaro}, {Giardino}, {Girard}, {Giuliano}, {Glassmire}, {Glauser}, {Glazer}, {Godfrey}, {Golimowski}, {Gollnitz}, {Gong}, {Gonzaga}, {Gordon}, {Gordon}, {Goudfrooij}, {Greene}, {Greenhouse}, {Grimaldi}, {Groebner}, {Grundy}, {Guillard}, {Gutman}, {Ha}, {Haderlein}, {Hagedorn}, {Hainline}, {Haley}, {Hami}, {Hamilton}, {Hammel}, {Hansen}, {Harkins}, {Harr}, {Hart}, {Hart}, {Hartig}, {Hashimoto}, {Haskins}, {Hathaway}, {Havey}, {Hayden}, {Hecht}, {Heller-Boyer}, {Henriques}, {Henry}, {Hermann}, {Hernandez}, {Hesman}, {Hicks}, {Hilbert}, {Hines}, {Hoffman}, {Holfeltz}, {Holler}, {Hoppa}, {Hott}, {Howard}, {Howard}, {Hunter}, {Hunter}, {Hurst}, {Husemann}, {Hustak}, {Ilinca Ignat}, {Illingworth}, {Irish}, {Jackson}, {Jahromi}, {Jakobsen}, {James}, {James}, {Januszewski}, {Jenkins}, {Jirdeh}, {Johnson}, {Johnson}, {Jones}, {Jones}, {Jones}, {Jones}, {Jordan}, {Jordan}, {Jurczyk}, {Jurling}, {Kaleida}, {Kalmanson},
  {Kammerer}, {Kang}, {Kao}, {Karakla}, {Kavanagh}, {Kelly}, {Kendrew}, {Kennedy}, {Kenny}, {Keski-kuha}, {Keyes}, {Kidwell}, {Kinzel}, {Kirk}, {Kirkpatrick}, {Kirshenblat}, {Klaassen}, {Knapp}, {Knight}, {Knollenberg}, {Koehler}, {Koekemoer}, {Kovacs}, {Kulp}, {Kumari}, {Kyprianou}, {La Massa}, {Labador}, {Labiano}, {Lagage}, {Lajoie}, {Lallo}, {Lam}, {Lamb}, {Lambros}, {Lampenfield}, {Langston}, {Larson}, {Law}, {Lawrence}, {Lee}, {Leisenring}, {Lepo}, {Leveille}, {Levenson}, {Levine}, {Levy}, {Lewis}, {Lewis}, {Libralato}, {Lightsey}, {Link}, {Liu}, {Lo}, {Lockwood}, {Logue}, {Long}, {Long}, {Loomis}, {Lopez-Caniego}, {Lorenzo Alvarez}, {Love-Pruitt}, {Lucy}, {Luetzgendorf}, {Maghami}, {Maiolino}, {Major}, {Malla}, {Malumuth}, {Manjavacas}, {Mannfolk}, {Marrione}, {Marston}, {Martel}, {Maschmann}, {Masci}, {Masciarelli}, {Maszkiewicz}, {Mather}, {McKenzie}, {McLean}, {McMaster}, {Melbourne}, {Mel{\'e}ndez}, {Menzel}, {Merz}, {Meyett}, {Meza}, {Miskey}, {Misselt}, {Moller}, {Morrison}, {Morse}, {Moseley},
  {Mosier}, {Mountain}, {Mueckay}, {Mueller}, {Mullally}, {Murphy}, {Murray}, {Murray}, {Mustelier}, {Muzerolle}, {Mycroft}, {Myers}, {Myrick}, {Nanavati}, {Nance}, {Nayak}, {Naylor}, {Nelan}, {Nickson}, {Nielson}, {Nieto-Santisteban}, {Nikolov}, {Noriega-Crespo}, {O'Shaughnessy}, {O'Sullivan}, {Ochs}, {Ogle}, {Oleszczuk}, {Olmsted}, {Osborne}, {Ottens}, {Owens}, {Pacifici}, {Pagan}, {Page}, {Park}, {Parrish}, {Patapis}, {Paul}, {Pauly}, {Pavlovsky}, {Pedder}, {Peek}, {Pena-Guerrero}, {Penanen}, {Perez}, {Perna}, {Perriello}, {Phillips}, {Pietraszkiewicz}, {Pinaud}, {Pirzkal}, {Pitman}, {Piwowar}, {Platais}, {Player}, {Plesha}, {Pollizi}, {Polster}, {Pontoppidan}, {Porterfield}, {Proffitt}, {Pueyo}, {Pulliam}, {Quirt}, {Quispe Neira}, {Ramos Alarcon}, {Ramsay}, {Rapp}, {Rapp}, {Rauscher}, {Ravindranath}, {Rawle}, {Regan}, {Reichard}, {Reis}, {Ressler}, {Rest}, {Reynolds}, {Rhue}, {Richon}, {Rickman}, {Ridgaway}, {Ritchie}, {Rix}, {Robberto}, {Robinson}, {Robinson}, {Robinson}, {Rock}, {Rodriguez}, {Rodriguez
  Del Pino}, {Roellig}, {Rohrbach}, {Roman}, {Romelfanger}, {Rose}, {Roteliuk}, {Roth}, {Rothwell}, {Rowlands}, {Roy}, {Royer}, {Royle}, {Rui}, {Rumler}, {Runnels}, {Russ}, {Rustamkulov}, {Ryden}, {Ryer}, {Sabata}, {Sabatke}, {Sabbi}, {Samuelson}, {Sapp}, {Sappington}, {Sargent}, {Sauer}, {Scheithauer}, {Schlawin}, {Schlitz}, {Schmitz}, {Schneider}, {Schreiber}, {Schulze}, {Schwab}, {Scott}, {Sembach}, {Shanahan}, {Shaughnessy}, {Shaw}, {Shawger}, {Shay}, {Sheehan}, {Shen}, {Sherman}, {Shiao}, {Shih}, {Shivaei}, {Sienkiewicz}, {Sing}, {Sirianni}, {Sivaramakrishnan}, {Skipper}, {Sloan}, {Slocum}, {Slowinski}, {Smith}, {Smith}, {Smith}, {Smith}, {Snyder}, {Soh}, {Sohn}, {Soto}, {Spencer}, {Stallcup}, {Stansberry}, {Starr}, {Starr}, {Stewart}, {Stiavelli}, {Straughn}, {Strickland}, {Stys}, {Summers}, {Sun}, {Sunnquist}, {Swade}, {Swam}, {Swaters}, {Swoish}, {Taylor}, {Taylor}, {Te Plate}, {Tea}, {Teague}, {Telfer}, {Temim}, {Thatte}, {Thompson}, {Thompson}, {Thomson}, {Tikkanen}, {Tippet}, {Todd}, {Toolan},
  {Tran}, {Trejo}, {Truong}, {Tsukamoto}, {Tustain}, {Tyra}, {Ubeda}, {Underwood}, {Uzzo}, {Van Campen}, {Vandal}, {Vandenbussche}, {Vila}, {Volk}, {Wahlgren}, {Waldman}, {Walker}, {Wander}, {Warfield}, {Warner}, {Wasiak}, {Watkins}, {Weaver}, {Weilert}, {Weiser}, {Weiss}, {Weissman}, {Welty}, {West}, {Wheate}, {Wheatley}, {Wheeler}, {White}, {Whiteaker}, {Whitehouse}, {Whiteleather}, {Whitman}, {Williams}, {Willmer}, {Willoughby}, {Wilson}, {Wirth}, {Wislowski}, {Wolf}, {Wolfe}, {Wolff}, {Workman}, {Wright}, {Wu}, {Wu}, {Wymer}, {Yates}, {Yeager}, {Yeates}, {Yerger}, {Yoon}, {Young}, {Yu}, {Zak}, {Zeidler}, {Zhou}, {Zielinski}, {Zincke}, \& {Zonak}}]{rigb23}
{Rigby}, J., {Perrin}, M., {McElwain}, M., {et~al.} 2023, \pasp, 135, 048001

\bibitem[{{Robson} {et~al.}(2014){Robson}, {Ivison}, {Smail}, {Holland}, {Geach}, {Gibb}, {Riechers}, {Ade}, {Bintley}, {Bock}, {Chapin}, {Chapman}, {Clements}, {Conley}, {Cooray}, {Dunlop}, {Farrah}, {Fich}, {Fu}, {Jenness}, {Laporte}, {Oliver}, {Omont}, {P{\'e}rez-Fournon}, {Scott}, {Swinbank}, \& {Wardlow}}]{robs14}
{Robson}, E.~I., {Ivison}, R.~J., {Smail}, I., {et~al.} 2014, \apj, 793, 11

\bibitem[{{Rodrigo} \& {Solano}(2020)}]{rodr20}
{Rodrigo}, C. \& {Solano}, E. 2020, in XIV.0 Scientific Meeting (virtual) of the Spanish Astronomical Society, 182

\bibitem[{{Rodrigo} {et~al.}(2012){Rodrigo}, {Solano}, \& {Bayo}}]{rodr12}
{Rodrigo}, C., {Solano}, E., \& {Bayo}, A. 2012, {SVO Filter Profile Service Version 1.0}, IVOA Working Draft 15 October 2012

\bibitem[{{Romano} {et~al.}(2021){Romano}, {Cassata}, {Morselli}, {Jones}, {Ginolfi}, {Zanella}, {B{\'e}thermin}, {Capak}, {Faisst}, {Le F{\`e}vre}, {Schaerer}, {Silverman}, {Yan}, {Bardelli}, {Boquien}, {Cimatti}, {Dessauges-Zavadsky}, {Enia}, {Fujimoto}, {Gruppioni}, {Hathi}, {Ibar}, {Koekemoer}, {Lemaux}, {Rodighiero}, {Vergani}, {Zamorani}, \& {Zucca}}]{roma21}
{Romano}, M., {Cassata}, P., {Morselli}, L., {et~al.} 2021, \aap, 653, A111

\bibitem[{{Rudy} {et~al.}(1989){Rudy}, {Cohen}, {Rossano}, {Puetter}, \& {Chapman}}]{rudy89}
{Rudy}, R.~J., {Cohen}, R.~D., {Rossano}, G.~S., {Puetter}, R.~C., \& {Chapman}, S.~C. 1989, \apj, 341, 120

\bibitem[{{Salim} {et~al.}(2018){Salim}, {Boquien}, \& {Lee}}]{sali18}
{Salim}, S., {Boquien}, M., \& {Lee}, J.~C. 2018, \apj, 859, 11

\bibitem[{{Salpeter}(1955)}]{salp55}
{Salpeter}, E.~E. 1955, \apj, 121, 161

\bibitem[{{S{\'e}rsic}(1963)}]{sers63}
{S{\'e}rsic}, J.~L. 1963, Boletin de la Asociacion Argentina de Astronomia La Plata Argentina, 6, 41

\bibitem[{{Speagle} {et~al.}(2014){Speagle}, {Steinhardt}, {Capak}, \& {Silverman}}]{spea14}
{Speagle}, J.~S., {Steinhardt}, C.~L., {Capak}, P.~L., \& {Silverman}, J.~D. 2014, \apjs, 214, 15

\bibitem[{{Strait} {et~al.}(2023){Strait}, {Brammer}, {Muzzin}, {Desprez}, {Asada}, {Abraham}, {Brada{\v{c}}}, {Iyer}, {Martis}, {Mowla}, {Noirot}, {Sarrouh}, {Sawicki}, {Willott}, {Gould}, {Grindlay}, {Matharu}, \& {Rihtar{\v{s}}i{\v{c}}}}]{stra23}
{Strait}, V., {Brammer}, G., {Muzzin}, A., {et~al.} 2023, \apjl, 949, L23

\bibitem[{{Tsukui} \& {Iguchi}(2021)}]{tsuk21}
{Tsukui}, T. \& {Iguchi}, S. 2021, Science, 372, 1201

\bibitem[{{{\"U}bler} {et~al.}(2023){{\"U}bler}, {Maiolino}, {Curtis-Lake}, {P{\'e}rez-Gonz{\'a}lez}, {Curti}, {Perna}, {Arribas}, {Charlot}, {Marshall}, {D'Eugenio}, {Scholtz}, {Bunker}, {Carniani}, {Ferruit}, {Jakobsen}, {Rix}, {Rodr{\'\i}guez Del Pino}, {Willott}, {Boeker}, {Cresci}, {Jones}, {Kumari}, \& {Rawle}}]{uble23}
{{\"U}bler}, H., {Maiolino}, R., {Curtis-Lake}, E., {et~al.} 2023, \aap, 677, A145

\bibitem[{{Umeda} {et~al.}(2023){Umeda}, {Ouchi}, {Nakajima}, {Harikane}, {Ono}, {Xu}, {Isobe}, \& {Zhang}}]{umed23}
{Umeda}, H., {Ouchi}, M., {Nakajima}, K., {et~al.} 2023, arXiv e-prints, arXiv:2306.00487

\bibitem[{{Veilleux} \& {Osterbrock}(1987)}]{veil87}
{Veilleux}, S. \& {Osterbrock}, D.~E. 1987, \apjs, 63, 295

\bibitem[{{Venemans} {et~al.}(2019){Venemans}, {Neeleman}, {Walter}, {Novak}, {Decarli}, {Hennawi}, \& {Rix}}]{vene19}
{Venemans}, B.~P., {Neeleman}, M., {Walter}, F., {et~al.} 2019, \apjl, 874, L30

\bibitem[{{Ventou} {et~al.}(2019){Ventou}, {Contini}, {Bouch{\'e}}, {Epinat}, {Brinchmann}, {Inami}, {Richard}, {Schroetter}, {Soucail}, {Steinmetz}, \& {Weilbacher}}]{vent19}
{Ventou}, E., {Contini}, T., {Bouch{\'e}}, N., {et~al.} 2019, \aap, 631, A87

\bibitem[{{Wagg} {et~al.}(2014){Wagg}, {Carilli}, {Aravena}, {Cox}, {Lentati}, {Maiolino}, {McMahon}, {Riechers}, {Walter}, {Andreani}, {Hills}, \& {Wolfe}}]{wagg14}
{Wagg}, J., {Carilli}, C.~L., {Aravena}, M., {et~al.} 2014, \apj, 783, 71

\bibitem[{{Wylezalek} {et~al.}(2022){Wylezalek}, {Vayner}, {Rupke}, {Zakamska}, {Veilleux}, {Ishikawa}, {Bertemes}, {Liu}, {Barrera-Ballesteros}, {Chen}, {Goulding}, {Greene}, {Hainline}, {Hamann}, {Heckman}, {Johnson}, {Lutz}, {L{\"u}tzgendorf}, {Mainieri}, {Maiolino}, {Nesvadba}, {Ogle}, \& {Sturm}}]{wyle22}
{Wylezalek}, D., {Vayner}, A., {Rupke}, D. S.~N., {et~al.} 2022, \apjl, 940, L7

\bibitem[{{Yang} {et~al.}(2018){Yang}, {Brandt}, {Vito}, {Chen}, {Trump}, {Luo}, {Sun}, {Xue}, {Koekemoer}, {Schneider}, {Vignali}, \& {Wang}}]{yang18}
{Yang}, G., {Brandt}, W.~N., {Vito}, F., {et~al.} 2018, \mnras, 475, 1887

\bibitem[{{Zakamska} \& {Greene}(2014)}]{zaka14}
{Zakamska}, N.~L. \& {Greene}, J.~E. 2014, \mnras, 442, 784

\end{thebibliography}

\begin{appendix}

\section{Astrometry correction}

\subsection{HST astrometry verification}\label{astrover}

In order to compare the JWST/NIRSpec IFU data with archival data, a common astrometric reference frame is required (i.e., Gaia DR3; \citealt{gaia16,gaia21}). We retrieve HST images from the STScI archive, but it is not clear if all of them have been aligned to the Gaia DR3 frame (i.e., some are lacking this comment in their headers). To verify that they have been correctly aligned, we retrieve the locations of the five closest objects in the Gaia archive\footnote{\url{https://gea.esac.esa.int/archive/}} (see Table \ref{gaialoc}) and examine the locations in each HST image. 

\begin{figure*}
\centering
\includegraphics[width=\textwidth]{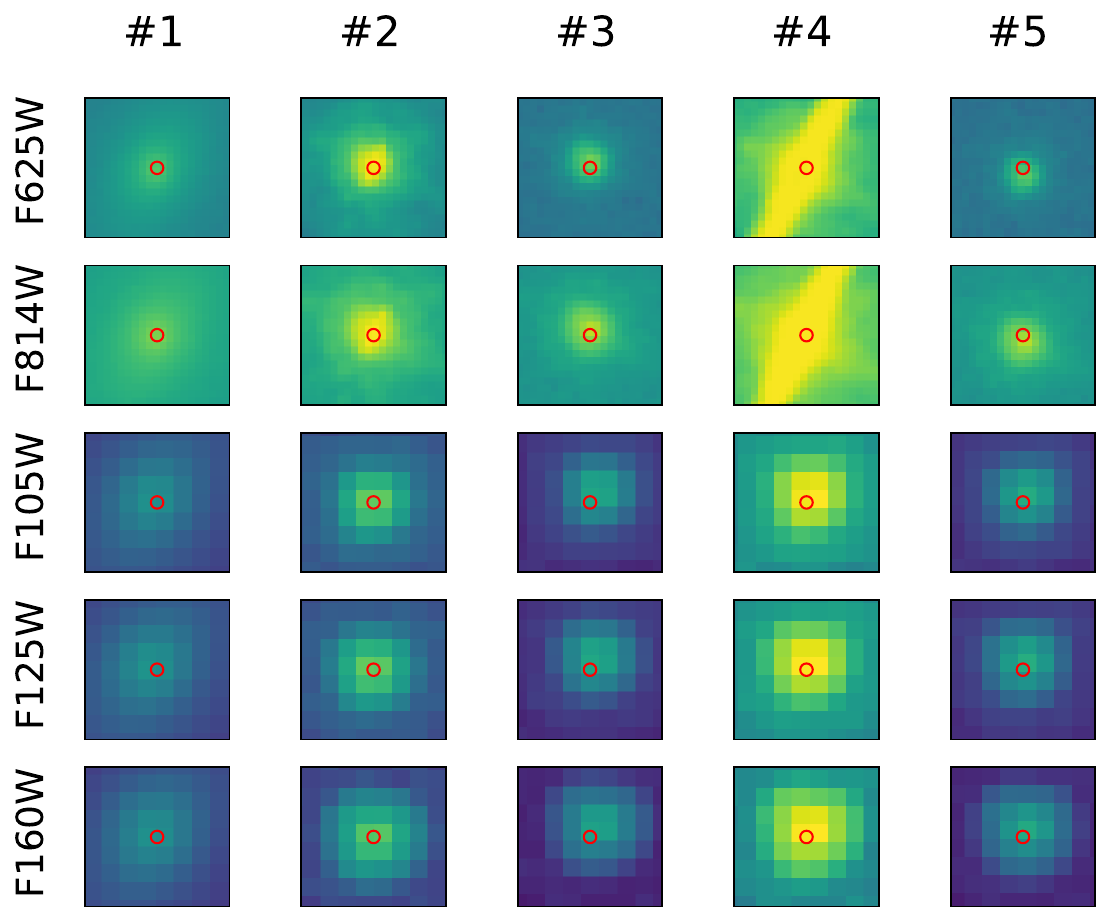}
\caption{Verification that HST images have been aligned to Gaia DR3 reference frame. Each panel contains a $1''\times1''$ view of an HST image (see labels for each row) that is focused on a given location from the Gaia archive (see labels for each column). The Gaia position is shown as a $0.1''$ diameter red circle. The locations of each Gaia source is listed in Table \ref{gaialoc}.}
\label{gaiademo}
\end{figure*}

As seen in Fig. \ref{gaiademo}, the HST images show significant emission at all five Gaia locations with small centroid offsets ($<0.2''$). From this, we conclude that the images have been properly aligned. 

\begin{table}
\centering
\caption{Positions of five objects in the Gaia archive that are closest to HFLS3.}
\begin{tabular}{c|cc}
Number & RA & Dec \\ \hline
1 & 256.7108238422430 & 58.767424792645000 \\
2 & 256.6918696535650 & 58.7610752035986\\
3 & 256.68468210287400 & 58.77599807468070\\
4 & 256.6942094613340 & 58.78483710562300\\
5 & 256.7115031731110 & 58.785864102389900\\
\end{tabular}

\label{gaialoc}
\end{table}

\subsection{Application of spatial shift}
\begin{figure*}
\centering
\includegraphics[width=\textwidth]{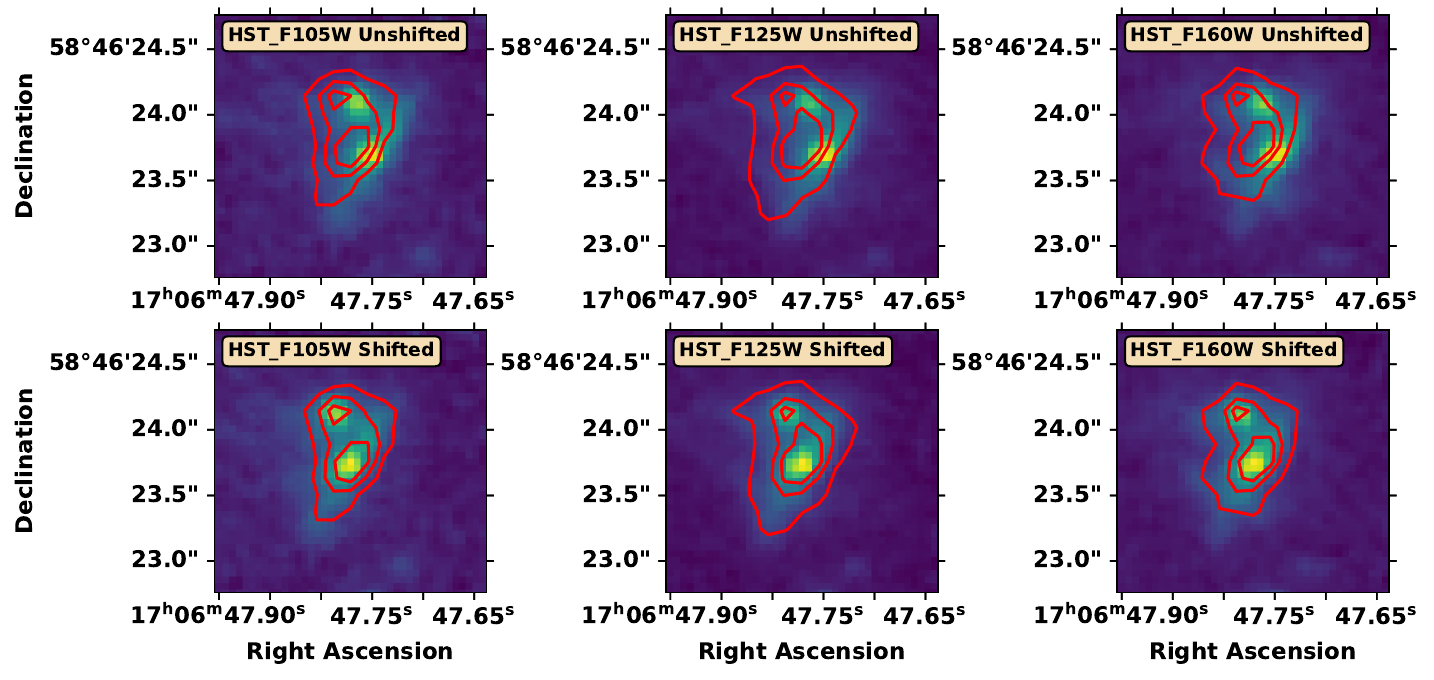}
\caption{JWST/NIRSpec IFU R100 data cubes integrated over the corresponding HST filter bandpasses: F105W (left), F125W (centre), and F160W (right). In each panel, the collapsed JWST emission is depicted as the background colours, while the HST data is shown by red contours. The JWST emission in the top row is shown without the astrometric correction, while the lower row includes the alignment to Gaia DR3.
}
\label{spatoff}
\end{figure*}

To shift the JWST/NIRSpec data cubes to the Gaia frame, we first create narrowband JWST images by convolving the R100 data cube with the transmission profiles of the three HST/WFC3 filters used by \citet{coor14} (see colour maps in Fig. \ref{spatoff})\footnote{Filter profiles were obtained from the SVO Filter Profile Service (\citealt{rodr12,rodr20}). We only use the WFC3 filters due to a higher S/N in the IFU data over the corresponding wavelengths.}. These maps were then compared with the corresponding HST images (see contours in Fig. \ref{spatoff}). An offset is estimated by finding the brightest spaxel in the JWST and HST maps for each filter and determining the distance between the centres of these spaxel. 

In all three cases, this offset is $0.17''$. Because the spaxel sizes are $0.05''$ (JWST) and $0.12825''$ (HST/WFC3-IR), we estimate an uncertainty on this offset based on half the spaxel sizes, added in quadrature ($\pm0.07''$). Thus, the offset is consistent with the pointing accuracy of $\sim0.1''$ (\citealt{rigb23}). We have verified that the R100 and R2700 cubes are aligned to the same initial reference frame, so throughout this work we apply this offset correction to the R100 and R2700 IFU data cubes.

We briefly note that \citet{coor14} aligned each image to the SDSS frame rather than the Gaia frame. Comparing objects near the HFLS3 field in SDSS DR9 and Gaia DR3, we find a negligible offset between these frames. While the previous HST data featured a small astrometric uncertainty ($\sim0.05''$), the Keck/NIRC2 and PdBI data have larger uncertainties ($\sim0.1''$ and $0.1''-0.3''$, respectively). When combined with the $\sim0.1''$ pointing uncertainty of our NIRSpec/IFU data, we find no disagreement between previous positions and our work.

\section{Additional candidate galaxies}\label{candsec}
The HFLS3 field contains multiple galaxies that are strongly detected in line and/or continuum emission. In addition, there are several areas of weak emission that we do not include in our analysis. We present additional details of each below.

\begin{figure}
\centering
\includegraphics[width=0.49\textwidth]{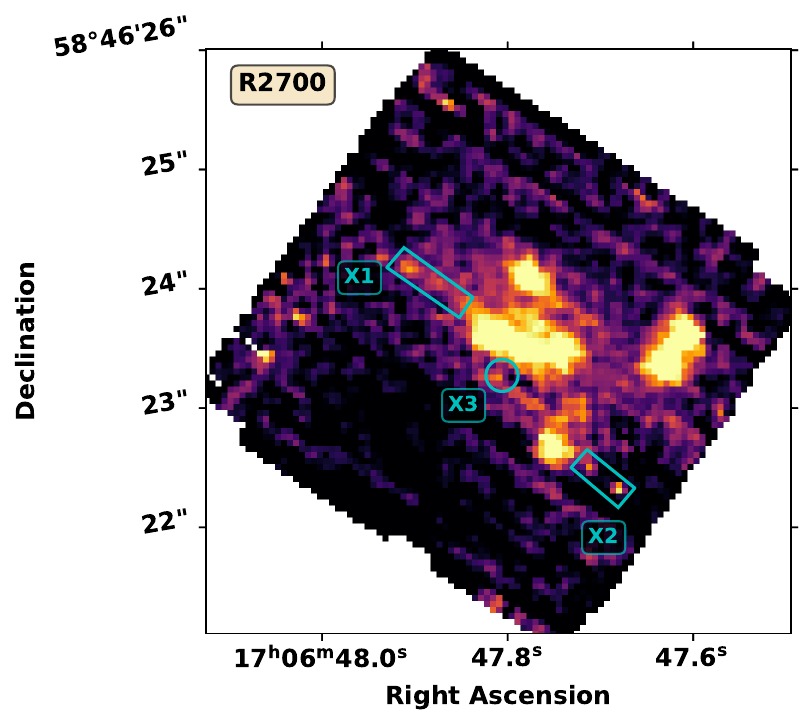}
\includegraphics[width=0.49\textwidth]{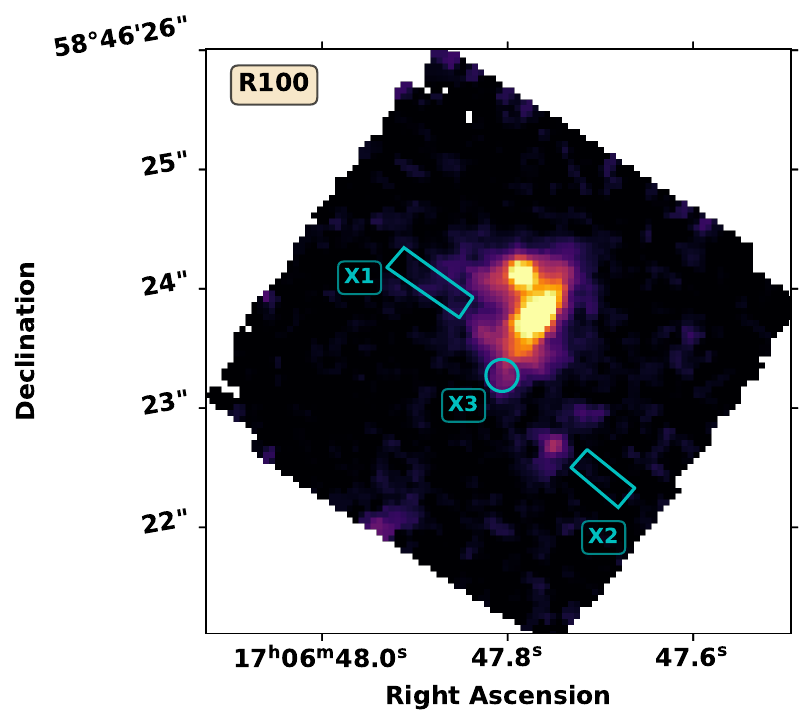}
\caption{Integrated emission of the HFLS3 field, using the same illustrative wavelength ranges as in Fig. \ref{white}: redshifted \ha for $z\sim6.34$ for the R2700 cube ($\lambda_{obs}=4.79954\,-\,4.84467\,\mu$m, left panel) and the approximate wavelength range of HST/WFC3 F160W for the R100 cube ($\lambda_{obs}=1.4\,-\,1.6\,\mu$m; right panel). The colour scale has been adjusted to highlight weak emission. Low-level candidate galaxies are shown with cyan boundaries. North is up and east is to the left.}
\label{white2}
\end{figure}

\subsection{Northeast extension (X1)}
The first candidate emission area is visible as an elongated stretch of emission in the left panel of Fig. \ref{white2} with a slightly brighter clump at the western edge. This morphology (a bar aligned with the axis of the IFU slices used to create the image cube) is similar to artefacts introduced by bright sources that saturate a detector slice (e.g., \citealt{boke22}), but our image is unsaturated. Instead, the high-contrast image of Fig. \ref{white2} shows that multiple slices show a speckled noise pattern, possibly reflecting a low-level calibration issue in this wavelength range.

\begin{figure*}
\centering
\begin{subfigure}{\textwidth}
\includegraphics[width=\textwidth]{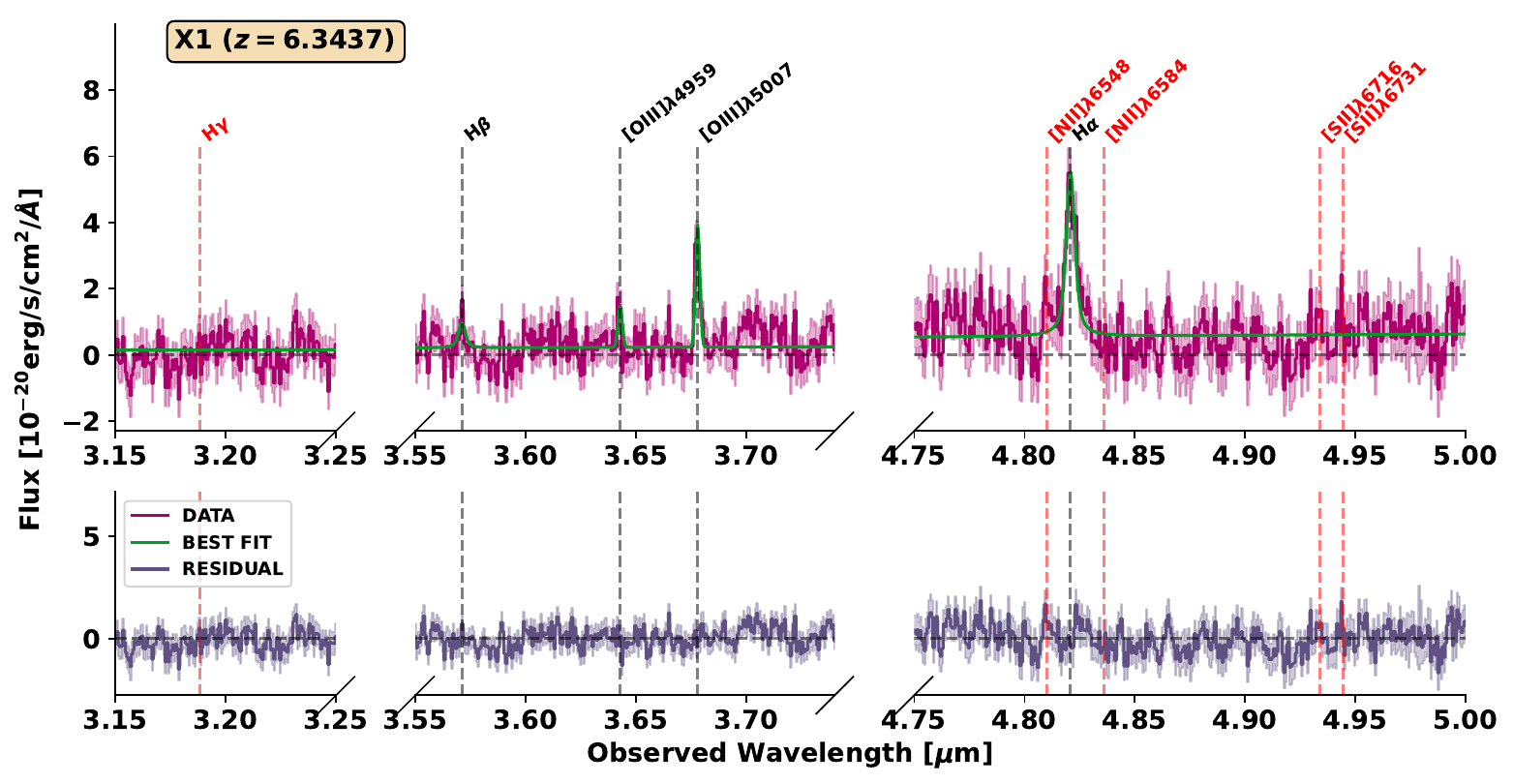}
\includegraphics[width=\textwidth]{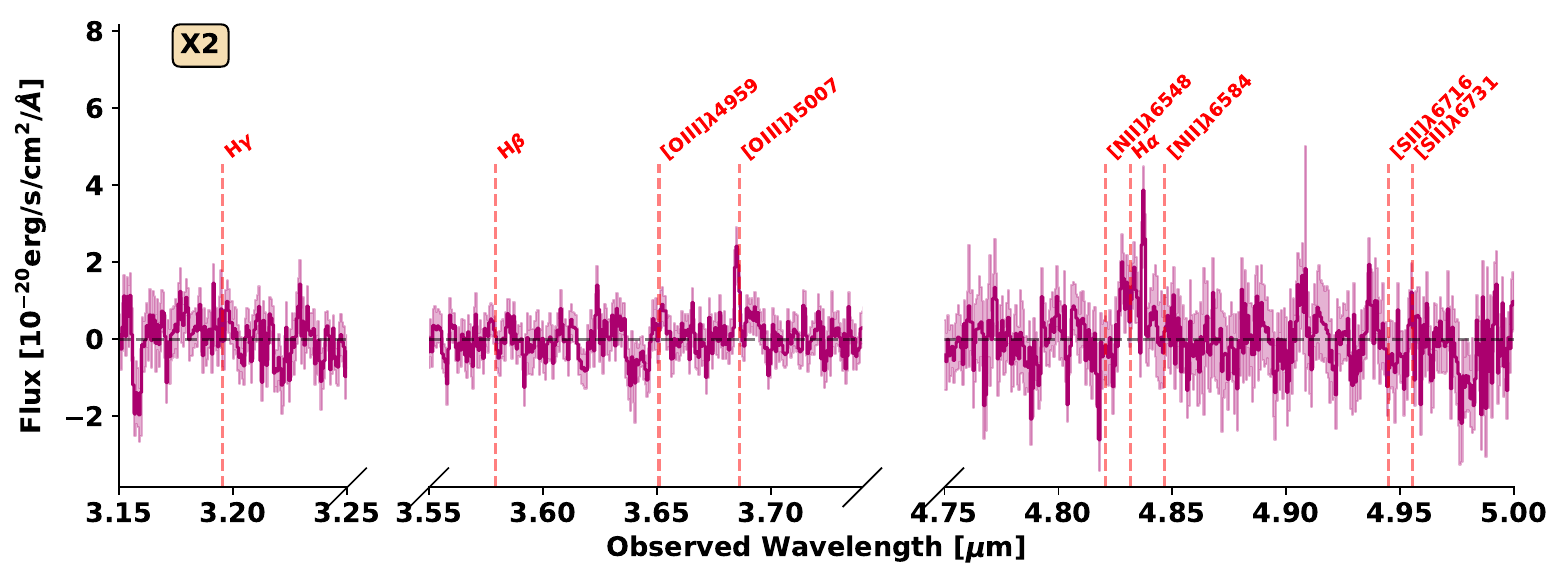}
\end{subfigure}
\caption{Integrated spectra of the R2700 cube using the masks of Fig. \ref{white2}, with $1\sigma$ errors from the associated error spectrum shown as shaded region. Best-fit models (line emission and continuum) are shown by green lines. The centroids of each line are depicted by dashed lines, with red lines indicating that the spectral line was not fit. The lower panel shows the residual.}
\label{xspec}
\end{figure*}

The western clump in X1 is brighter than this noise pattern, but is substantially weaker than the primary components (e.g., C, W). An integrated spectrum over this rectangular region returns strong detections in \hb, \oiiiAB, and \ha, yielding a redshift of $z_{X1}=6.3438\pm0.0001$ (see Table \ref{specfittab} and Fig. \ref{xspec}). The unique redshift and line properties suggest that this represents a true galaxy rather than a noise artefact or a lensed image of another galaxy. However, the weak detection of \hb and the non-detection of \niiAB and \siiAB limits our ability to characterise the source. Its detection suggests that the HFLS3 field contains other low-level galaxy that may contribute to the star formation activity but are undetected by our analysis, as suggested by cosmological simulations (e.g., \citealt{pall19}). 

\subsection{Southwest clumps (X2)}
Fig. \ref{white2} shows two low-level clumps of emission to the southwest of component S in the collapsed R2700 image. These points lie along the same IFU slice, and a spectrum extracted over a region containing both reveals moderate emission that may correspond to \ha emission at $z\sim6.36$. However, the amplitude of this emission is quite low ($<2\sigma$) with the exception of two narrow peaks that do not align with the centres of lines. Because of the low amplitude and artefact-like morphology, we do not investigate this region further.

\subsection{G2 extension (X3)}
When the R100 cube is collapsed over a wavelength range that is dominated by continuum emission from the $z<6$ components ($\lambda_{obs}=0.8-1.1\,\mu$m; see upper left panel of Fig. \ref{r100_lens}), there is a circular source of emission to the southeast of G2 (hereafter X3). This emission is not captured by the S\'{e}rsic mode that we fit to the data (see residual in upper right panel of Fig. \ref{r100_lens}). There is a significant emission line in this region in the R100 data at $\lambda_{obs}=4.822\,\mu$m, but a moment zero image shows that this is simply low-level H$\alpha$ from the outskirts of component C. Our lens modelling analysis does not show an image-plane component to the south, so X3 is likely not a lensed image of C. This extension of G2 also appears in archival HST images (red contours of Fig. \ref{spatoff}), suggesting that it is not an artefact of our JWST data. Further high spatial resolution observations are required for a spectral redshift of this source, in order to determine its nature.

\section{Pseudo-Voigt profile fitting}

\subsection{Profile details}\label{morevoigt}
A standard Voigt profile is a convolution of a Gaussian and Lorentz profile. To simplify the computation of this profile, we adopt the pseudo-Voigt profile as implemented in \lmfit. This begins with a normalised Gaussian:
\begin{equation}
G(\lambda)=\frac{1}{\sigma \sqrt{2\pi}}e^{-(\lambda-\lambda_C)^2/(2\sigma^2)}
\end{equation}
which is centred on $\lambda_C$ and features a FWHM=$2\sqrt{2\ln(2)}\sigma$. Next, we consider a normalised Lorentzian with the same centroid and FWHM:
\begin{equation}
L(\lambda)=\frac{1}{\pi}\frac{\sqrt{2\ln(2)}\sigma}{(\lambda-\lambda_C)^2+2\ln(2)\sigma^2}
\end{equation}
A pseudo-Voigt profile with integrated area $A$ is then given by:
\begin{equation}
V(\lambda)=(1-\alpha)AG(\lambda)+\alpha A L(\lambda)
\end{equation}
where $\alpha$ is the fraction of emission from the Lorentzian component. Note that this form is flexible, with the ability to encapsulate Gaussian profiles ($\alpha=0$), Lorentzian profiles ($\alpha=1$), and all combinations of these profiles such that the integrated intensity, centroid, and FWHM are conserved ($0<\alpha<1$).

\subsection{Morpho-kinematic map creation}\label{momsec}
In order to examine the morpho-kinematics of each source and line, we may produce maps of integrated intensity, velocity offset, and velocity dispersion. These are usually derived by extracting spectra from each spaxel ($\rm f(v)$) and measuring a normalised cumulative velocity distribution ($\rm F(v)=\int_{-\infty}^v f(v') dv' / \int_{-\infty}^{\infty} f(v') dv'$; e.g., \citealt{zaka14}), which is used to calculate the velocity at which N$\%$ of the flux is captured ($\rm v_N$; e.g., $\rm v_{50}$) as well as associated widths (e.g., $\rm w_{80}\equiv v_{90}-v_{10}$). Since there are overlapping lines in some spectra, we will adopt a pseudo-Voigt-based approach. 

For each spaxel, we extract a spectrum and fit a model containing a flat continuum and one or more pseudo-Voigt lines using \lmfit with a least-squares minimiser. Lines that are closely associated or related through a flux ratio are fit concurrently (i.e., \oiiiA/\oiiiB, \niiA/\ha/\niiB, and \siiA/\siiB) with identical continuum values. Line pairs (i.e., \oiiiA/\oiiiB, \niiA/\niiB, and \siiA/\siiB) are assumed to have identical kinematics. We adopt the standard assumptions of \niiB/\niiA$=2.94$ (e.g., \citealt{dojc23}) and \oiiiB/\oiiiA$=2.98$ (e.g., \citealt{dimi07}) for each fit. The LSF is accounted for when calculating the linewidths. Continuum maps are generated using the best-fit constant continuum value for each fit and spaxel. Only fits with $r^2$ (i.e., the coefficient of determination) values $>0.5$ are presented. 

Using the best-fit parameters of each model pseudo-Voigt profile (Integrated intensity $A$ [erg\,s\,cm$^{-2}$], centroid wavelength $\lambda_C$ $[\mu$m], FWHM $[\mu$m], and Lorentzian fraction $\alpha$), we may generate morpho-kinematic maps. Since the Gaussian and Lorentzian components of each pseudo-Voigt profile are normalised, the integrated intensity map is simply the best-fit $A$:
\begin{equation}
\frac{I(x,y)}{\rm erg\,s\,cm^{-2}}=A(x,y)
\end{equation}
The velocity field ($\rm v_{50}$) represents the dominant line-of-sight velocity for a given spaxel. Since the Gaussian and Lorentzian components of the pseudo-Voigt profile are symmetric around the same centroid, $\rm v_{50}$ may be expressed for a redshifted rest-frame wavelength of $\lambda_o$ as:
\begin{equation}
\frac{v_{50}(x,y)}{\rm km\,s^{-1}}=c\left(  \frac{\lambda_C(x,y)}{\lambda_o} -1 \right)
\end{equation}
where $c[$km\,s$^{-1}]$ is the speed of light.

The velocity dispersion is more complex. While the pseudo-Voigt profile is constructed such that its FWHM is constant with respect to the Lorentzian fraction $\alpha$, the non-parametric width $w_{80}$ is dependent on $\alpha$. To calculate this value, we generate a cumulative velocity distribution of each best-fit model spectrum and extract $v_{10}$ and $v_{90}$. Using these:
\begin{equation}
\frac{w_{80}(x,y)}{\rm km\,s^{-1}}=v_{90}-v_{10}
\end{equation}
In this process, we assume that each spectral line may be fit with a single {pseudo-Voigt profile}. While this is motivated by the lack of an obvious broad component (as seen in e.g., \citealt{mars23}), it is possible that a spaxel may contain contributions from narrow and broad emission. Since there are no strong AGN (from previous observations) included in the HFLS3 field, this is expected to be a small effect.

\section{Best-fit lens models of components}\label{BFLMC}

\begin{figure}
\includegraphics[width=0.24\textwidth]{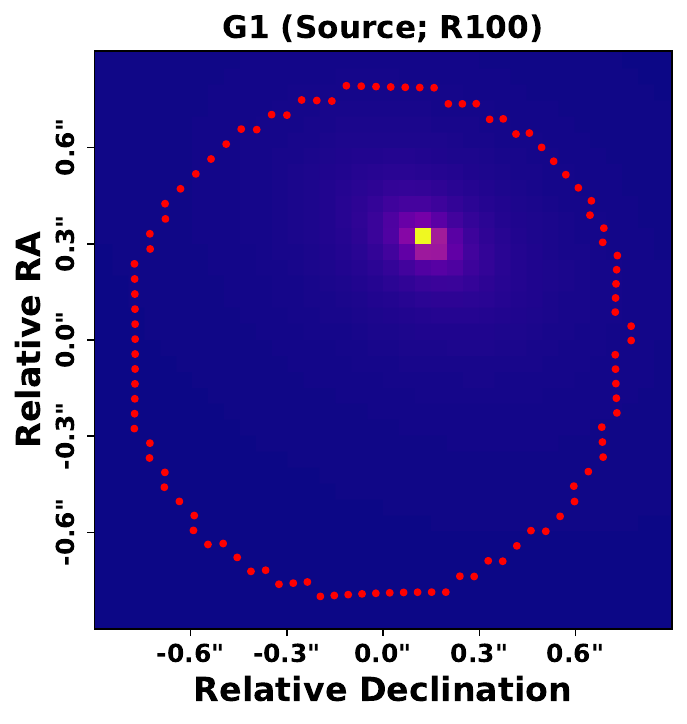}
\includegraphics[width=0.24\textwidth]{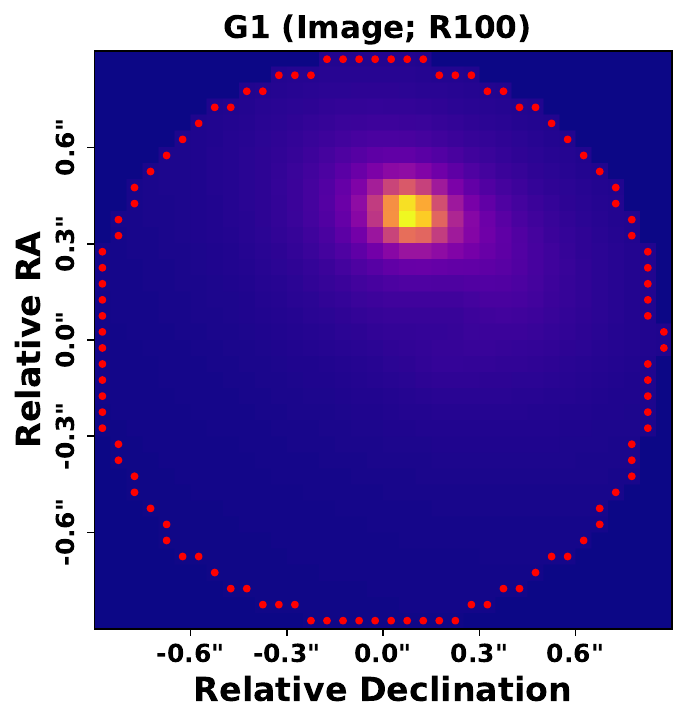}
\includegraphics[width=0.24\textwidth]{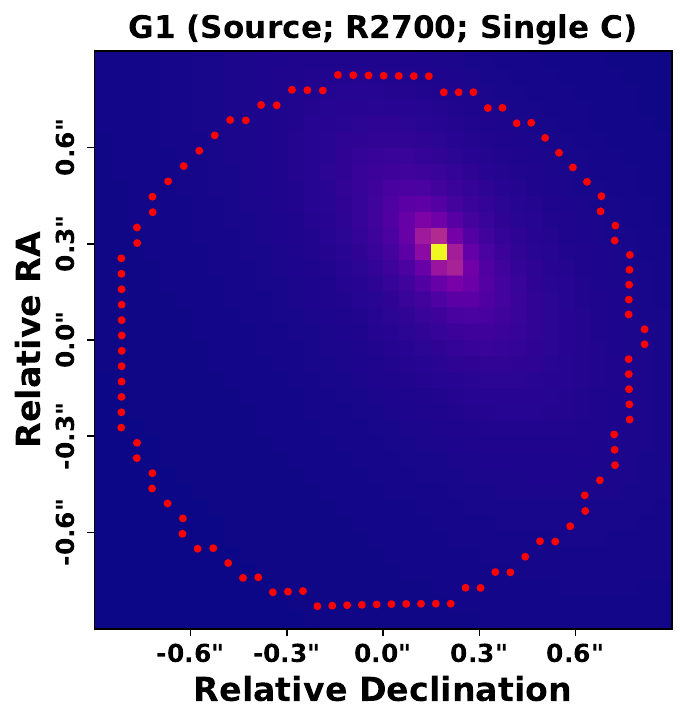}
\includegraphics[width=0.24\textwidth]{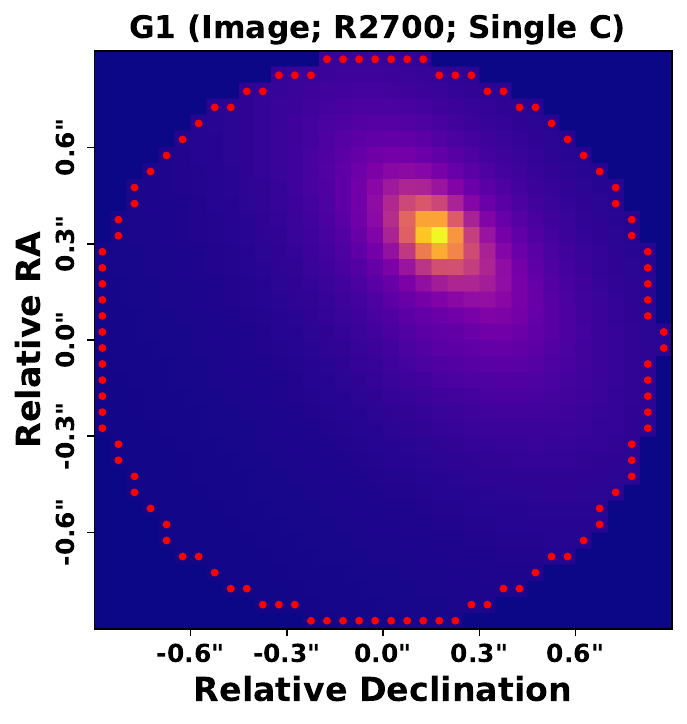}
\includegraphics[width=0.24\textwidth]{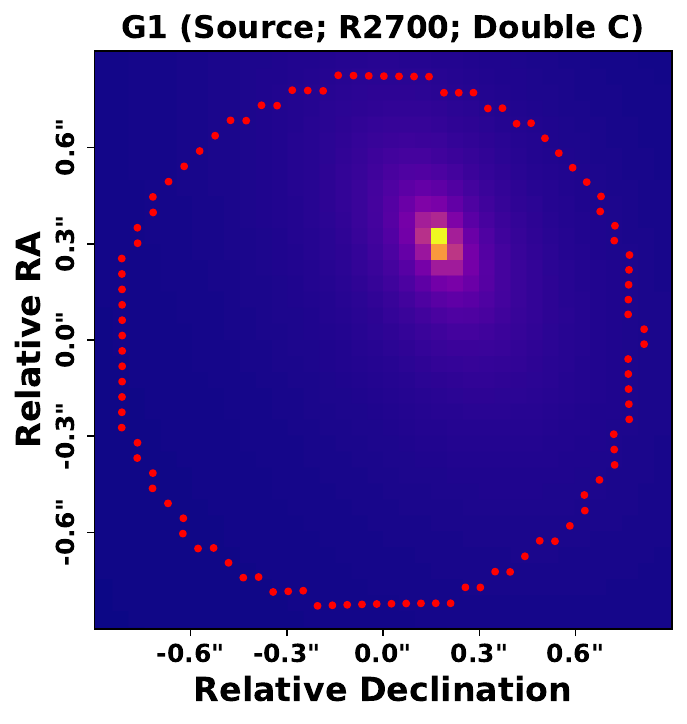}
\includegraphics[width=0.24\textwidth]{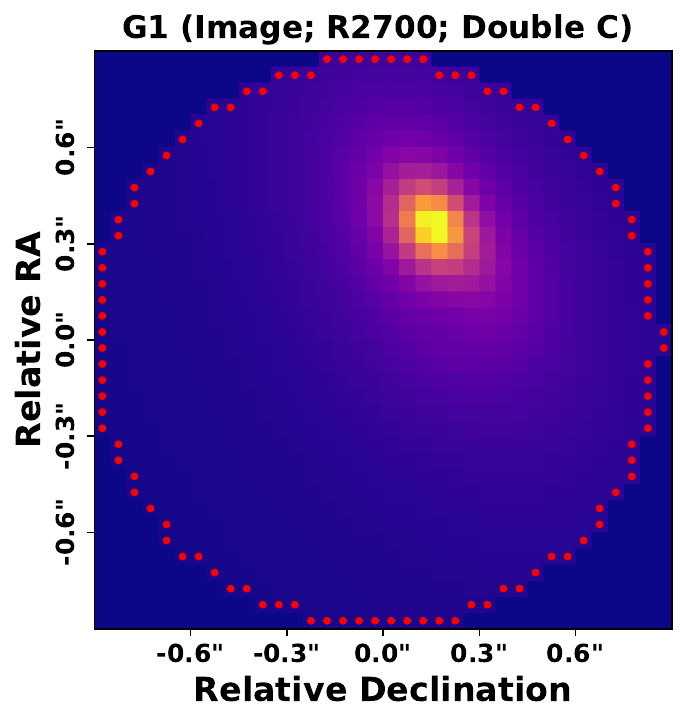}
\caption{Best-fit source-plane (left column) and image-plane \PAL models (right column) of G1 using the R100 cube ($\lambda=0.8-1.1\,\mu$m; top row) and the R2700 cube ($\lambda=4.80-4.85$) assuming that C is composed of a single component (middle row) or two components (lower row). Mass and light profiles are assumed to be S\'{e}rsic and isothermal ellipse profiles, respectively. The outline of the spatial mask is shown by red markers.}
\label{G1_source}
\end{figure}

\begin{figure}
\centering
\includegraphics[width=0.3\textwidth]{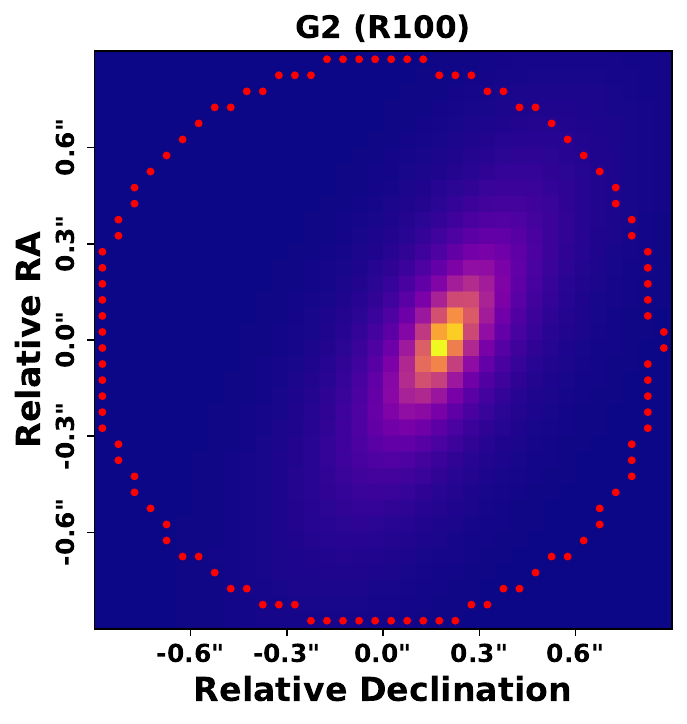}
\caption{Best-fit \PAL model of the unlensed source G2 using the R100 cube ($\lambda_{obs}=0.8-1.1\,\mu$m). Mass and light profiles are assumed to be S\'{e}rsic and isothermal ellipse profiles, respectively. The outline of the spatial mask is shown by red markers.}
\label{G2_source}
\end{figure}

\begin{figure*}
\includegraphics[width=0.33\textwidth]{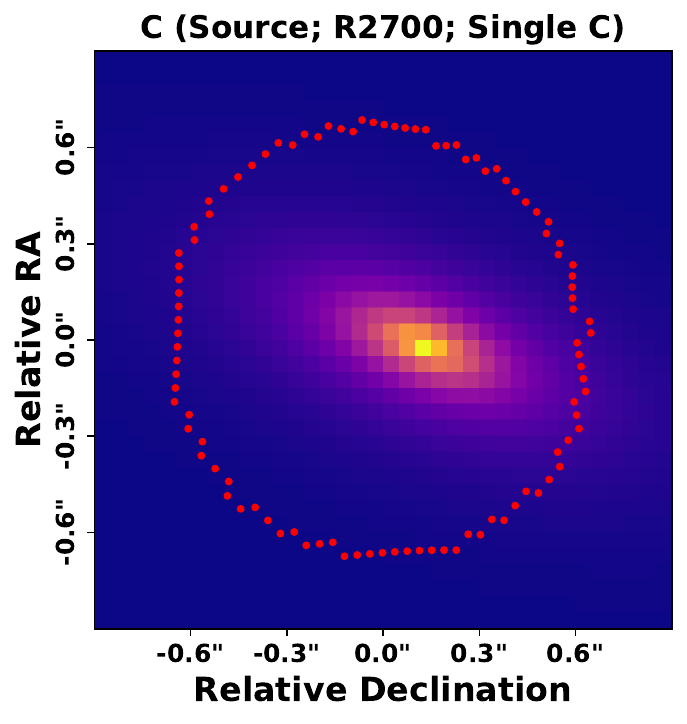}
\includegraphics[width=0.33\textwidth]{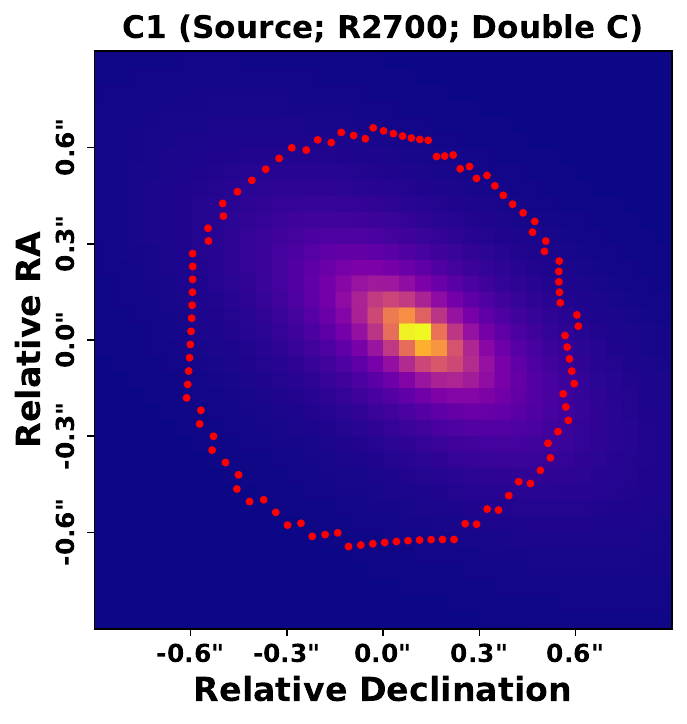}
\includegraphics[width=0.33\textwidth]{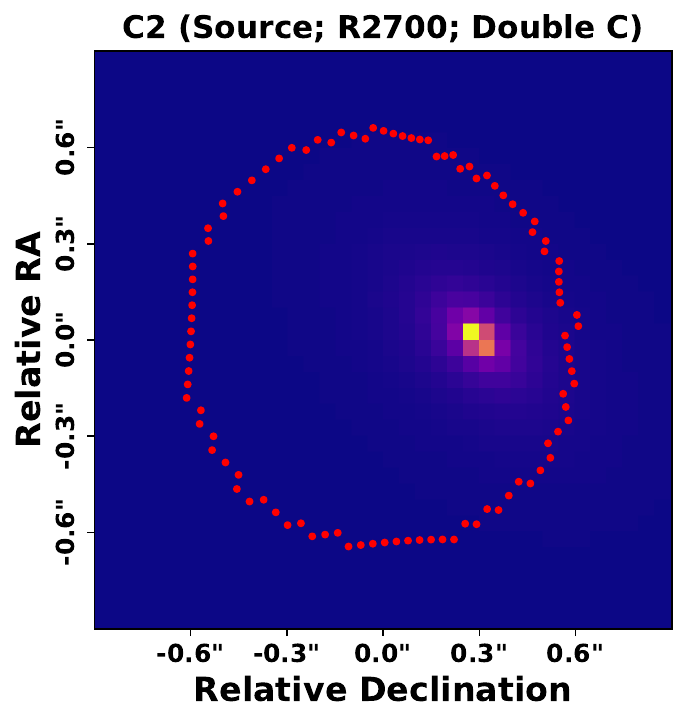}
\includegraphics[width=0.33\textwidth]{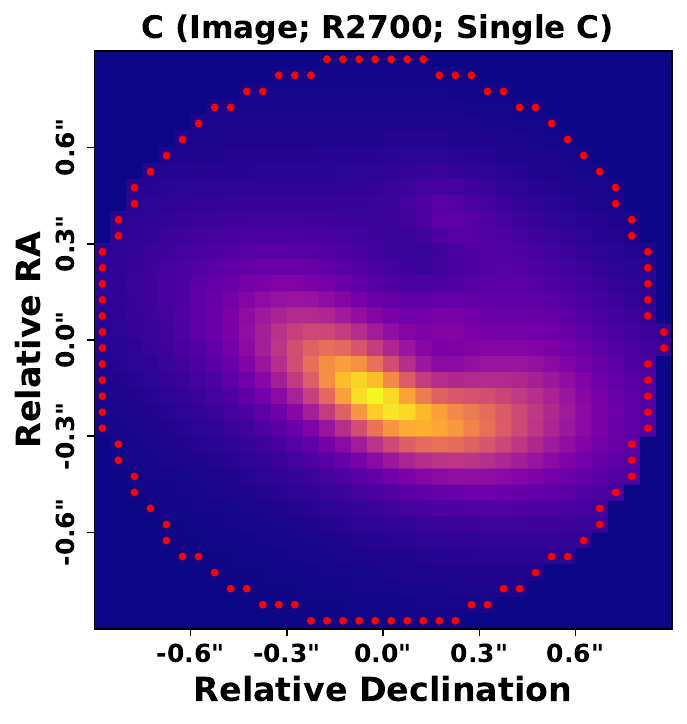}
\includegraphics[width=0.33\textwidth]{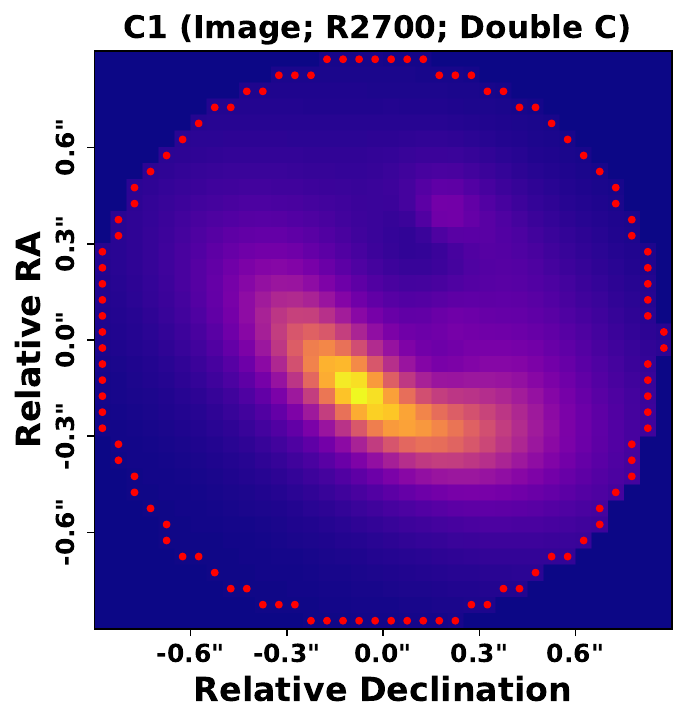}
\includegraphics[width=0.33\textwidth]{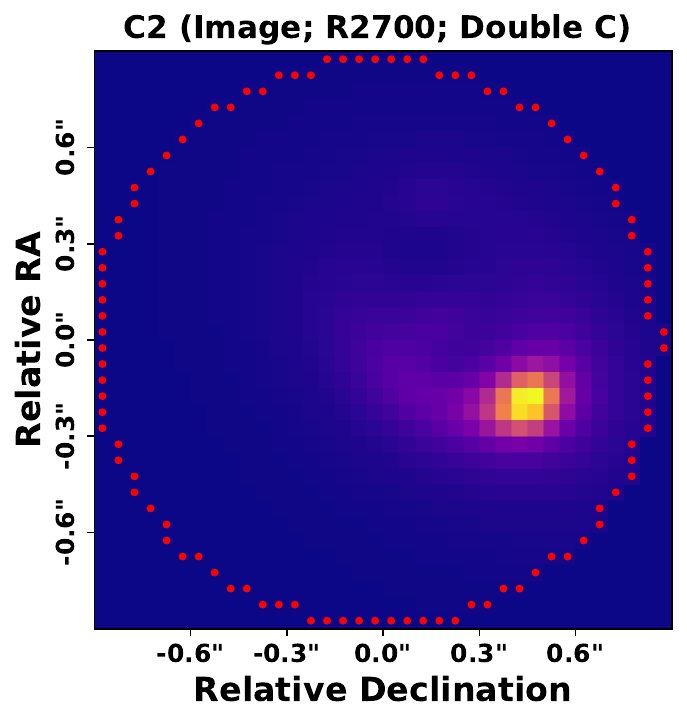}
\caption{Best-fit source-plane (top row) and image-plane \PAL models (bottom row) of component C using the R2700 cube ($\lambda_{obs}=4.80-4.85$) assuming that it is composed of a single component (left column) and the individual models of the two-component model (centre and right columns). Mass and light profiles are assumed to be S\'{e}rsic and isothermal ellipse profiles, respectively. The outline of the spatial mask is shown by red markers.}
\label{C_source}
\end{figure*}

In Section \ref{gravlenssec}, we used \PAL to examine the gravitational lensing effect created by the $z<6$ sources (G1 and G2) on the C source. The observed maps were compared with the best-fit image-plane model. Here, we show source- and image-plane maps of each individual component: G1 (Fig. \ref{G1_source}), G2 (Fig. \ref{G2_source}), and C (Fig. \ref{C_source}).

\section{Effects of shocks on line ratio diagnostics}\label{shocksec}
\begin{figure}
\centering
\includegraphics[width=0.5\textwidth]{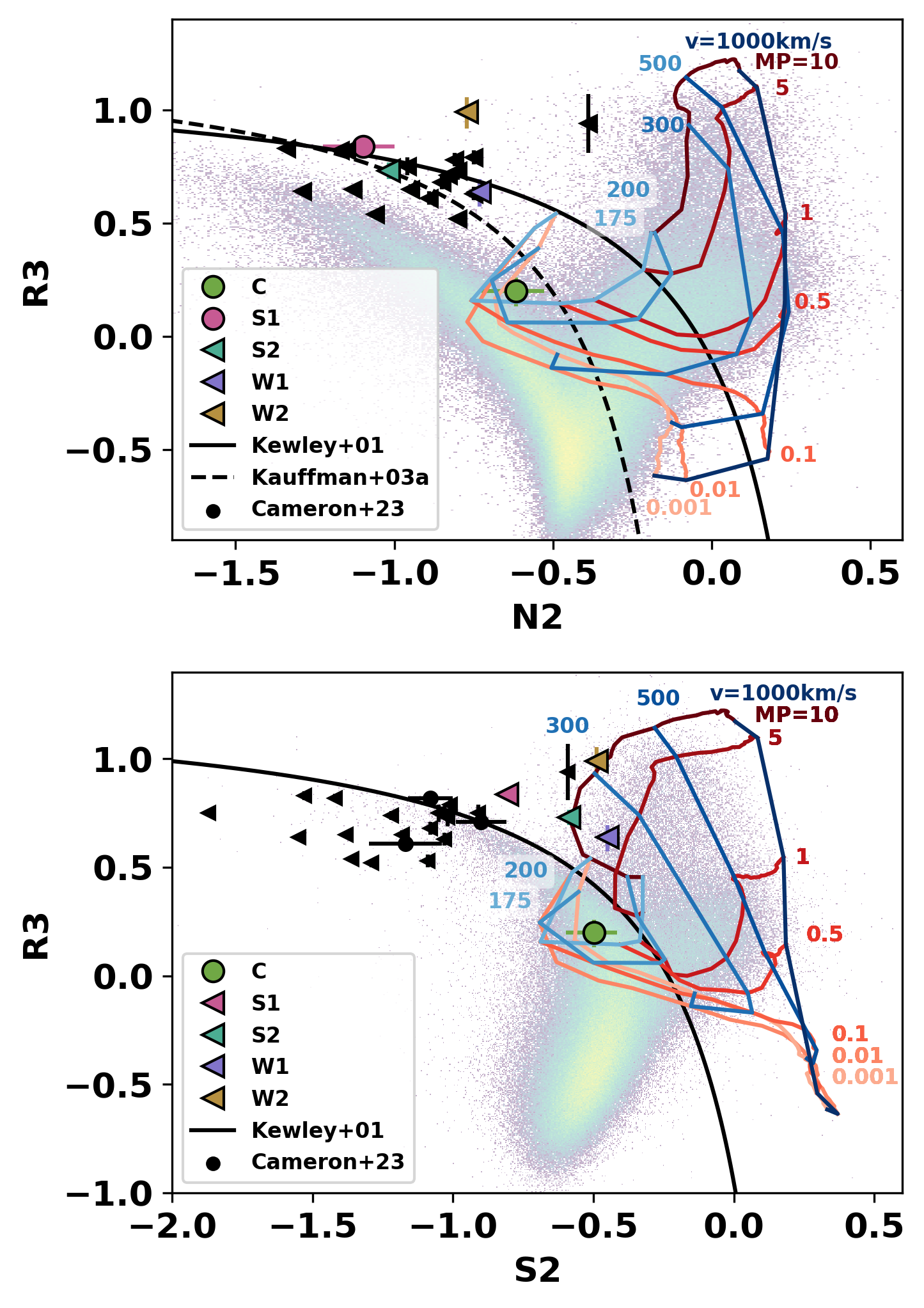}
\caption{[SII]-VO87 (top) and [NII]-BPT (lower) plots created using best-fit line fluxes for each source (see values in Table \ref{specfittab}), as seen in Figure \ref{BPT}. We now include the model grids from MAPPINGS III \citep{alle08}. Red-shaded lines display models with constant magnetic parameter (MP$=B/\sqrt{n}$ in units of $\mu$G\,cm$^{3/2}$), while blue-shaded lines represent models with constant shock velocity. We assume solar metallicity, low density ($\rm n\sim0.1\,cm^{-3}$), and no precursor.}
\label{BPTsh}
\end{figure}

In Section \ref{linerat}, we briefly discuss the applicability of low-redshift demarcation lines for each high-redshift source. Here, we explore the effects of shocks, as done for the local galaxy group Stephan's Quintet \citep{duar21}. This previous work applied the MAPPINGS III \citep{alle08} shock models, which contain expected line ratios for a variety of shock velocities, initial density, and magnetic field strength. The resulting model grids with our data are shown in Figure \ref{BPTsh}, adopting the same assumptions of solar metallicity, low density ($\rm n\sim0.1\,cm^{-3}$), and no shock precursor.

It is clear that the high-R3 sources (S, W) are not captured by these models, while component C falls within the low-velocity ($v\sim200$\,km\,s$^{-1}$), low-magnetic field ($\rm MP\sim0.001\,\mu$G\,cm$^{3/2}$) portion of the grid. Many emission line systems in Stephan's Quintet also fall into this region \citep{duar21}, which is interpreted as evidence for the presence of shocks. While this may also be true for component C of HFLS3, we note that these models were derived for relatively high metallicity ($Z'=1$), and lower metallicity models may be required for the high-redshift galaxies.

\end{appendix}

\label{lastpage}
\end{document}